\documentclass[reprint,prx,amsmath,amssymb,aps,floatfix]{revtex4-2}

\usepackage[dvipsnames]{xcolor}
\usepackage{booktabs}
\usepackage{tabularx}
\usepackage{dcolumn}
\usepackage{bm}
\usepackage{physics}
\usepackage{graphicx}
\usepackage[caption=false]{subfig}
\usepackage{placeins}
\usepackage[colorlinks=true]{hyperref}

\begin{document}

\title{Quantum Key Distribution via Charge Teleportation}
\author{Amir Yona}
\email[]{yonaamir@gmail.com}
\author{Yaron Oz}
\email[]{yaronoz@tauex.tau.ac.il}
\affiliation{School of Physics and Astronomy, Tel Aviv University, Ramat Aviv 69978, Israel}

\date{\today}

\begin{abstract}

We demonstrate that charge teleportation serves as a superior observable for Quantum Energy Teleportation (QET)-based cryptographic primitives. While following the LOCC protocol structure of earlier proposals~\cite{QKDbyQET}, we show that decoding key bits via local charge rather than energy provides exact bit symmetry and enhanced robustness: by  Local Operations and Classical Communication (LOCC) on an entangled many-body ground state, Alice’s one-bit choice steers the sign of a local charge shift at Bob, which directly encodes the key bit. Relative to energy teleportation schemes, the charge signal is bit-symmetric, measured in a single basis, and markedly more robust to realistic noise and model imperfections. We instantiate the protocol on transverse-field Ising models, star-coupled and one-dimensional chain,  obtain closed-form results for two qubits, and for larger systems confirm performance via exact diagonalization, circuit-level simulations, and a proof-of-principle hardware run. We quantify resilience to classical bit flips and local quantum noise, identifying regimes where sign integrity, and hence key correctness, is preserved. These results position charge teleportation as a practical, \textbf{low-rate QKD primitive} compatible with near-term platforms.

\end{abstract}

\maketitle

\section{Introduction}
\label{sec:introduction}

\if{
Quantum key distribution (QKD) promises information-theoretic security grounded in the laws of quantum mechanics \cite{Bennett1984, Ekert1991}. While canonical protocols such as BB84 \cite{Bennett1984, Bennett1984Original, Bennett1992, Shor2000, Mayers2001, Scarani2009, Bennett1992_exp} rely on transmitting and measuring single-qubit states, recent theory and experiments have opened alternative cryptographic primitives. Among these is quantum energy teleportation (QET)—the remote extraction of energy using only local operations and classical communication (LOCC) on a shared entangled ground state \cite{Hotta2011}. QET does not teleport a quantum state but rather the expectation value of a local observable. Demonstrations on superconducting hardware \cite{Ikeda2023} have established feasibility and, more broadly, motivated extensions from energy to other globally conserved quantities such as charge and current \cite{IkedaTeleportingCharge}—an approach we refer to as quantum observable teleportation.

Building on this idea, it was proposed to encode key bits in the sign of a teleported observable’s expectation value, controlled by a single classical bit chosen by Alice
\cite{QKDbyQET}. Such a mechanism offers a general framework for secure signaling: Alice deterministically controls a binary outcome at Bob’s location while using only LOCC. Because the required resource is a carefully prepared many-body ground state, the approach is naturally suited to applications where unconditional security is prioritized over throughput—precisely the trade-off emphasized in QKD. 

In this work we develop and analyze a QKD protocol based on charge teleportation. We study two implementations using transverse-field Ising models (TFIM): a star-coupled architecture in which Alice interacts with all other spins, and a nearest-neighbor one-dimensional chain with finite separation between Alice and Bob. Using numerical simulations and Qiskit-level circuit models, we compare charge vs energy teleportation and find that charge provides superior bit-symmetry, statistical stability, and scalability with system size, making it a strong candidate for practical QKD. We also quantify robustness to classical and quantum noise channels, identifying operating regimes in which secure key distribution remains viable.

\paragraph{Organization:} Section~\ref{sec:protocol} reviews the minimal observable-teleportation for a generic system and reintroduces the protocol designed in QKD by QET~\cite{QKDbyQET}. Section~\ref{sec:hamiltonians} presents the analysis for the two Hamiltonians we use for our model and the optimal control angles. Section~\ref{sec:numerical_sim} discusses numerical simulations of the protocol for both models, while Section~\ref{sec:qiskit_sim} provides validation of the numerical simulations using quantum circuits and presents results from execution on real quantum hardware. Section~\ref{sec:noise-errors} analyzes the protocol's resilience to different noise and error models, followed by a discussion and outlook in Section~\ref{sec:summary}.

}\fi

Quantum Key Distribution (QKD) promises unconditionally secure communication by leveraging the fundamental principles of quantum mechanics \cite{Bennett1984, Ekert1991}. While traditional protocols like the BB84 protocol~\cite{Bennett1984, Bennett1984Original, Bennett1992, Shor2000, Mayers2001, Scarani2009, Bennett1992_exp, Pirandola2020} rely on the transmission and measurement of single-qubit states, recent theoretical and experimental advances have opened new avenues for cryptographic tasks. Among these is Quantum Energy Teleportation (QET), a protocol that enables the extraction of energy from a distant location by performing only local operations and classical communication (LOCC) on a shared entangled ground state, without violating local energy conservation \cite{Hotta2011, Amico2008}. This process does not teleport a quantum state, but rather the expectation value of a local observable.

Recent work has demonstrated the experimental feasibility of QET on superconducting quantum hardware \cite{Ikeda2023} and generalized the protocol beyond energy to other globally conserved quantities, such as charge and current \cite{IkedaTeleportingCharge}. This generalization, termed Quantum Observable Teleportation, provides a versatile toolkit for manipulating local properties of many-body systems. Building on this, QKD based on QET~\cite{QKDbyQET} proposed a general framework where information is encoded in the sign of the teleported observable's expectation value. In this work, we demonstrate that charge teleportation serves as a qualitatively superior observable for this primitive. While retaining the established LOCC protocol structure, we show that decoding key bits via local charge provides exact bit symmetry, single-operator readout, and significantly enhanced robustness against model imperfections compared to energy-based schemes. More broadly, this mechanism establishes a general framework for secure quantum signaling, where Alice can deterministically control a binary outcome at Bob's location. However, given that the protocol's resource is a meticulously prepared many-body ground state, its practical application is best suited for tasks where unconditional security is prioritized over high data rates. QKD is the canonical example of such a task and thus forms the central focus of our investigation.

Focusing on this primary application, we develop and analyze a QKD protocol based on the teleportation of a local charge observable. We investigate the protocol's performance under two distinct interaction models for the underlying many-body system: a star-coupled Transverse Field Ising Model (TFIM), $H^{(1)} = J\sum_{k=1}^{N}X_{0}X_{k}+h\sum_{k=0}^{N}Z_{k}$, where a central qubit (Alice) interacts with all others, and a 1D nearest-neighbor TFIM, $H^{(2)} = J\sum_{k=1}^{N}X_{k-1}X_{k}+h\sum_{k=0}^{N}Z_{k}$. Through numerical simulations and Qiskit-based quantum circuit analysis~\cite{Amir_QKD_by_Charge_Teleportation}, we compare the efficacy and robustness of charge teleportation against energy teleportation for these models. Our findings show that charge teleportation exhibits superior symmetry, stability against statistical noise, and scalability with system size, making it a more promising candidate for practical QKD implementations. Furthermore, we analyze the protocol's resilience to various noise channels, identifying the operational regimes where secure key distribution remains viable.

We emphasize that this is a model-dependent, low-rate primitive suitable for short-range communication in specific many-body ground states, rather than a universal replacement for standard QKD protocols like BB84.

\paragraph{Organization:} Section~\ref{sec:protocol} reviews the minimal observable-teleportation for a generic system and reintroduces the protocol designed in~\cite{QKDbyQET}. Section~\ref{sec:hamiltonians} presents the analysis for the two Hamiltonians we use for our model. Section~\ref{sec:numerical_sim} discusses numerical simulations of the protocol, while Section~\ref{sec:qiskit_sim} provides validation of the numerical simulations using quantum circuits. Section~\ref{sec:noise-errors} analyzes the protocol's resilience to different noise and error models. Section~\ref{sec:security_analysis} then provides a cryptographic security analysis \cite{Renner2005}, quantifying the secret key rate under noise. The paper concludes with a summary and outlook in Section~\ref{sec:summary}.

\section{Protocol Definition}
\label{sec:protocol}

\subsection{Observables Teleportation:\texorpdfstring{\\}{line}General Framework}
\label{sec:generic_teleportation}

Quantum charge teleportation (QCT) extends Quantum Energy Teleportation (QET)~\cite{Hotta2011,Ikeda2023} from energy to symmetry charges. We consider two closely related $(N\!+\!1)$-site quantum systems prepared in a non-degenerate ground state $\ket{gs}$ of a Hamiltonian $H$, with Alice at site $0$ and Bob at site $N$~\cite{QKDbyQET,IkedaTeleportingCharge}. In QCT, the teleported quantity is a global conserved charge \footnote{The conserved charge can also be discrete, for which there is no continuity equation in the usual Noether sense.}—i.e., the generator of an exact symmetry with an associated continuity equation—so that Alice’s local measurement followed by classical communication (LOCC) enables Bob to apply a conditional local unitary that deterministically activates a charge signal at his site while the total charge remains conserved. This symmetry structure is essential: unlike arbitrary local observables, only globally conserved charges admit nonlocal redistribution via QET-style feedback without intermediate transport. Building on this principle, we analyze how Alice’s measurement basis and Bob’s conditional operation steer the sign and magnitude of Bob’s local charge, thereby realizing charge teleportation suitable for cryptographic primitives based on QET.

\subsubsection{Protocol Steps}
We explicitly define the protocol steps as follows~\cite{QKDbyQET}:

\begin{enumerate}
    \item \textbf{Ground State Preparation:} 
    The global system is initialized in the entangled ground state density matrix $\rho_{gs}=\ketbra{gs}{gs}$.

    \item \textbf{Alice’s Local Measurement:}
    Alice performs a local projective measurement on site $0$ given by:
    \begin{equation}
    P_A(b,\sigma_A)=\frac{1}{2}\left( 1-(-1)^b\sigma_A \right),\quad b\in\{0,1\},
    \end{equation}
    where $\sigma_A=\hat{n}\cdot\vec{\sigma}_A$ is randomly selected from a predefined set (for instance, $\{X_0,Y_0\}$), obtaining measurement outcome $b$.

    \item \textbf{Classical Communication:}
    Alice decides if she communicates her true measurement result ($b$) or its opposite ($b\oplus 1$). This decision is captured by a classical bit $a\in\{0,1\}$, where $a=0$ indicates the true result transmission, and $a=1$ its opposite. Alice communicates the classical bit $c=b\oplus a$ and the chosen measurement basis $\sigma_A$ to Bob. Note that Bob can deduce neither Alice's secret bit $a$ nor her raw measurement outcome $b$ from the public communication of $c$ and $\sigma_A$ alone.

    \item \textbf{Bob’s Conditional Rotation:}
    Based on the classical information received, Bob applies a conditional rotation on his subsystem at site $N$:
    \begin{equation}
    U_B(c,\sigma_B)=e^{-i\theta(-1)^c\sigma_B},
    \end{equation}
    with the operator $\sigma_B$ selected appropriately to match Alice's measurement basis, ensuring maximal teleportation efficiency.

    \item \textbf{Bob’s Observable Measurement:}
    Bob evaluates the expectation shift in his local observable $O_B$ resulting from the protocol:
    \begin{equation}
    \langle\Delta O_B\rangle=\Tr[\rho_B O_B]-\Tr[\rho_{gs} O_B],
    \end{equation}
    where the final density matrix at Bob's site is:
    \begin{align}
    \rho_B=\sum_{b}&{}U_B(b\oplus a,\sigma_B)P_A(b,\sigma_A)\rho_{gs} \nonumber\\
    &P_A(b,\sigma_A)U_B^\dagger(b\oplus a,\sigma_B)
    \end{align}
\end{enumerate}

\subsubsection{Generic Observable Analysis Including Bit \texorpdfstring{$a$}{a}}

Considering a generic observable $O_B$ commuting with Alice’s measurement, i.e., $[P_A,O_B]=0$, we derive explicitly:
\begin{equation}
\label{eq:deltaO-general}
\langle\Delta O_B\rangle=\frac{1}{2}\xi(1-\cos 2\theta)-\frac{1}{2}(-1)^a\eta\sin 2\theta,
\end{equation}
with parameters:
\begin{align}
&{}\xi=\Tr[\rho_{gs}\sigma_B O_B \sigma_B]-\Tr[\rho_{gs}O_B] \nonumber\\
&\eta=i\Tr[\rho_{gs}\sigma_A[O_B,\sigma_B]]
\end{align}

Bob chooses the optimal rotation angle assuming Alice sent her true result ($a=0$), thus maximizing the magnitude of $\langle\Delta O_B\rangle$:
\begin{equation}
\tan(2\theta)=\frac{\eta}{\xi}.
\end{equation}

This leads to the maximal teleported observable expectation shift:
\begin{equation}
\langle\Delta O_B\rangle=\frac{1}{2}\xi-\frac{1}{2}\frac{\xi^2+(-1)^a\eta^2}{\sqrt{\xi^2+\eta^2}}.
\end{equation}

This explicit dependence on $a$ demonstrates the crucial role of Alice's communication decision in secure quantum key distribution (QKD) protocols, and for $a=0$ it is simplified into the known form of:

\begin{equation*}
\langle\Delta O_B\rangle=\frac{1}{2}\xi-\frac{1}{2}\sqrt{\xi^2+\eta^2}.
\end{equation*}

\subsubsection{Complex Sum–Constructed Charge Operator}
\label{sec:complex-sum-constructed-operator}

In what follows we assume, as demanded earlier, that the teleported quantity is a \emph{global conserved charge}
\begin{equation}
[O_B,H]=0,
\end{equation}
with $O_B$ acting at (or supported on) Bob’s site and generated by an exact symmetry of $H$. We further allow $O_B$ to be represented as a complex sum of not-necessarily commuting components, $O_B=\sum_i O_i$, where the $O_i$ are physical observables but, in general, are \emph{not} themselves conserved charges: typically $[O_i,H]\neq 0$, and mutual non-commutativity $[O_i,O_j]\neq 0$ is permitted. Consequently, individual $O_i$ cannot be teleported as charges on their own; rather, the protocol teleports the conserved \emph{sum} $O_B$.

Hence, in order to teleport the charge $O_B$, we analyze each component $O_i$ under this same $\theta$, optimizing $O_B$'s teleportation. The component-wise teleported expectation shift is then
\begin{equation}
\label{eq:deltaOi}
\langle \Delta O_i \rangle
= \frac{1}{2}\,\xi_i
- \frac{1}{2}\,\frac{\xi_i\,\xi + (-1)^a\,\eta_i\,\eta}{\sqrt{\xi^2+\eta^2}},
\end{equation}
with
\begin{align}
\xi_i &= \Tr\!\left[\rho_{gs}\,\sigma_B O_i \sigma_B\right]-\Tr\!\left[\rho_{gs}\,O_i\right], \\
\eta_i &= i\,\Tr\!\left[\rho_{gs}\,[O_i,\sigma_B]\right].
\end{align}
Equation~\eqref{eq:deltaOi} makes explicit that (i) the protocol’s optimization is performed at the level of the conserved $O_B$ (not per $O_i$), and (ii) it reveals the inherent complexity in measuring and analyzing teleported observables composed of non-commuting terms, emphasizing the necessity of separate measurements for each operator component. Such complexity must be carefully considered in practical implementations and QKD protocol designs leveraging charge teleportation~\cite{QKDbyQET,IkedaTeleportingCharge}.

\subsubsection{Discussion and Implications}

The general teleportation protocol for observables discussed herein significantly extends the utility of quantum teleportation beyond energy, demonstrating broad applications for charge, current, and other observables~\cite{IkedaTeleportingCharge}. It lays essential foundations for novel quantum cryptographic techniques, such as QKD schemes, wherein the transmitted classical information (captured by the decision bit $a$) critically influences security and robustness.

Moreover, the presented analysis clarifies the conditions required for successful and optimal teleportation, highlighting the roles of measurement basis selection, classical feedback control, and operator commutation relations. It provides a unified framework facilitating future research and experiments in quantum information, quantum cryptography, and condensed matter physics employing quantum many-body systems~\cite{Hotta2011,Ikeda2023,QKDbyQET}.

\subsection{QKD with Charge teleportation}

We now apply the general observable teleportation protocol to the charge operator. The charge at Bob’s site (site $N$) is defined as:
\begin{equation}
Q_B = \frac{1}{2}(I + Z_N),
\end{equation}
which acts as a projector onto the logical \( \ket{0} \) state. This observable is Hermitian, diagonal in the computational basis, and naturally suited for logical bit interpretation in a QKD scheme.

From the general analysis in Section~\ref{sec:generic_teleportation}, we specialize the result for commuting observables to the case of $Q_B$, yielding:
\begin{equation}
\langle \Delta Q_B \rangle = -\frac{1}{2} \langle Z_N \rangle - \frac{1}{2} \frac{\langle Z_N \rangle^2 + (-1)^a \langle O_A O_B \rangle^2}{\sqrt{\langle Z_N \rangle^2 + \langle O_A O_B \rangle^2}},
\end{equation}
where the classical bit \(a \in \{0,1\}\) encodes whether Alice sent her true measurement result (\(a=0\)) or its complement (\(a=1\)).

The operator pair \((O_A, O_B)\) depends on the chosen measurement and feedback basis:
\begin{equation}
(O_A, O_B) =
\begin{cases}
(X_0, X_N) & \text{for } (\sigma_A, \sigma_B) = (X_0, Y_N), \\
(Y_0, Y_N) & \text{for } (\sigma_A, \sigma_B) = (Y_0, X_N).
\end{cases}
\end{equation}

The teleported expectation value $\langle \Delta Q_B \rangle$ is positive or negative depending on $a$, which enables secure key distribution:
\begin{equation}
\langle \Delta Q_B \rangle
\begin{cases}
< 0 & \Rightarrow \text{logical bit } 1, \\
> 0 & \Rightarrow \text{logical bit } 0.
\end{cases}
\end{equation}

This charge-based observable was specifically chosen for this protocol. As we will demonstrate, its perfectly symmetric response to Alice's classical bit $a$ provides a clean, unambiguous mapping to a logical key bit, forming the basis for a more robust and reliable QKD protocol compared to other observables like energy.

This behavior is at the heart of QKD via observable teleportation: Bob cannot distinguish which logical value was intended unless he knows Alice's classical bit $a$. The security stems from the fact that the expectation value's sign flips based on $a$, yet neither the quantum state nor the classical communicated bit alone is sufficient to decode the logical value.

\section{Hamiltonians}
\label{sec:hamiltonians}

\subsection{Star-Coupled TFIM Hamiltonian}

\subsubsection{Star-Coupled TFIM Hamiltonian Definition}
\label{sec:alice_hamiltonian}

We consider the first type of Hamiltonian, denoted \( H^{(1)} \), in which a central site (denoted as site $0$ - Alice's site) interacts locally with all other spins via transverse Ising-type couplings (star interaction). The system is defined as:

\begin{equation}
H^{(1)} = J \sum_{k=1}^{N} X_0 X_k + h \sum_{k=0}^{N} Z_k,
\label{eq:H1}
\end{equation}

where the global state resides in the Hilbert space \( \mathcal{H} = (\mathbb{C}^2)^{\otimes (N+1)} \). The site \( 0 \) is assigned to Alice, while site \( N \) is assigned to Bob. The corresponding localized Hamiltonian on Bob’s side, derived from the global Hamiltonian by isolating terms involving site \( N \), is:

\begin{equation}
H_B^{(1)} = h Z_N + J X_0 X_N.
\end{equation}

\subsubsection{Teleportation Protocol with \texorpdfstring{\( \sigma_A = X_0 \)}{sigmaAX0}}

In this configuration, Alice measures her qubit (site 0) using the Pauli operator \( X_0 \), and sends the result \( b \in \{0,1\} \) to Bob. The measurement corresponds to the projector:
\begin{equation}
P_A(b) = \frac{1}{2} \left( I - (-1)^b X_0 \right).
\end{equation}
Upon receiving the classical bit \( b \), Bob applies the conditional rotation:
\begin{equation}
U_B(b) = \exp\left( -i \theta (-1)^b Y_N \right),
\end{equation}
where \( \theta \) is a protocol-dependent angle chosen to optimize the extraction of the target observable.

The choice \( \sigma_A = X_0 \) is dictated by a structural requirement of the teleportation protocol. In order for Bob’s observable to exhibit a nontrivial response following Alice’s measurement, it is necessary that the measurement projector \( P_A(b) \) commute with Bob’s local Hamiltonian:
\begin{equation}
[H_B, P_A(b)] = 0.
\end{equation}
But in the current case, Bob's local Hamiltonian is
\begin{equation*}
H_B^{(1)} = h Z_N + J X_0 X_N.
\end{equation*}
This Hamiltonian involves the operator \( X_0 \), and thus commutes with the projector \( P_A(b) \) if and only if it is defined using \( \sigma_A = X_0 \). If we were to choose, for example, \( \sigma_A = Y_0 \), then \( P_A(b) \) would no longer commute with \( H_B^{(1)} \), violating a core requirement of the protocol and destroying the condition for coherent observable transfer.

This commutation condition ensures that Alice’s measurement introduces no direct disturbance into Bob’s local energy configuration beyond what is mediated by classical communication and pre-existing entanglement. It is therefore essential for enabling observable teleportation in this setting.

\subsubsection{Evaluation of Energy Extraction for \texorpdfstring{\( H_B^{(1)} \)}{HB1}}

We analyze the teleportation of Bob’s local energy observable \( H_B^{(1)} = h Z_N + J X_0 X_N \). Following standard derivations, the relevant quantities for computing the extracted energy are:

\begin{equation}
    \xi = -2h \langle Z_N \rangle - 2J \langle X_0 X_N \rangle = -2\langle H_B^{(1)} \rangle \\
\end{equation}
\begin{equation}
    \eta = 2h \langle X_0 X_N \rangle - 2J \langle Z_N \rangle
\end{equation}

Hence, assuming \( a = 0 \), i.e., Bob uses the correct bit from Alice (no flip), the expected value of Bob’s energy change, optimized over \( \theta \), becomes:

\begin{equation}
\label{eq:delta_HB1}
    \begin{aligned}
        \langle \Delta H_B^{(1)} \rangle =&{} -\left( h\langle Z_N\rangle + J\langle X_0X_N\rangle\right) \nonumber\\
        & - \sqrt{\left( h^2 + J^2\right)\left( \langle Z_N\rangle^2 + \langle X_0X_N\rangle^2 \right)}
    \end{aligned}
\end{equation}

\subsubsection{Interpretation and Observable Behavior}

Based on the properties of the ground state for \( H^{(1)} \), it holds that \( \langle X_0 \rangle = \langle X_N \rangle = 0 \), and \( \langle Z_N \rangle \in [-1, 1] \). Hence, all contributions to observables teleportation depend on the correlations \( \langle X_0 X_N \rangle \) and \( \langle Z_N \rangle \). Notably, in the limit \( J \ll h \), we recover:

\begin{equation*}
\langle \Delta H_B^{(1)} \rangle \to 0, \quad \langle \Delta Q_B \rangle \to 0,
\end{equation*}

and in the strong coupling limit \( J \gg h \), we again obtain:

\begin{equation*}
\langle \Delta H_B^{(1)} \rangle \to 0, \quad \langle \Delta Q_B \rangle \to -\frac{1}{2} (-1)^a,
\end{equation*}

where \( a \in \{0,1\} \) is chosen by Alice according to the described protocol.

This confirms that although energy teleportation vanishes in the asymptotic regimes, charge teleportation may still retain bit-dependent signatures, making it more suitable for QKD applications. These observations match the simulations presented in Sections \ref{sec:numerical_sim} and \ref{sec:qiskit_sim} of this work.

\subsection{Nearest-Neighbors Hamiltonian}
\label{sec:nn_hamiltonian}

\subsubsection{Nearest Neighbors Hamiltonian}

We now consider the second type of Hamiltonian, denoted \( H^{(2)} \), in which each spin interacts only with its nearest neighbors, forming a 1D transverse field Ising chain. The system Hamiltonian is given by:

\begin{equation}
H^{(2)} = J \sum_{k=1}^{N} X_{k-1} X_k + h \sum_{k=0}^{N} Z_k
\label{eq:H2}
\end{equation}

where, again, site \( 0 \) is assigned to Alice and site \( N \) to Bob. The Hilbert space is \( \mathcal{H} = (\mathbb{C}^2)^{\otimes (N+1)} \). The localized Hamiltonian for Bob’s site (site \( N \)) includes the interaction with its nearest neighbor and is given by:

\begin{equation}
H_B^{(2)} = h Z_N + J X_{N-1} X_N.
\end{equation}

\paragraph{Note on the Case \( N = 1 \):} For a system with $2$ sites, the global Hamiltonian \( H^{(2)} \) reduces to:

\begin{equation*}
H^{(2)} = J X_0 X_1 + h (Z_0 + Z_1),
\end{equation*}

which is identical to \( H^{(1)} \). In this case, the measurement basis must again be \( \sigma_A = X_0 \) to satisfy the commutation requirement. Therefore, the flexibility of using \( \sigma_A = Y_0 \) is only applicable when \( N \geq 2 \).

\subsubsection{Teleportation Protocol with \texorpdfstring{\( \sigma_A = X_0 \)}{sigmaAX0}}

For \( N \geq 2 \), Alice may choose \( \sigma_A = X_0 \), leading to the following measurement and rotation steps:

\begin{equation}
P_A^{(X)}(b) = \frac{1}{2} \left( I - (-1)^b X_0 \right),
\end{equation}
\begin{equation}
U_B^{(X)}(b) = \exp\left( -i \theta (-1)^b Y_N \right),
\end{equation}

where \( b \in \{0,1\} \) is the classical bit sent from Alice to Bob. The choice \( \sigma_A = X_0 \) is compatible with \( H_B^{(2)} \) because the operator \( X_0 \) appears only indirectly via the entanglement structure of the ground state, and the measurement projector \( P_A^{(X)}(b) \) commutes with \( H_B^{(2)} \) due to the spatial separation.

\subsubsection{Teleportation Protocol with \texorpdfstring{\( \sigma_A = Y_0 \)}{sigmaAY0}}

Alternatively, Alice may choose \( \sigma_A = Y_0 \) (such that \( \sigma_B = X_N \) is chosen to correspond with Alice's basis), using the measurement projector:

\begin{equation}
P_A^{(Y)}(b) = \frac{1}{2} \left( I - (-1)^b Y_0 \right),
\end{equation}
\begin{equation}
U_B^{(Y)}(b) = \exp\left( -i \theta (-1)^b X_N \right).
\end{equation}

In contrast to the \( H^{(1)} \) case, this choice is valid here because \( H_B^{(2)} \) does not contain \( X_0 \) explicitly. Thus, both \( P_A^{(X)}(b) \) and \( P_A^{(Y)}(b) \) commute with \( H_B^{(2)} \), and the protocol remains structurally consistent.

\subsubsection{Evaluation of Energy Extraction for \texorpdfstring{\( H_B^{(2)} \)}{HB2}}

\subsubsection*{Case 1: \texorpdfstring{\( \sigma_A = X_0 \)}{sigmaAX0}}

In this configuration, the protocol is defined by Alice’s measurement using \( X_0 \) and Bob’s rotation with \( Y_N \). The protocol analysis yields the following:

\begin{align}
\label{eq:HB2_xi_eta}
\xi &= -2h \langle Z_N \rangle - 2J \langle X_{N-1} X_N \rangle = -2 \langle H_B^{(2)} \rangle \\
\eta &= 2h \langle X_0 X_N \rangle - 2J \langle X_0 X_{N-1} Z_N \rangle
\end{align}

These expectation values are taken in the ground state of the global Hamiltonian. The teleported energy is then given by the expression in \eqref{eq:delta_HB2}.

\begin{widetext}
    \centering
    {\scriptsize
    \begin{equation}
    \label{eq:delta_HB2}
    \begin{aligned}
      &\langle \Delta H_B^{(2)} \rangle = -h\langle Z_N \rangle - J\langle X_{N-1} X_N \rangle \\
            & - \frac{
                h^2\left(\langle Z_N \rangle^2 + (-1)^a\langle X_0X_N \rangle^2\right)
                - 2hJ\left(\langle Z_N \rangle\langle X_{N-1}X_N \rangle - (-1)^a\langle X_0X_N \rangle\langle X_0X_{N-1}Z_N \rangle\right)
                + J^2\left(\langle X_0X_N \rangle^2 + (-1)^a\langle X_0X_{N-1}Z_N \rangle^2\right)
            }{
            \sqrt{
                h^2\left(\langle Z_N \rangle^2 + \langle X_0X_N \rangle^2\right)
                - 2hJ\left(\langle Z_N \rangle\langle X_{N-1}X_N \rangle - \langle X_0X_N \rangle\langle X_0X_{N-1}Z_N \rangle\right)
                + J^2\left(\langle X_0X_N \rangle^2 + \langle X_0X_{N-1}Z_N \rangle^2\right)
            }}
    \end{aligned}
    \end{equation}
    }
    Energy extracted by Bob for $H^{(2)}$ for $\sigma_A = X_0$
\end{widetext}

\subsubsection*{Case 2: \texorpdfstring{\( \sigma_A = Y_0 \)}{sigmaAY0}}

When Alice measures using \( Y_0 \), the analysis proceeds similarly but with distinct correlation terms:

\begin{align}
\xi &= -2h \langle Z_N \rangle \\
\eta &= -2h \langle Y_0 X_N \rangle
\end{align}

So the optimal teleported energy becomes:

\begin{equation}
\label{eq:delta_HB2_Y}
    \begin{aligned}
        \langle \Delta H_B^{(2)} \rangle &{}= -h\langle Z_N \rangle - h\frac{\langle Z_N \rangle^2 + (-1)^a\langle Y_0Y_N \rangle^2}{\sqrt{\langle Z_N \rangle^2 + \langle Y_0Y_N \rangle^2}} \nonumber\\
        &\xrightarrow[]{a=0} -h\langle Z_N \rangle - h\sqrt{\langle Z_N \rangle^2 + \langle Y_0Y_N \rangle^2}
    \end{aligned}    
\end{equation}

\subsubsection*{Summary and Remarks}

Unlike the star-coupled Hamiltonian \( H^{(1)} \), where only \( \sigma_A = X_0 \) is valid due to commutation constraints, the structure of \( H^{(2)} \) permits both \( X_0 \) and \( Y_0 \) as valid measurement observables. The corresponding expressions show distinct operator correlations, enabling energy teleportation through different symmetry channels in the entangled ground state. This is important for security reasons in order to prevent an attacker (Eve) from learning, using weak measurements, about imperfections of the shared state between Alice and Bob.

These exact expressions form the basis for evaluating and comparing the performance of QET or QKD protocols across different measurement strategies and coupling regimes.

\subsubsection{Interpretation and Observable Behavior}

In contrast to \( H^{(1)} \), the Hamiltonian \( H^{(2)} \) generates indirect entanglement between Alice and Bob only through intermediate sites. As a result, correlations are weaker but non-zero for \( N \geq 2 \). This permits observable teleportation using either basis \( X_0 \) or \( Y_0 \), although the efficiency of the protocol depends on the specific entangled correlations in the ground state.

In particular, numerical simulations indicate that the two bases lead to qualitatively similar but quantitatively distinct energy and charge profiles as functions of the coefficients \( J/h \).

\subsection{Analytical Analysis}
\subsubsection{Feasibility of Analytical Solutions}

As discussed earlier, the Hilbert space for the system scales as \( 2^{N+1} \), making analytical treatment rapidly intractable as \( N \) increases. Specifically, for \( N = 2 \) (i.e., a 3-site system), one must diagonalize an \( 8 \times 8 \) Hamiltonian matrix. Although symmetry considerations allow decomposition into smaller blocks—e.g., a \( 2 \times 2 \) and a \( 6 \times 6 \) block—there is no closed-form solution for the ground state for arbitrary values of \( h \) and \( J \) when \( N \geq 2 \).

Therefore, complete analytical derivation of the ground state and the corresponding teleportation dynamics is feasible only for \( N = 1 \), corresponding to a 2-site system. This serves as a fundamental benchmark for validating numerical simulations and establishing intuition for the energy and charge teleportation mechanisms.

\subsubsection{\texorpdfstring{$N = 1$}{N1} Case: Exact Ground State and Teleportation Dynamics}

For the 2-site system (\( N = 1 \)), the Hamiltonian takes the form:
\begin{equation}
    H = h(Z_0 + Z_1) + J X_0 X_1.
\end{equation}

Importantly, in this minimal model, the two types of interaction Hamiltonians coincide:
\begin{equation*}
    H^{(1)}_{N=1} = H^{(2)}_{N=1}.
\end{equation*}

This symmetry ensures that the protocol and its analytical consequences are identical in both cases. Following the derivation in \cite{Hotta2011,Ikeda2023}, the exact ground state is:
\begin{equation}
    \ket{gs} = \frac{1}{\sqrt{2E_0}} \left( -\sqrt{E_0 - h} \ket{00} + \sqrt{E_0 + h} \ket{11} \right),
\end{equation}
where the auxiliary variables are defined as:
\begin{equation}
    J = 2k, \quad E_0 = \sqrt{h^2 + k^2}, \quad r = \frac{k}{E_0}.
\end{equation}

Applying the teleportation protocol described in previous sections, the analytically calculated expectation values for the change in Bob’s local energy and charge observables are:
\begin{align}
    \langle \Delta H_B \rangle &= h + Jr - \frac{ \left( h^2 + (-1)^a J^2 \right) \left( 1 + (-1)^a r^2 \right) }{ \sqrt{ (h^2 + J^2)(1 + r^2) } }, \\
    \langle \Delta Q_B \rangle &= \frac{1}{2} \left( 1 - \frac{ 1 + (-1)^a r^2 }{ \sqrt{1 + r^2} } \right).
\end{align}

These expressions explicitly capture the dependence of the teleportation process on the coupling ratio \( J/h \). The charge teleportation result \( \langle \Delta Q_B \rangle \) is dimensionless and solely determined by this ratio, while the energy teleportation \( \langle \Delta H_B \rangle \) is expressed in natural energy units, which can be scaled using either \( h \) or \( J \) depending on context.

\subsubsection{Interpretation and Implications}

The analytical solution for \( N=1 \) provides crucial insight into the behavior of both energy and charge teleportation in the presence of entanglement. While the model is too simple to capture long-range correlations present in larger systems, it nonetheless illustrates:

\begin{itemize}
    \item The nontrivial dependence of teleportation on the measurement bit \( a \), evident in the sign alternation terms \( (-1)^a \).
    \item The distinct scaling behaviors of energy versus charge teleportation, suggesting potential for protocol optimization depending on the physical observable of interest.
    \item That charge teleportation may exhibit larger bit-dependent contrast even in the strong-coupling regime \( J \gg h \), a key feature for QKD applicability.
\end{itemize}

This analytical benchmark not only validates the broader numerical framework used for larger systems (\( N \geq 2 \)), but also provides a reference point for identifying operational regimes and for calibrating hardware implementations.

\section{Numerical Simulation}
\label{sec:numerical_sim}

\subsection{Simulation Results: Star-Coupled Hamiltonian}

In Section~\ref{sec:alice_hamiltonian}, we introduced the star-type interaction Hamiltonian \( H^{(1)} \), where Alice interacts uniformly with all other sites via transverse Ising-type couplings (equation~\ref{eq:H1}).
This Hamiltonian structure enables a central-node interaction pattern ideal for parallel entanglement distribution and multi-party quantum key distribution (QKD).

We now turn to the numerical results for both energy and charge teleportation protocols implemented under \( H^{(1)} \), evaluating Bob's extracted observable expectation values as a function of the coupling strength \( J \), with fixed \( h = 1 \). The extracted values are compared against the analytical predictions derived earlier (see Appendix~\ref{appendix:charge_teleportation}).

\begin{figure}[ht]
    \centering
    \subfloat[\( N = 1 \)]{
        \includegraphics[width=0.45\columnwidth]{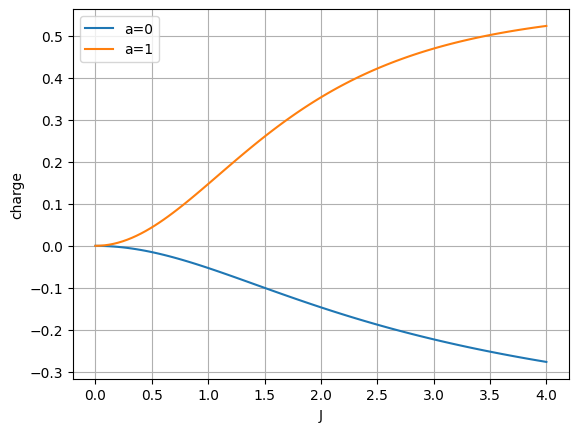}
    }
    \hfill
    \subfloat[\( N = 2 \)]{
        \includegraphics[width=0.45\columnwidth]{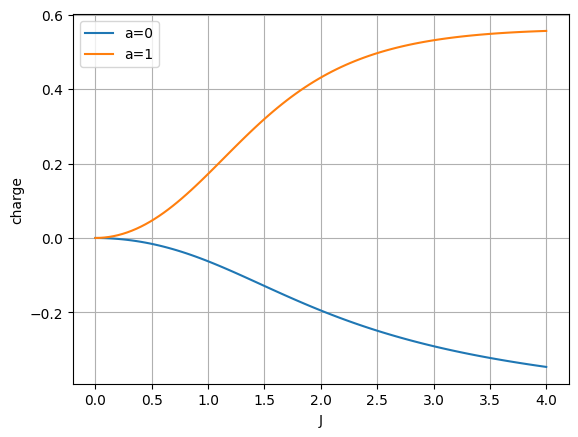}
    }

    \medskip
    
    \subfloat[\( N = 3 \)]{
        \includegraphics[width=0.45\columnwidth]{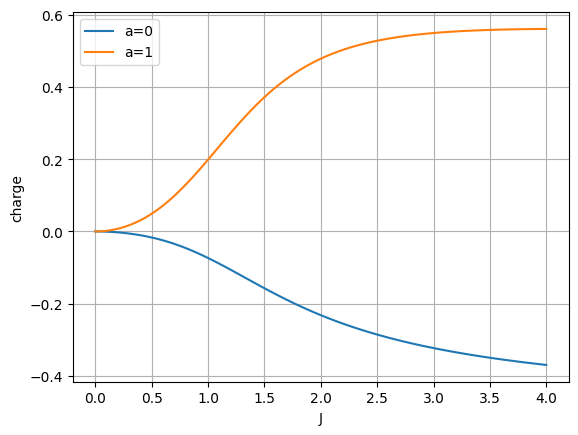}
    }
    \hfill
    \subfloat[\( N = 4 \)]{
        \includegraphics[width=0.45\columnwidth]{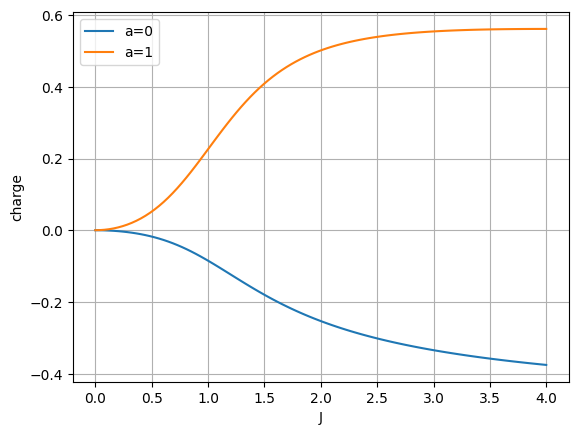}
    }
    \caption{\( Q_B \) vs. \( J \), for the star-coupled model \( H^{(1)} \), with \( h=1 \) and \( \sigma_A = X_0 \).}
    \label{fig:alice_x_vs_j}
\end{figure}

\subsubsection{Teleportation Behavior vs Coupling Strength \texorpdfstring{\( J \)}{J}}

Figure~\ref{fig:alice_x_vs_j} shows the simulated expectation values of Bob’s measured observables for the charge protocol, plotted against the interaction strength \( J \). In accordance with theoretical predictions, we observe a clear distinction between the behavior under energy and charge teleportation:
\begin{itemize}
    \item In the \textbf{energy teleportation protocol}, Bob's observable exhibits a nontrivial and asymmetric profile, with maximal negative expectation value at an intermediate value of \( J \), consistent with the optimal teleportation regime identified in Refs.~\cite{IkedaTeleportingCharge, QKDbyQET} (see Appendix Fig.~\ref{fig:alice_vs_hJ} for the full energy data).
    \item In the \textbf{charge teleportation protocol}, the observable expectation values are symmetric and exhibit the expected mirror-image pattern under bit flipping, aligning with the predictions in the teleportation formalism.
\end{itemize}

\subsubsection{Energy vs. Charge Teleportation: Comparative Discussion}

Although energy and charge teleportation protocols share the same structure of local operations and classical communication (LOCC), they differ substantially in their physical implementation and resulting behavior:

\begin{itemize}
    \item \textbf{Energy teleportation} involves local injection and extraction of energy from the entangled ground state. Its effectiveness depends sensitively on the commutation relations between local Hamiltonians, often leading to asymmetric behavior under bit-flip operations. This asymmetry, as seen in our simulations, may pose challenges for extending the protocol to multi-party or long-range quantum networks.
    
    \item \textbf{Charge teleportation}, in contrast, exhibits more symmetric and robust behavior. The observable expectation values at Bob’s site respond linearly and predictably to Alice's bit, enabling straightforward mapping to key bits in QKD. This symmetry not only improves interpretability but also offers resilience to noise and misalignment in measurement bases.
    
    \item \textbf{Scalability of extracted signal—} Importantly, since both energy and charge observables were rescaled to arbitrary units, we may compare their relative magnitudes. Our simulations reveal that the charge expectation value at Bob’s site remains approximately constant as the system size \( N \) increases, whereas the extracted energy diminishes. This suggests that charge teleportation may be more favorable for large-scale implementations, especially in distributed QKD architectures.
    This suggests that charge teleportation may be more favorable for large-scale implementations, especially in distributed QKD architectures. Furthermore, the perfect symmetry of the charge signal is a decisive advantage for cryptography; it ensures that the logical '0' and '1' are equally distinguishable, preventing any potential bias in the raw key that an eavesdropper might exploit.
\end{itemize}

These findings corroborate the observations in~\cite{IkedaTeleportingCharge, QKDbyQET}, where the symmetry and noise robustness of charge-like observables were highlighted. Furthermore, since energy teleportation can depend on subtle inter-site Hamiltonian partitionings (cf. Eq.~(8) in~\cite{QKDbyQET}), it might be less generalizable than charge teleportation for QKD networks involving multiple recipients.

\begin{figure}[ht]
    \centering
    \subfloat[\( N = 2 \), \( \sigma_A = X_0 \)]{
        \includegraphics[width=0.45\columnwidth]{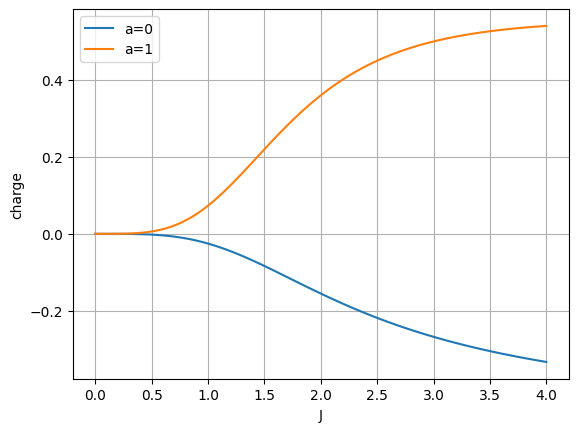}
    }
    \hfill
    \subfloat[\( N = 2 \), \( \sigma_A = Y_0 \)]{
        \includegraphics[width=0.45\columnwidth]{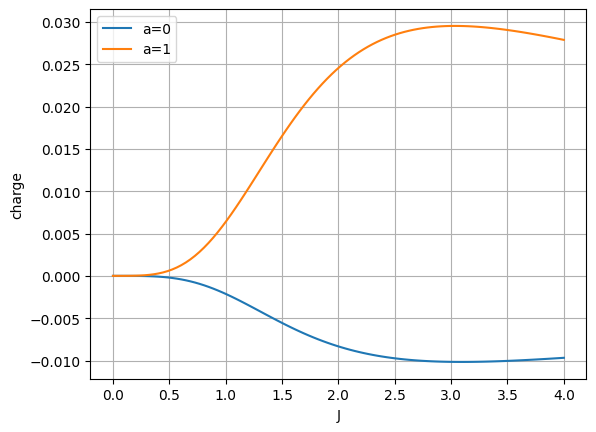}
    }

    \medskip

    \subfloat[\( N = 3 \), \( \sigma_A = X_0 \)]{
        \includegraphics[width=0.45\columnwidth]{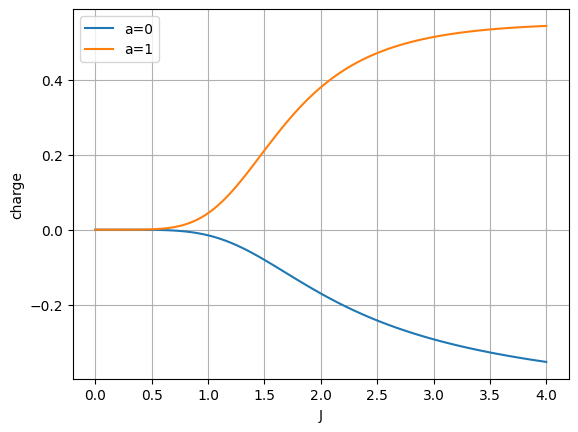}
    }
    \hfill
    \subfloat[\( N = 3 \), \( \sigma_A = Y_0 \)]{
        \includegraphics[width=0.45\columnwidth]{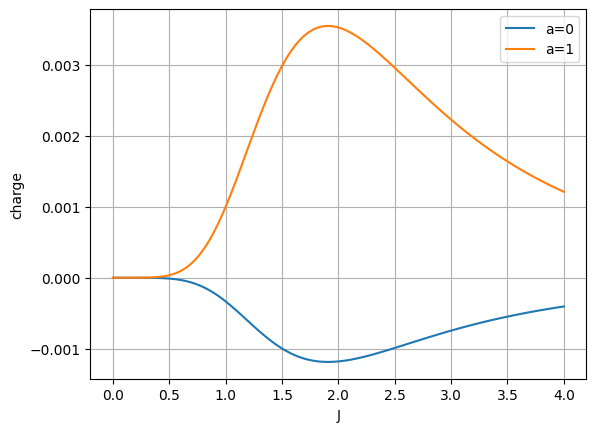}
    }

    \medskip

    \subfloat[\( N = 4 \), \( \sigma_A = X_0 \)]{
        \includegraphics[width=0.45\columnwidth]{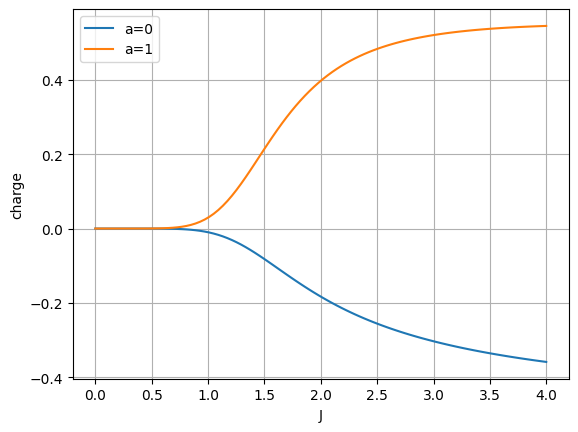}
    }
    \hfill
    \subfloat[\( N = 4 \), \( \sigma_A = Y_0 \)]{
        \includegraphics[width=0.45\columnwidth]{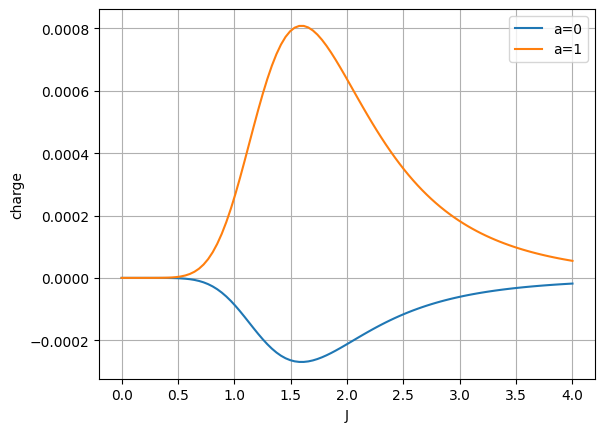}
    }

    \caption{\( Q_B \) vs. \( J \), for the nearest-neighbors model \( H^{(2)} \), with \( h=1 \).}
    \label{fig:nn_vs_j}
\end{figure}

\subsubsection{Importance of Coupling Ratio and Energy Scale}

As discussed in Section~\ref{sec:alice_hamiltonian}, the essential quantity is the ratio \( h/J \), as the overall energy scale can be rescaled without loss of generality. Therefore, while Fig.~\ref{fig:alice_x_vs_j} presents results as a function of \( J \) with fixed \( h \), the conclusions remain valid under constant ratio rescaling.

Detailed simulations varying \( h \) at fixed \( J \), and the corresponding dependence of teleportation fidelity on field strength, are presented in Appendix~\ref{appendix:numerical-simulation}.

\subsection{Simulation Results: Nearest-Neighbor Hamiltonian}

We now present simulation results for the teleportation protocol under the nearest-neighbor interaction Hamiltonian \( H^{(2)} \), (as defined in equation ~\ref{eq:H2}). Unlike the star-type model, this Hamiltonian restricts Alice’s influence to her immediate neighbor, resulting in significantly reduced and more localized entanglement propagation.

\subsubsection{Protocol Implementation and Basis Choices}

For this interaction pattern, the case \( N = 1 \) is equivalent to the previous star model and is therefore omitted. From \( N = 2 \) and beyond, we simulate both energy and charge teleportation protocols, using two distinct measurement bases for Alice: \( \sigma_A = X \) and \( \sigma_A = Y \).

\subsubsection{Extracted Observables vs. Coupling Strength \texorpdfstring{\( J \)}{J}}

Figure~\ref{fig:nn_vs_j} shows the extracted expectation values at Bob’s site as a function of the coupling \( J \), for both basis choices. As expected, the magnitude of the teleported observable diminishes with increasing system size, reflecting the locality of interaction.

Interestingly, while both energy and charge teleportation degrade as \( N \) increases, the \textbf{energy protocol appears more robust to the choice of measurement basis}. This is clearly seen by the different amplitudes and axis scale in Figure~\ref{fig:nn_vs_j} compared to the energy case.

\subsubsection{Implications for Teleportation Robustness and QKD}

The reduced sensitivity of the energy protocol to the choice of measurement basis is advantageous for realistic applications, especially where full control of Alice’s basis is not guaranteed (e.g., under noise or adversarial disturbances). In contrast, the strong basis-dependence in charge teleportation highlights its utility primarily in well-controlled scenarios.

These findings echo and expand upon the discussion in~\cite{IkedaTeleportingCharge, QKDbyQET}, where energy teleportation was shown to tolerate less structured measurement randomness while still extracting meaningful energy. Our simulations validate this robustness in larger systems with nearest-neighbor coupling.

\section{Qiskit Simulation}
\label{sec:qiskit_sim}

In this section, we analyze the implementation of the quantum teleportation protocols using real quantum circuits simulated via Qiskit. Our aim is to bridge the gap between ideal numerical simulations, presented in earlier sections, and the practical realizations of quantum teleportation of observables on both simulators and real quantum hardware \cite{Hotta2009_SpinChain, Hotta2009_Ions, Nambu2010, Yusa2011}.

\begin{figure}[ht!]
    \centering
    \includegraphics[width=0.95\columnwidth]{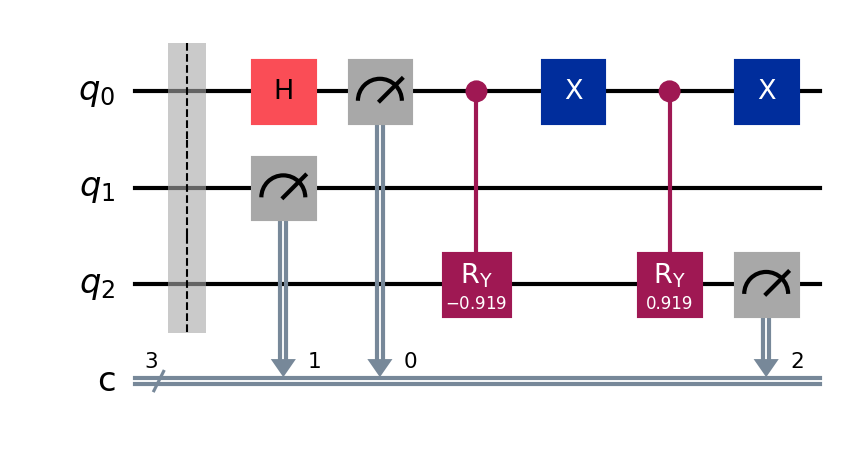}
    \caption{Quantum circuit used to extract Bob's charge expectation value in the charge teleportation protocol.}
    \label{fig:charge_qc}
\end{figure}

\subsection{Circuit Implementations and Simulation Strategy}

The teleportation protocols for both energy and charge observables were implemented as quantum circuits tailored to the specific measurement settings discussed throughout the thesis. 
Figure~\ref{fig:charge_qc} shows the charge-teleportation circuit \(Q_B\) measured at Bob’s site. The energy components \(H_B\) and \(V\) are shown in the appendix. The circuit closely follows the implementation used by Ikeda in the quantum energy teleportation experiment on superconducting hardware~\cite{Ikeda2023}.

Each circuit represents a specific measurement setting derived for a chosen value of \( J \) and \( h \), resulting in a corresponding optimal rotation angle \( \theta \) applied at Bob's site. The full derivation of these angles, and the basis in which Alice measures, are described in detail in Appendix~\ref{appendix:charge_teleportation}. These circuits are initialized to the entangled ground state of the relevant TFIM model by directly loading either the density matrix or the state itself into the circuit using a Qiskit initializer.

\subsubsection{Comparison with Ideal Numerical Simulations and Hardware Execution}

Unlike the ideal numerical simulations presented in earlier sections—which compute the exact expectation values from the full many-body ground state—the Qiskit-based approach simulates the entire teleportation protocol circuit including:
\begin{itemize}
    \item Preparation of the ground state as a density matrix (in simulation),
    \item Explicit measurements on Alice's qubit(s),
    \item Classical communication and conditional rotations,
    \item Measurement of Bob's observable.
\end{itemize}

The \textbf{simulated circuit execution} using Qiskit's `aer simulator` allows for a more realistic modeling of the teleportation process, but still assumes idealized, noise-free conditions. In contrast, results from \textbf{real quantum hardware} (discussed in Section~\ref{sec:hardware_results}) are subject to noise, decoherence, and gate errors, providing valuable insight into the feasibility of implementing QET-based protocols in practice.

This comparison between three levels of abstraction—analytical prediction, noiseless quantum circuit simulation, and real hardware execution—forms the foundation for assessing the practicality, robustness, and scalability of QET and charge teleportation protocols.

\subsection{Star-Coupled Hamiltonian Results}

\begin{figure*}
    \centering
    \subfloat[Energy, \( N = 1 \)]{
        \includegraphics[width=0.3\textwidth]{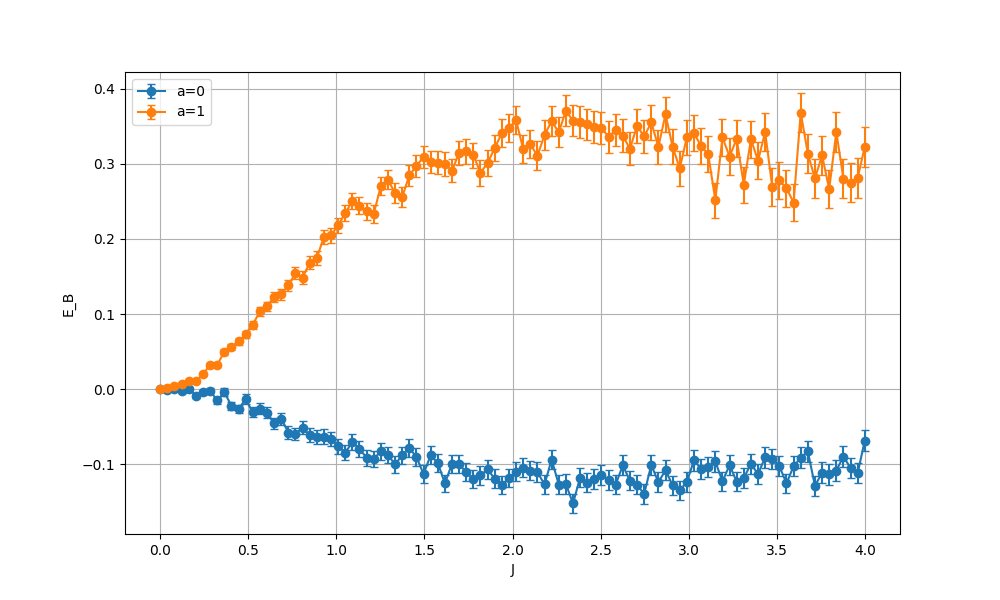}
    }
    \hfill
    \subfloat[Energy, \( N = 2 \)]{
        \includegraphics[width=0.3\textwidth]{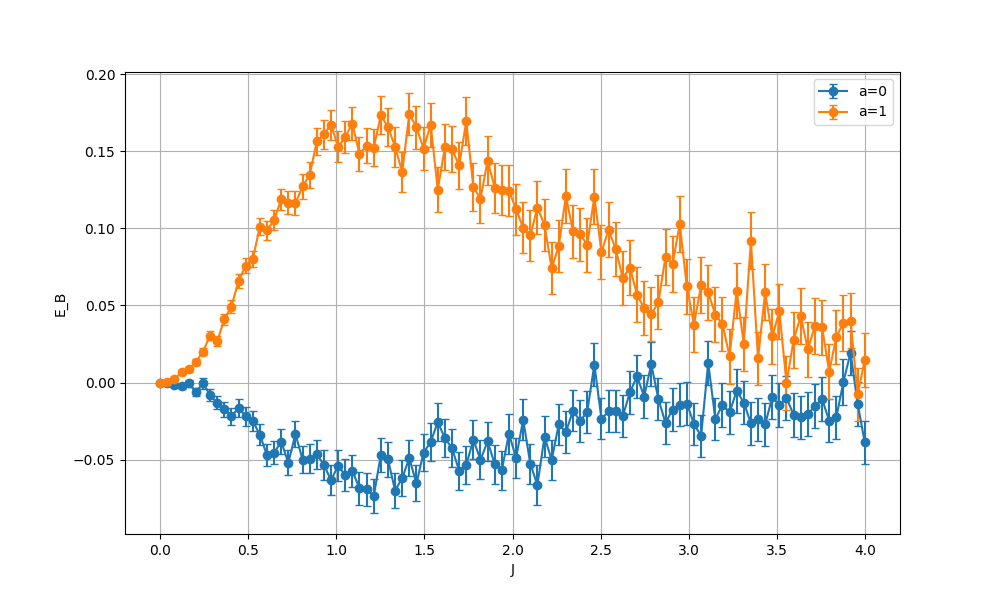}
    }
    \hfill
    \subfloat[Energy, \( N = 3 \)]{
        \includegraphics[width=0.3\textwidth]{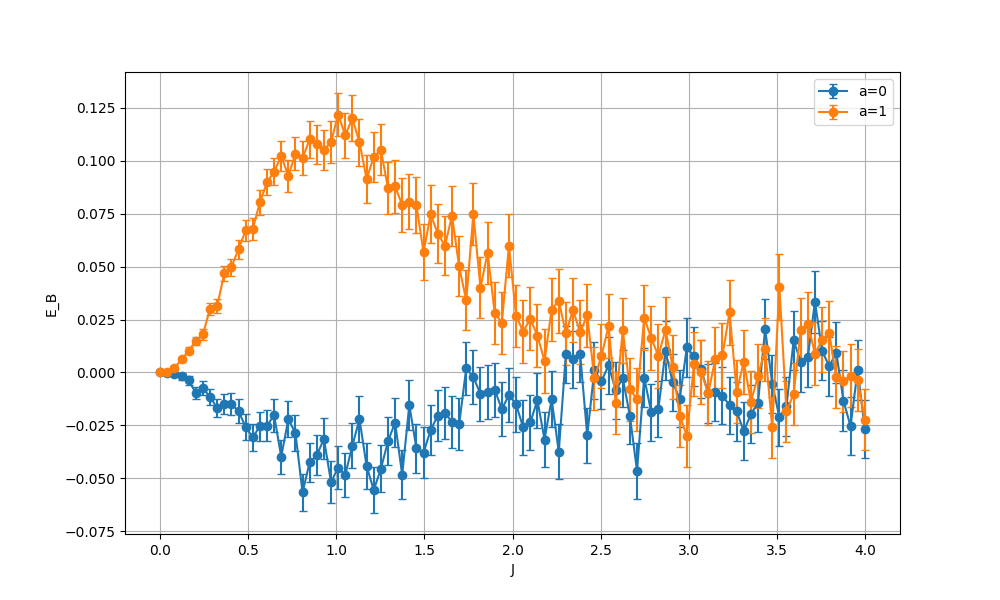}
    }

    \vspace{0.5em}

    \subfloat[Charge, \( N = 1 \)]{
        \includegraphics[width=0.3\textwidth]{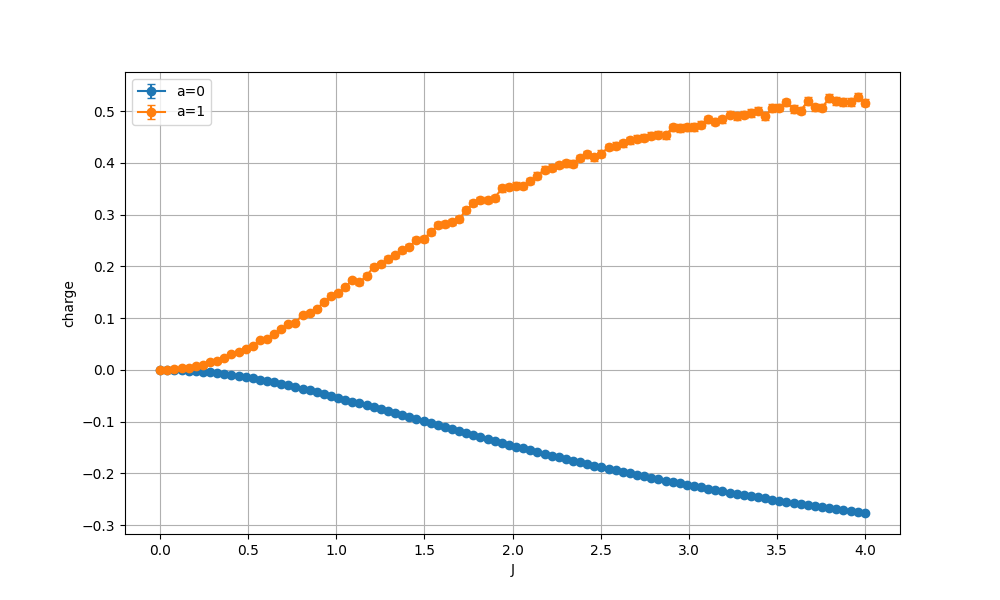}
    }
    \hfill
    \subfloat[Charge, \( N = 2 \)]{
        \includegraphics[width=0.3\textwidth]{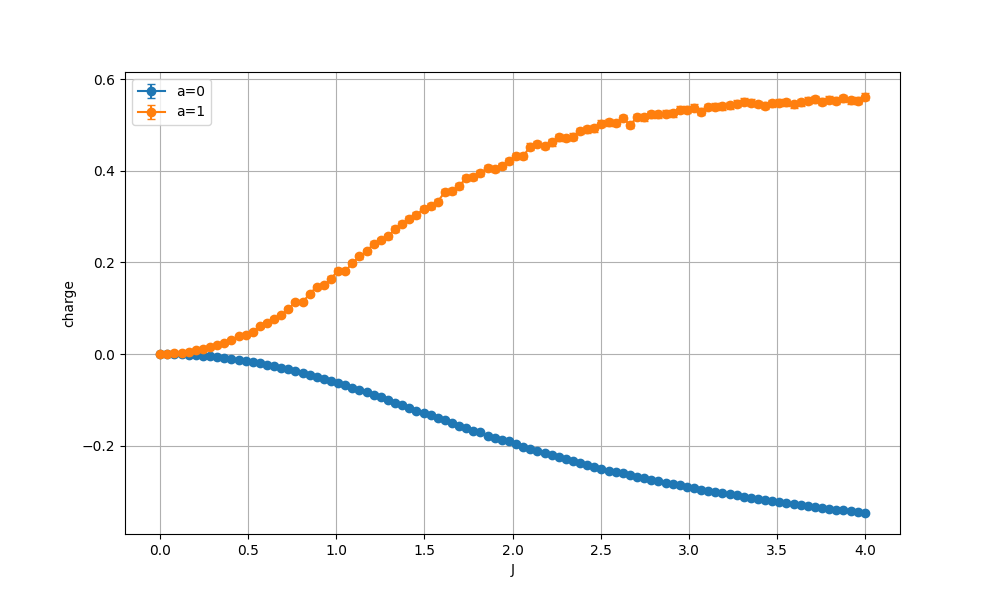}
    }
    \hfill
    \subfloat[Charge, \( N = 3 \)]{
        \includegraphics[width=0.3\textwidth]{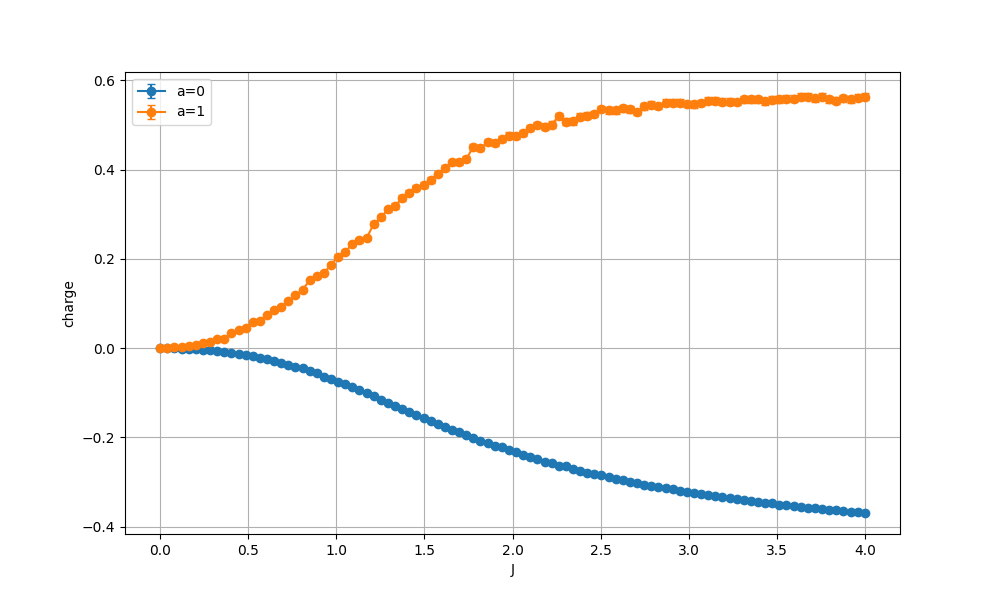}
    }

    \caption{Qiskit simulation results for Energy (top) and Charge (bottom) teleportation protocol using the Star-Coupled Hamiltonian \( H^{(1)} \). All results are averaged over 10{,}000 shots.}
    \label{fig:qiskit_star_results}
\end{figure*}

We begin by examining the teleportation protocol under the star-type interaction Hamiltonian \( H^{(1)} \), where Alice interacts simultaneously with all other sites. Qiskit simulations were executed for both energy and charge teleportation protocols at system sizes \( N = 1, 2, 3 \), using the circuits introduced in the previous subsection. All simulations were averaged over 10{,}000 shots to approximate expectation values from discrete measurement outcomes.

Figure~\ref{fig:qiskit_star_results} presents the results of these simulations: the top row shows Bob’s extracted energy values, and the bottom row shows the corresponding charge values, all plotted against the coupling strength \( J \).

The data points in the figure represent the expectation values of the target observables, calculated by averaging over a large number of experimental runs, or "shots" ($n_{\text{shots}}$). The error bars correspond to the statistical uncertainty of this average, quantified by the standard error of the mean (SEM). The SEM is derived from the standard deviation of a single measurement and the total number of shots. A complete derivation of the statistical framework and the specific formulas used to calculate the SEM for both charge and energy observables is provided in Appendix~\ref{appendix:stats-analysis}.

\subsubsection*{Analysis and Comparison with Numerical Results}

As system size \( N \) increases, the \textbf{statistical noise} in the energy teleportation results becomes more pronounced. This effect is especially visible in panel (c), where the observable fluctuations for \( N = 3 \) exhibit substantial variance due to the low signal-to-noise ratio of energy teleportation at larger \( N \). Charge teleportation, in contrast, remains smooth and symmetric across all system sizes (Fig.~\ref{fig:qiskit_star_results}, bottom row), highlighting its superior robustness. This stands in clear opposition to the energy results, which show significant statistical fluctuations (Fig.~\ref{fig:qiskit_star_results}, top row).

Despite the increased variance, the Qiskit results align well with the ideal numerical simulations discussed in Section~\ref{sec:alice_hamiltonian}, both in structure and value range. Key features preserved include:
\begin{itemize}
    \item The \textbf{asymmetry} of energy teleportation with respect to Alice’s measurement result (bit \( b \)),
    \item The \textbf{mirror symmetry} in charge teleportation, where the two branches (for \( b = 0 \) and \( b = 1 \)) are nearly perfect reflections,
    \item The locations of the \textbf{maximal extracted values} near intermediate \( J \) values (around \( J \sim 1 \)).
\end{itemize}

In summary, these results demonstrate that Qiskit simulations capture the essential physics of the QET protocol under the Star-Coupled Hamiltonian. While averaging over many shots mitigates sampling noise, the energy protocol remains more sensitive to shot noise, especially at larger system sizes, whereas the charge protocol proves more reliable and scalable in this regime.

\subsection{Nearest Neighbors Hamiltonian Results}

\begin{figure*}[t!]
    \centering
    \subfloat[Energy, \( N = 2 \), \( \sigma_A = X_0 \)]{
        \includegraphics[width=0.225\textwidth]{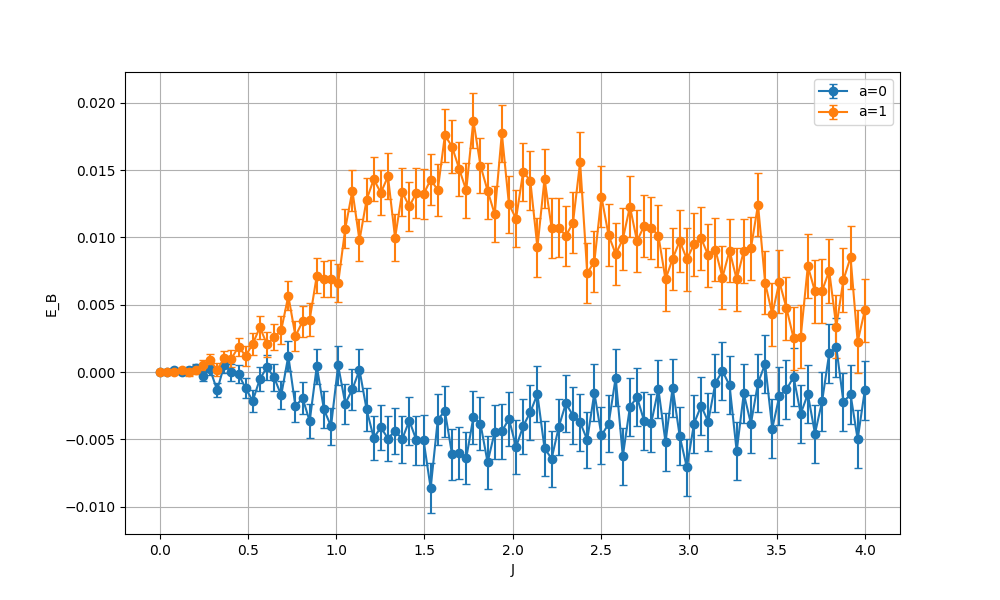}
    }
    \hfill
    \subfloat[Energy, \( N = 3 \), \( \sigma_A = X_0 \)]{
        \includegraphics[width=0.225\textwidth]{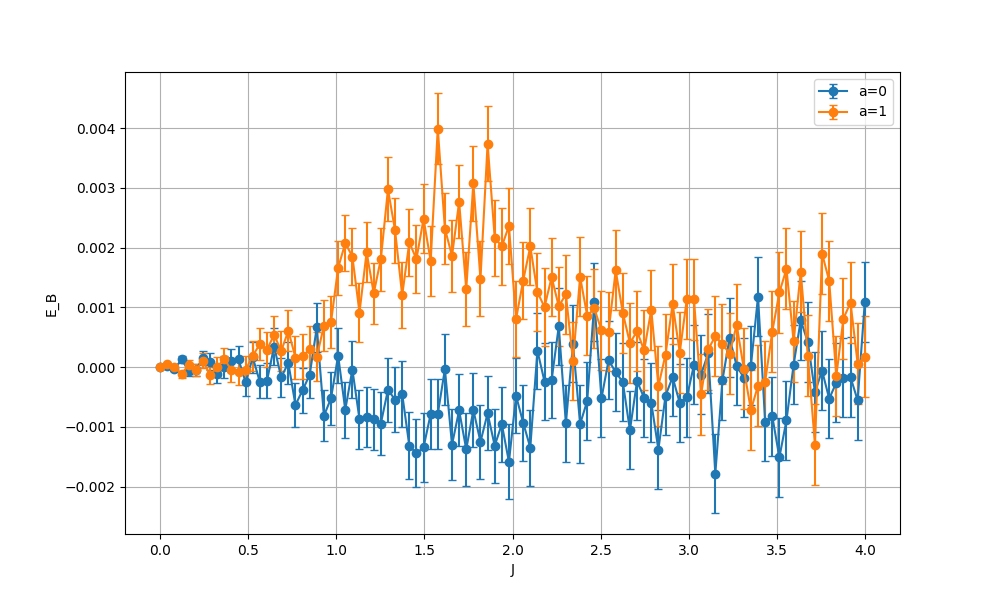}
    }
    \hfill
    \subfloat[Energy, \( N = 2 \), \( \sigma_A = Y_0 \)]{
        \includegraphics[width=0.225\textwidth]{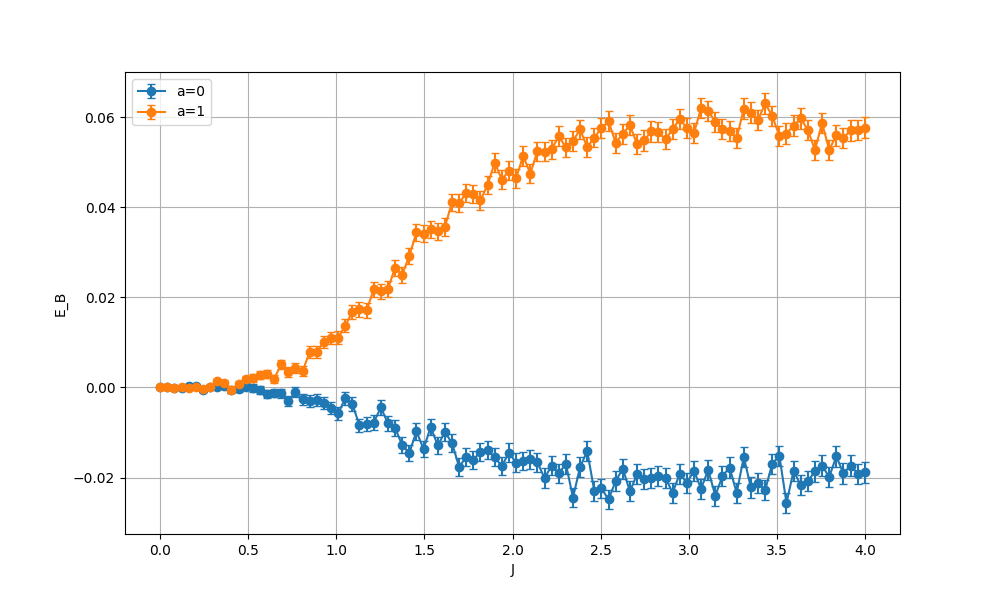}
    }
    \hfill
    \subfloat[Energy, \( N = 3 \), \( \sigma_A = Y_0 \)]{
        \includegraphics[width=0.225\textwidth]{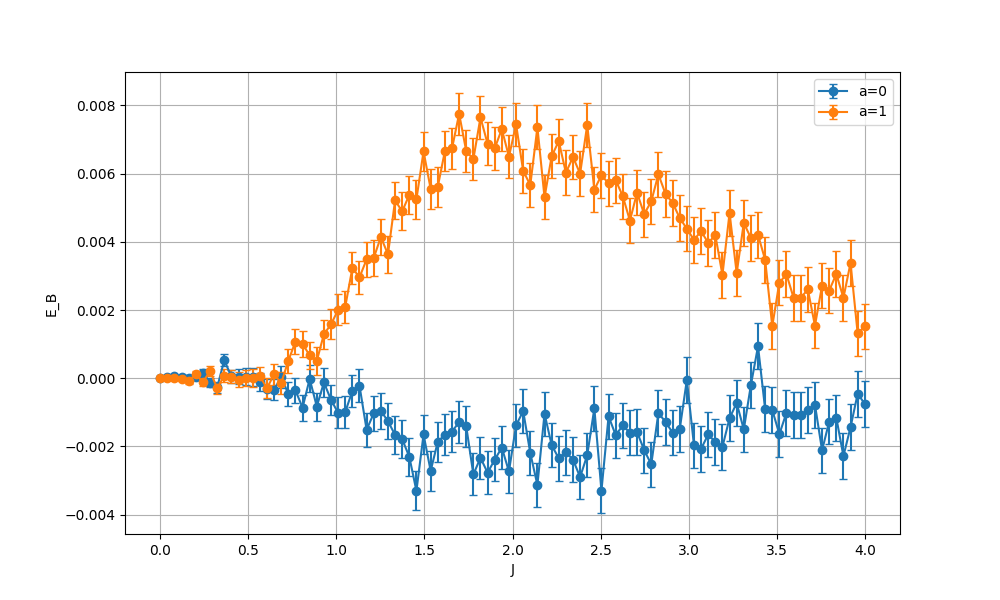}
    }

    \vspace{0.5em}

    \subfloat[Charge, \( N = 2 \), \( \sigma_A = X_0 \)]{
        \includegraphics[width=0.225\textwidth]{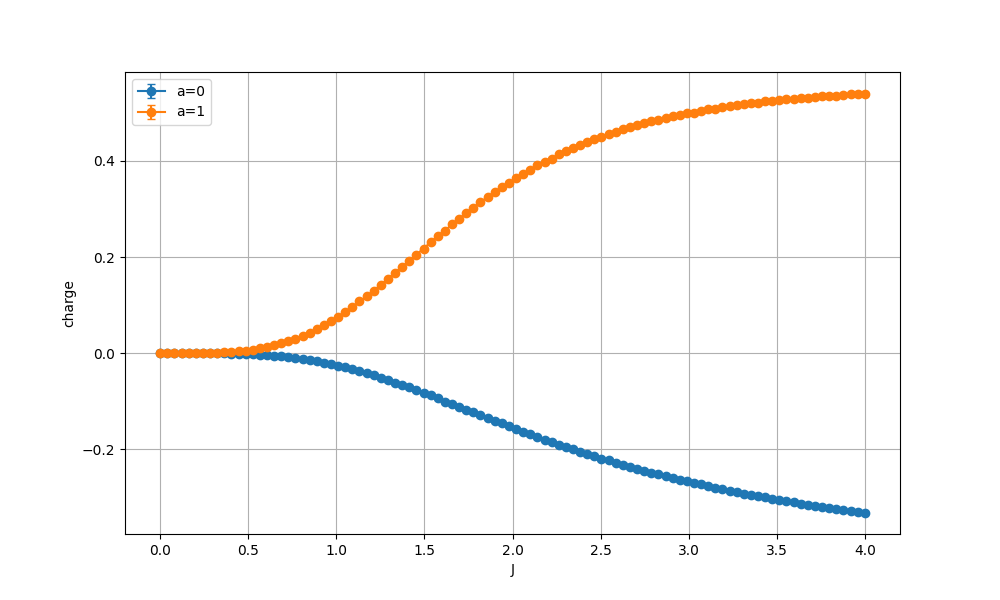}
    }
    \hfill
    \subfloat[Charge, \( N = 3 \), \( \sigma_A = X_0 \)]{
        \includegraphics[width=0.225\textwidth]{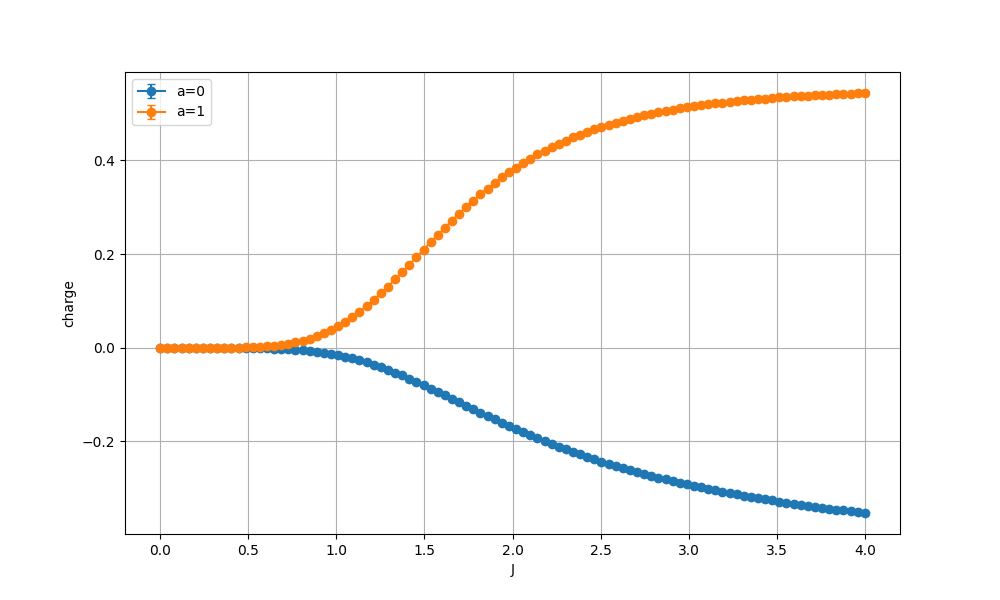}
    }
    \hfill
    \subfloat[Charge, \( N = 2 \), \( \sigma_A = Y_0 \)]{
        \includegraphics[width=0.225\textwidth]{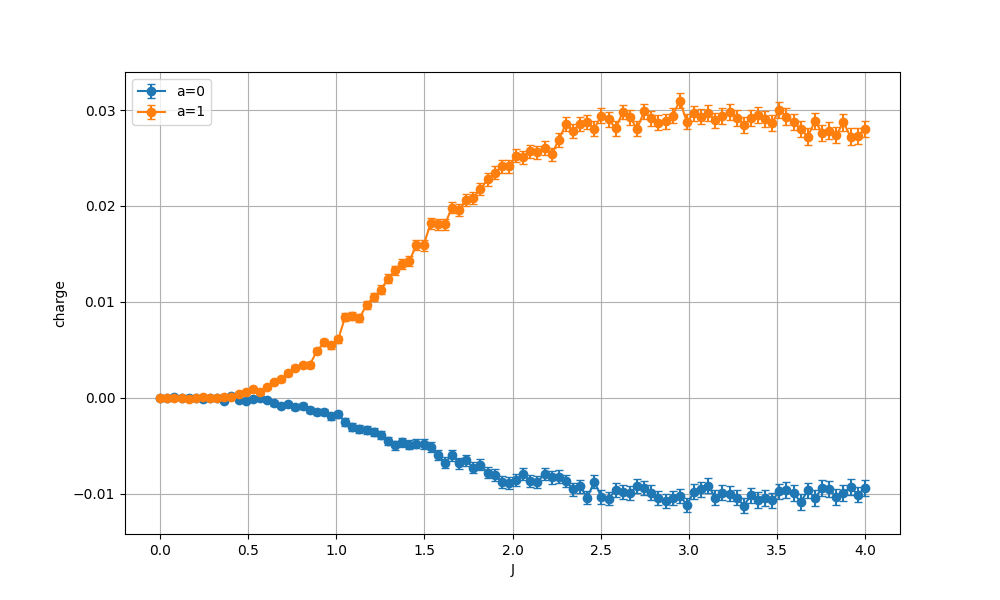}
    }
    \hfill
    \subfloat[Charge, \( N = 3 \), \( \sigma_A = Y_0 \)]{
        \includegraphics[width=0.225\textwidth]{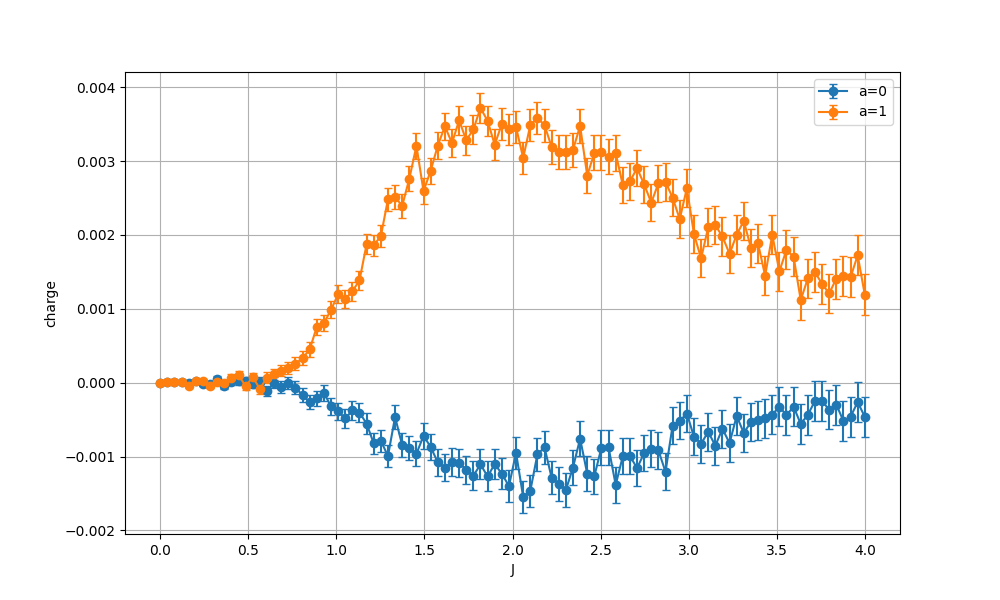}
    }
    
    \caption{Qiskit simulation results for Energy (top) and Charge (bottom) teleportation using the nearest-neighbor Hamiltonian for both Alice's bases choices.}
    \label{fig:qiskit_nn}
\end{figure*}

We now analyze the behavior of the teleportation protocol under the nearest-neighbors Hamiltonian \( H^{(2)} \), where Alice interacts only with her immediate neighbor. As described in Section~\ref{sec:hamiltonians}, this modifies the communication channel between Alice and Bob from a direct interaction to one mediated by intermediate qubits. As in the previous case, we perform Qiskit simulations for both energy and charge teleportation, but now for system sizes \( N = 2 \) and \( N = 3 \) (since \( N = 1 \) reduces to the star interaction case). We consider two different measurement bases for Alice: \( \sigma_A = X_0 \) and \( \sigma_A = Y_0 \).

All simulations are averaged over a large number of shots to mitigate statistical noise: for \( N = 2 \), we average over 500{,}000 shots; for \( N = 3 \), we use 5{,}000{,}000 shots for the \( \sigma_A = X_0 \) basis and 500{,}000 for \( \sigma_A = Y_0 \). This aggressive sampling strategy is necessary due to the increasing weakness of the signal as the system grows in size and complexity.

\subsubsection*{Discussion}

The results shown in Figure~\ref{fig:qiskit_nn} present a significantly noisier behavior compared to the Star-Coupled Hamiltonian case. This increased noisiness arises from multiple sources:

\begin{itemize}
    \item \textbf{Propagation through intermediate sites:} In the nearest-neighbors model, Alice and Bob are no longer directly coupled. The teleportation signal must propagate through a chain of intermediate qubits, weakening the effect as \( N \) increases.
    \item \textbf{Statistical sampling noise:} The extracted signal, particularly for energy, is significantly smaller in magnitude. As a result, many more shots are required to resolve expectation values. This is clearly seen in the \( N = 3 \) case, where energy plots fluctuate heavily even after 5{,}000{,}000 shots.
    \item \textbf{Circuit separation:} Since the two components of Bob's energy — \( H^{(1)} \) and \( V \) — do not commute, they must be measured in separate circuits. This lack of coherence between components introduces additional statistical noise in the energy reconstruction.
\end{itemize}

It is also noteworthy that the \textbf{charge teleportation protocol is less sensitive} to these complications. Across both bases, the extracted charge shows a more stable and consistent pattern with respect to the coupling strength \( J \). Although the Y-basis appears to be a bit more sensitive to noise, which may result also from the small absolute values extracted by Bob. This emphasizes the importance of careful basis selection in realistic implementations.

These results underscore the relative robustness of charge-based teleportation in hardware-limited scenarios, as well as the practical challenges in implementing QET for energy observables in extended systems.

The increased statistical variance observed in the energy protocols is a direct consequence of error propagation: since the energy operator terms do not commute, they must be measured in separate circuit executions, summing their variances (see Appendix~\ref{appendix:stats-analysis} for derivation).

\subsection{Qiskit Experiment on Real Quantum Hardware}
\label{sec:hardware_results}

\begin{figure*}[t!]
    \centering
    \subfloat[Charge teleportation]{
        \includegraphics[width=0.45\textwidth]{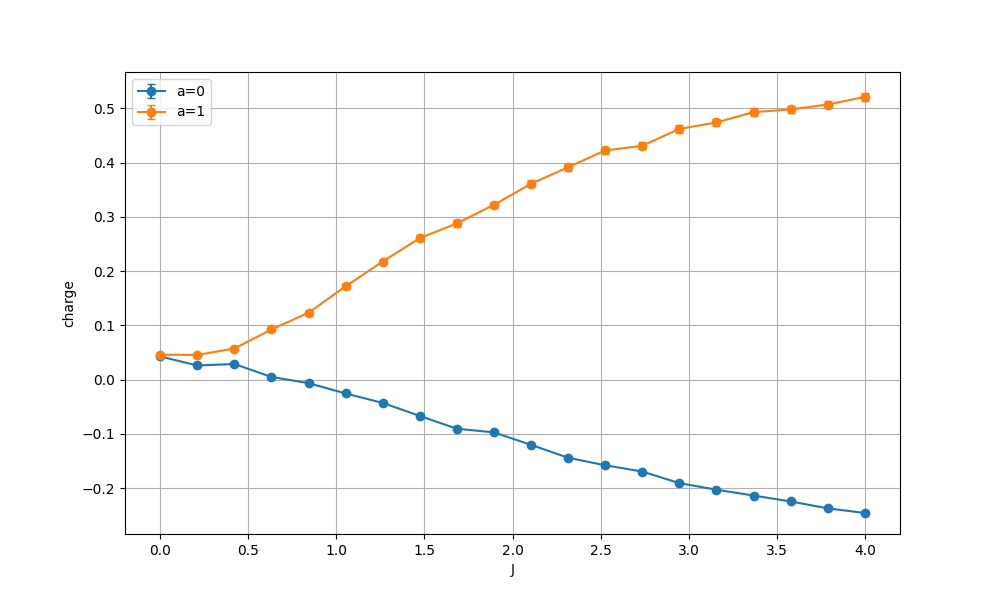}
        \label{subfloat:real_hw_charge_results}
    }
    \hfill
    \subfloat[Energy teleportation]{
        \includegraphics[width=0.45\textwidth]{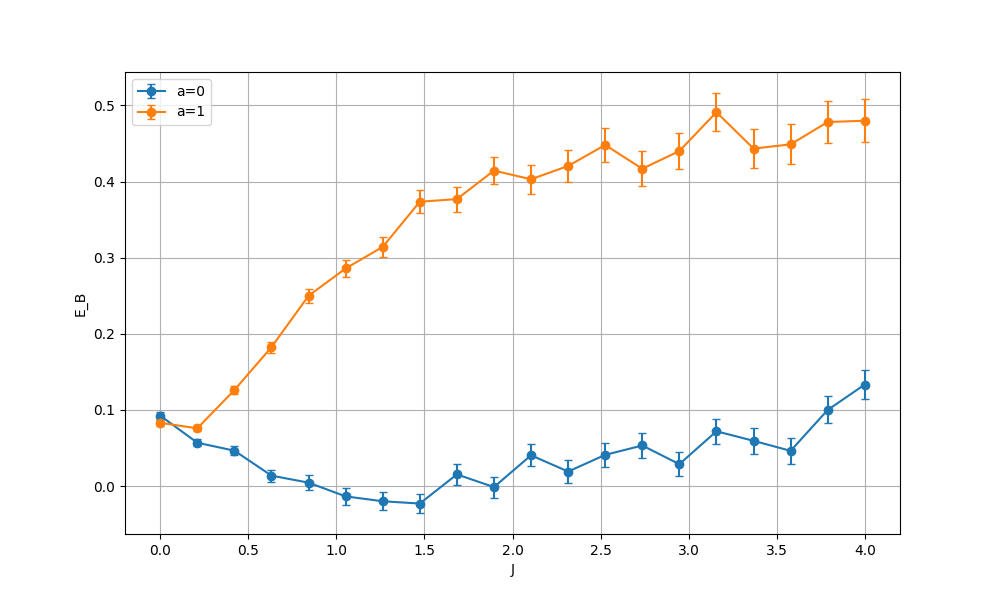}
        \label{subfloat:real_hw_energy_results}
    }
    \caption{Experimental protocol results obtained from real quantum hardware for \( N = 1 \), using Alice’s \( \sigma_A = X_0 \) basis. Each point represents an independent execution for a fixed value of coupling \( J \).}
    \label{fig:real_hw_results}
\end{figure*}

To complete the analysis, we implement the teleportation protocol on an actual quantum processor, the \texttt{ibm\_torino} QPU, via the \texttt{IBM Quantum Platform}. The experiment was conducted for the minimal configuration with \( N = 1 \) using Alice's basis \( \sigma_A = X_0 \), which corresponds to both the \( H^{(1)} \) and \( H^{(2)} \) models in this simple setting. The charge and energy observables were measured separately using the circuits introduced earlier.

The results, shown in Figure~\ref{fig:real_hw_results}, provide crucial insight into the protocol's performance in the presence of realistic device noise. A key observation is a significant positive offset in the measured expectation values compared to the ideal simulations. This is particularly evident at \( J=0 \), where both the charge and energy signals are positive, regardless of Alice's classical bit choice (\( a=0 \) or \( a=1 \)). Such a systematic shift could be caused by various factors, including an imperfectly prepared ground state, decoherence, or measurement readout errors.

For the energy teleportation protocol, this offset proves to be critical. As seen in Figure~\ref{subfloat:real_hw_energy_results}, the expectation values for the \( a=0 \) branch remain positive across the entire range of \( J \). Since the protocol's security relies on encoding key bits in the sign of the measured observable, the persistent positive sign means the protocol fails to produce a distinguishable key. Without sophisticated error correction to compensate for this offset, energy teleportation is not viable in this noisy environment.

The charge teleportation protocol demonstrates its enhanced robustness. While it is also affected by the initial offset at \( J=0 \), the signal strength for the \( a=0 \) branch grows sufficiently negative as \( J \) increases. As shown in Figure~\ref{subfloat:real_hw_charge_results}, the signal overcomes the systematic positive bias, re-establishing the clear sign separation between the \( a=0 \) and \( a=1 \) branches. This confirms that for a large enough coupling \( J \), the charge protocol remains functional and can reliably encode key bits.

This experiment provides practical validation that observable teleportation is feasible on near-term quantum hardware using only LOCC. More importantly, it highlights that while the underlying mechanism works, the choice of observable is paramount. The pronounced resilience of charge teleportation against systematic hardware noise reinforces our central claim that it is a more practical and robust candidate for QKD applications.

\section{Noise and Errors}
\label{sec:noise-errors}

This section analyzes the impact of noise on protocol variants that teleport either energy or charge. We model imperfections in the classical communication channel together with noise processes acting on the entangled resource state, specifically bit-flip, phase-flip, and contamination from excited states. Robustness is assessed under realistic error budgets via numerical studies and device-level simulations performed in \texttt{Qiskit}. Our emphasis is comparative: we contrast energy- and charge-teleportation across the star-coupled ($H^{(1)}$) and nearest-neighbor ($H^{(2)}$) models, referencing extended data in Appendix~\ref{appendix:noise-errors}.

A primary metric for protocol failure is the sign reversal of the teleported expectation value $\langle \Delta O_B \rangle$. Since the protocol encodes key bits in this sign, a flip from negative to positive constitutes a bit-error, rendering the key insecure. We analyze the error probability threshold at which such reversals occur, identifying the operational limits for reliable information transmission.

\subsection{Classical Communication Error}
\label{subsec:classical-comm}

\begin{figure}[ht]
  \centering
  \subfloat[\(H^{(1)}\) (Star-Coupling interaction)]{
        \includegraphics[width=0.8\columnwidth]{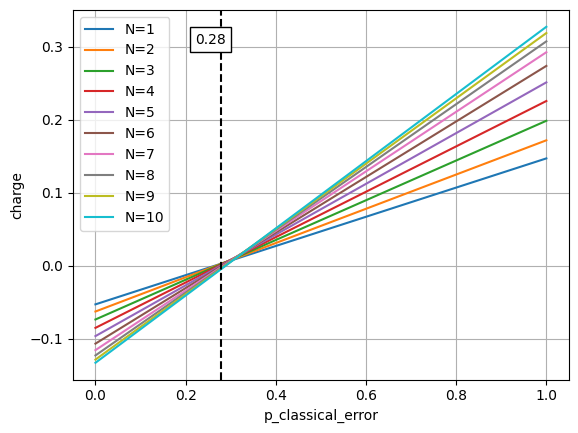}
    }
  \vspace{0.5em}
  \subfloat[\(H^{(2)}\) (Nearest-Neighbors interaction), \(\sigma_A=X_0\)]{
        \includegraphics[width=0.8\columnwidth]{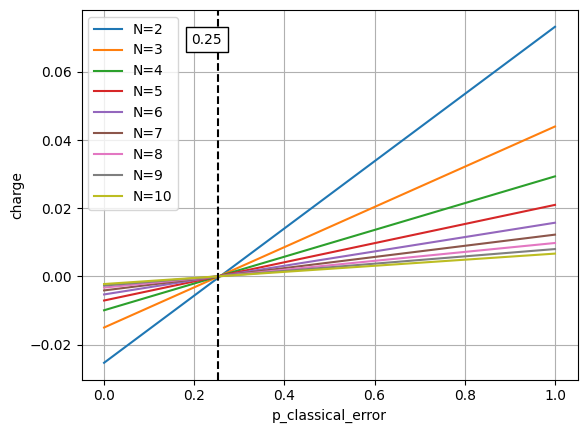}
    }
  \vspace{0.5em}
  \subfloat[\(H^{(2)}\) (Nearest-Neighbors interaction), \(\sigma_A=Y_0\)]{
        \includegraphics[width=0.8\columnwidth]{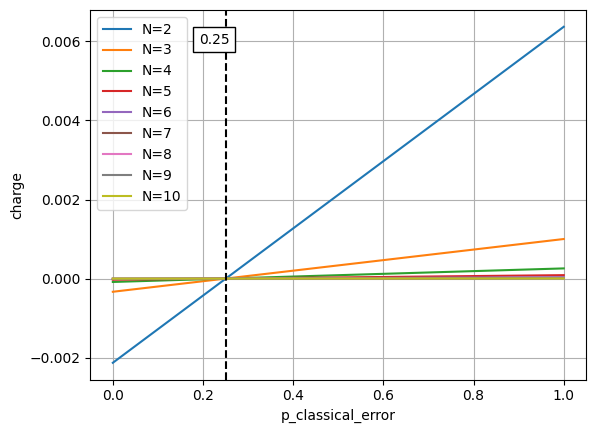}
    }
  \caption{Classical communication error: numerical \(\langle \Delta O_B\rangle\) vs. classical communicated bit-flip probability \(p\) across multiple system sizes \(N\) for Charge. We see a consistent failure threshold around $p \approx 0.25-0.28$}
  \label{fig:cls-Ns-main}
\end{figure}

An error in the classical channel, where the bit Alice sends is flipped with probability $p$, causes Bob to apply the incorrect unitary operation. This effectively creates a statistical mixture of the intended ($a=0$) and flipped ($a=1$) outcomes. As analyzed in \cite{QKDbyQET}, this leads to a linear degradation of the expected signal: $\langle\Delta O_B\rangle_p = (1-p)\langle\Delta O_B\rangle_{a=0} + p\langle\Delta O_B\rangle_{a=1}$. Since $\langle\Delta O_B\rangle_{a=0}$ is negative and $\langle\Delta O_B\rangle_{a=1}$ is positive, the signal is attenuated and moves towards zero.

The numerical results in Figure~\ref{fig:cls-Ns-main} show this linear decay for all simulated configurations. For the star-interaction model $H^{(1)}$, a clear sign-flip threshold appears around $p \approx 0.25$, consistent with the findings in \cite{QKDbyQET}. While the absolute magnitude of the energy signal is larger than that of the charge signal in this model, both are similarly vulnerable to this error. However, for the more realistic nearest-neighbor model ($H^{(2)}$), the energy signal diminishes rapidly with system size $N$, whereas the charge signal remains more stable (a trend clearly visible in the energy data in Appendix Fig.~\ref{fig:energy_vs_classical_err_for_n}). This highlights that for larger distances, the charge protocol's signal is more resilient, even though the fundamental error threshold remains similar.

\begin{figure}[ht]
  \centering
  \includegraphics[width=0.8\columnwidth]{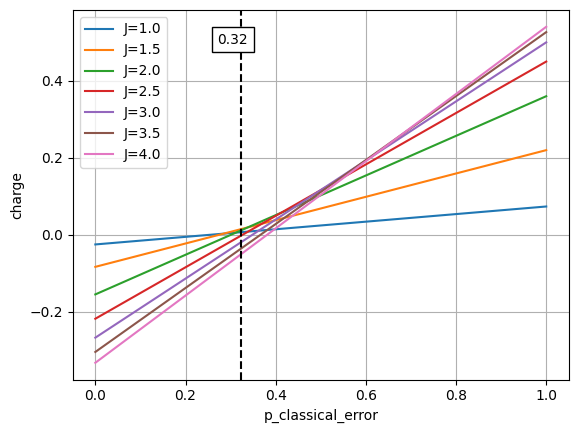}
  \caption{Classical communication error for the $H^{(2)}$ model ($N=2, \sigma_A=X_0$). The sign-flips are
not prominent at low error probabilities.}
  \label{fig:classical-main}
\end{figure}

From Figure~\ref{fig:classical-main} one can see that error flipping the classical bit communicated from Alice to Bob changes the extracted value sign at Bob's site at high probabilities. This is the expected result since by the protocol's design and from simulations of the protocol in previous sections we showed that flipping this bit inverts the sign measured by Bob. Comparing to energy from the equivalence simulation in~\cite{QKDbyQET} (see also Figures~\ref{fig:classical_err_nn_ham_x} and~\ref{fig:classical_err_nn_ham_y} in Appendix), the charge is somewhat more robust to such error where the sign flips around \(p_{err} \approx  0.32\) instead of \(p_{err} \approx  0.25\) for energy, a threshold introduce in~\cite{QKDbyQET} and is confirmed in our own energy teleportation simulations (Appendix Fig.~\ref{fig:classical_err_nn_ham_x} and Fig.~\ref{fig:classical_err_nn_ham_y}).

\subsection{Mixture with an Excited State}
\label{subsec:mixture}

\begin{figure}[ht!]
    \centering
    \subfloat[Numerical, \(H^{(2)}\), \(N{=}2\), \(\sigma_A=X_0\)]{
        \includegraphics[width=0.8\columnwidth]{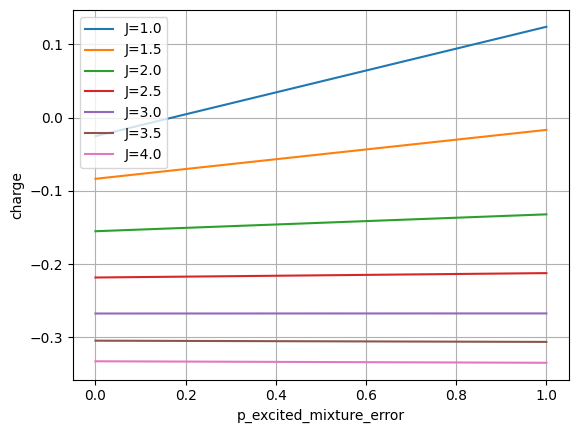}
    }
    \vspace{0.5em}
    \subfloat[Qiskit, \(H^{(2)}\), \(N{=}2\), \(\sigma_A=X_0\)]{
        \includegraphics[width=0.9\columnwidth]{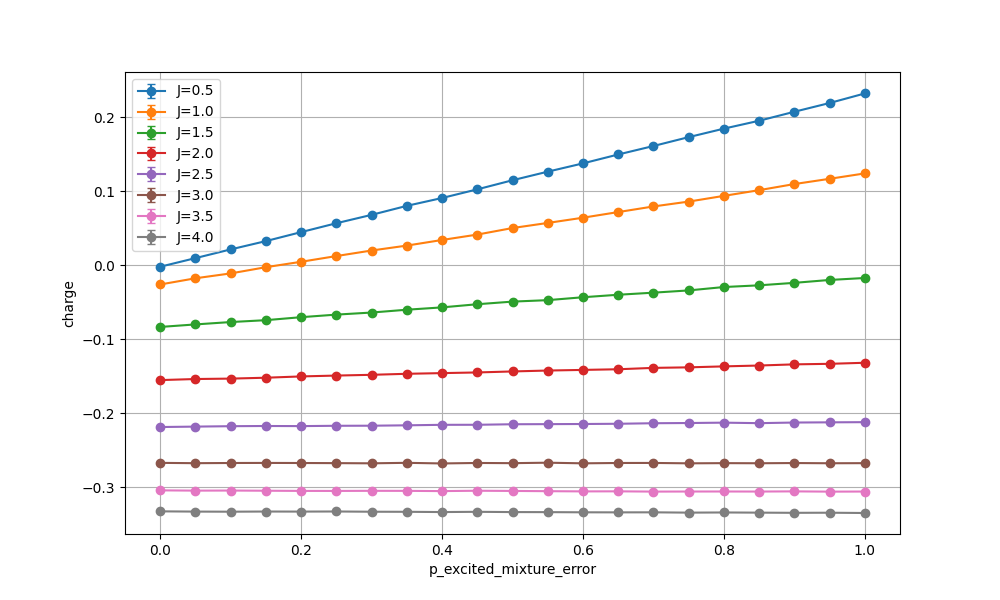}
    }
  
  \caption{Mixture with an excited state for the $H^{(2)}$ model ($N=2, \sigma_A=X_0$). For most \(J\) values, the signal remains negative, demonstrating its robustness. Qiskit simulations (bottom) confirm the trend but show higher statistical variance.}
  \label{fig:mixture-main}
\end{figure}

Perfect ground state preparation is experimentally challenging; a common imperfection is a residual population of excited states, modeled as a classical mixture $\rho = (1-p)\rho_{gs} + p\rho_{excited}$. The teleported signal becomes a weighted average of the contributions from each state.

Figure~\ref{fig:mixture-main} reveals a critical performance difference between energy and charge teleportation.
This difference in robustness stems from the nature of the observables themselves. The energy teleportation signal is highly sensitive to the local energy expectation value $\langle H_B\rangle$. Low-lying excitations contribute a large positive energy offset, which can easily overwhelm the negative teleported signal originating from the ground state component, causing a premature sign-flip fatal to the QKD protocol.

\begin{figure*}[htbp]
    \centering
    \subfloat[Numerical, \(H^{(2)}\), \(N{=}2\), \(\sigma_A=X_0\)]{
        \includegraphics[width=0.8\columnwidth]{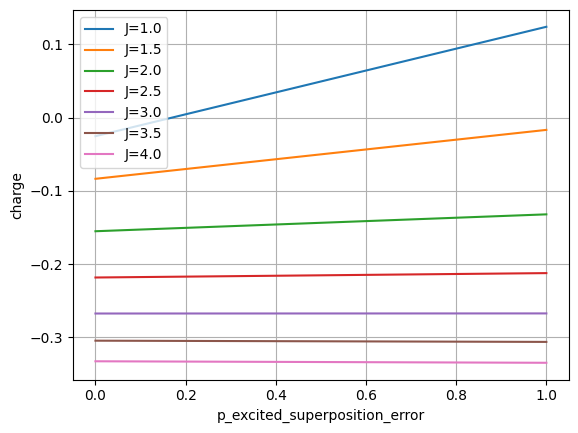}
    }
    \hfill
    \subfloat[Qiskit, \(H^{(2)}\), \(N{=}2\), \(\sigma_A=X_0\)]{
        \includegraphics[width=0.9\columnwidth]{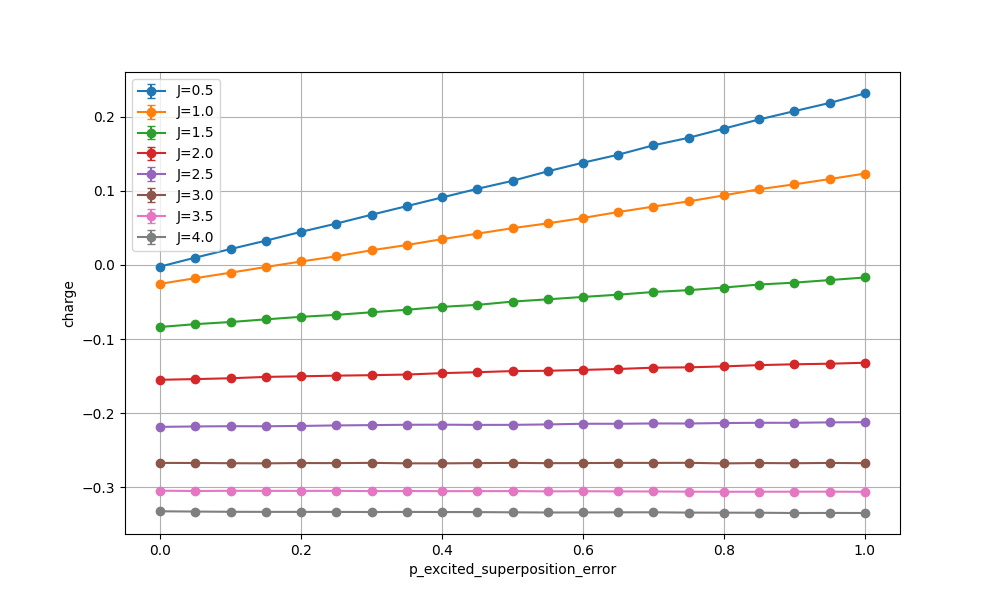}
    }
  
  \caption{Coherent superposition with an excited state for the $H^{(2)}$ model ($N=2, \sigma_A=X_0$). Similar to the classical mixture, the charge protocol proves its resilient.}
  \label{fig:superposition-main}
\end{figure*}

In contrast, the charge observable $Q_B$, which is related to local spin parity, is less sensitive to these absolute energy shifts. The quantum correlations responsible for generating the signal's sign (the $\eta_Q$ term) often retain their structure across different low-energy states. As a result, the contribution from the excited state does not introduce a large positive offset, and the teleported charge signal remains negative over a much wider range of error probabilities, merely decreasing in magnitude.
While both signals degrade as the mixture probability $p$ increases, the energy signal frequently crosses zero and becomes positive at moderate error levels ($p \approx 0.15-0.25$) \cite{QKDbyQET}, a critical vulnerability that is evident in our energy simulations shown in Appendix Fig.~\ref{fig:nn_X_num_mixture_error_appendix} and Fig.~\ref{fig:nn_Y_num_mixture_error_appendix}. In contrast, the charge signal often remains negative even at high error probabilities, merely decreasing in magnitude. This superior robustness of charge stems from the nature of the observables - because charge is related to symmetry which is preserved in low-lying excitations, whereas energy is directly corrupted by the excitation gap. Low-lying excitations can significantly alter the local energy expectation value $\langle H_B \rangle$, contributing a large positive offset term ($\xi_H$) that overwhelms the negative teleported signal from the ground state. The charge observable $Q_B$, which is tied to local spin parity, is less sensitive to such energy shifts. The correlations responsible for charge teleportation ($\eta_Q$) often retain their sign across different low-energy manifolds, preserving the integrity of the key bit.

The Qiskit simulations (top panels of Fig.~\ref{fig:mixture-main}) corroborate these findings while also illustrating the practical measurement challenges. The energy plots exhibit significant statistical variance, a direct result of the weak signal and the error propagation from measuring non-commuting terms (Appendix~\ref{appendix:stats-analysis}). The charge plots are markedly smoother, underscoring not only its theoretical robustness but also its superior statistical stability in a measurement context. This trend holds across all tested Hamiltonians and system sizes, as detailed in Appendix~\ref{appendix:noise-errors}.

\begin{figure}[ht!]
    \centering
    \subfloat[Numerical, \(H^{(2)}\), \(N{=}2\), \(\sigma_A=X_0\)]{
        \includegraphics[width=0.8\columnwidth]{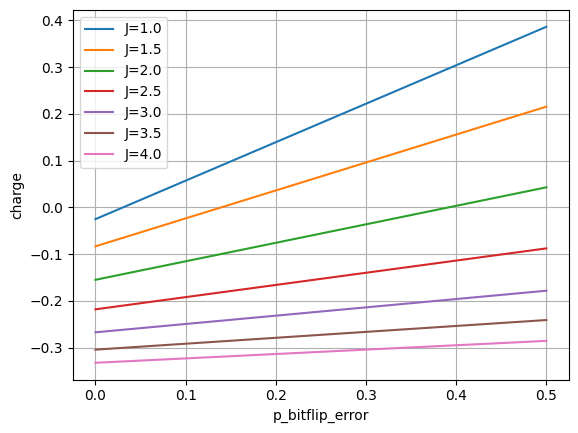}
    }
    \vspace{0.5em}
    \subfloat[Qiskit, \(H^{(2)}\), \(N{=}2\), \(\sigma_A=X_0\)]{
        \includegraphics[width=0.9\columnwidth]{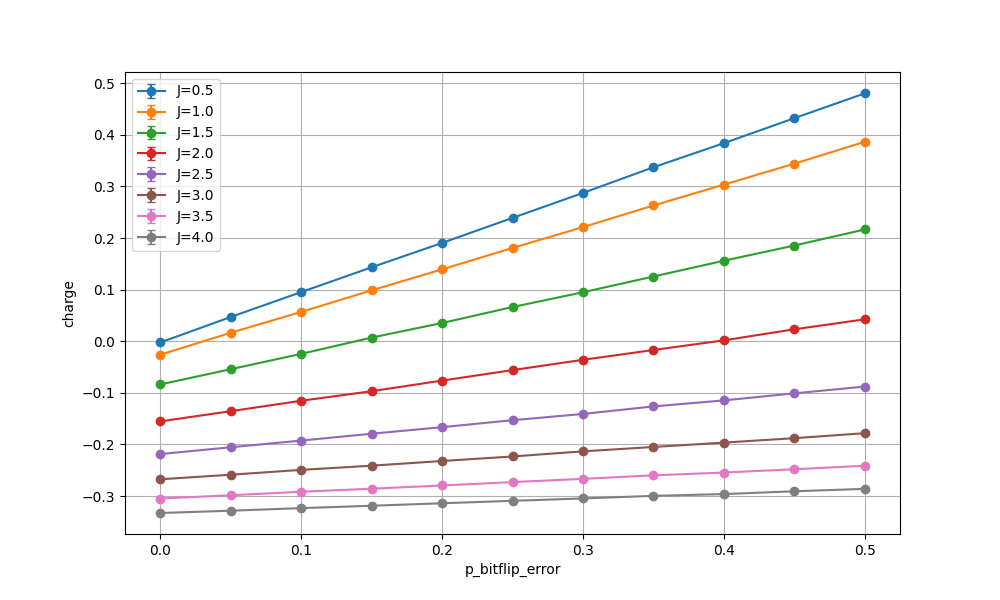}
    }
  
  \caption{Bit-flip error on the resource state for the $H^{(2)}$ model ($N=2, \sigma_A=X_0$). We see a smooth, gradual degradation without a sharp sign-crossing threshold, indicating relative robustness against this error type.}
  \label{fig:bitflip-main}
\end{figure}

\begin{figure*}[ht!]
    \centering
    \subfloat[Numerical, Alice's site]{
        \includegraphics[width=0.4\textwidth]{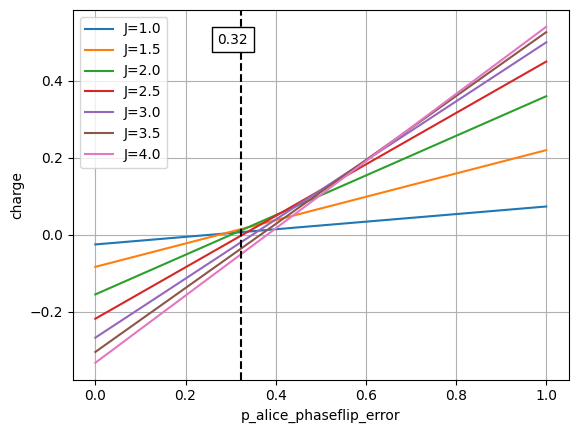}
    }
    \hfill
    \subfloat[Qiskit, Alice's site]{
        \includegraphics[width=0.55\textwidth]{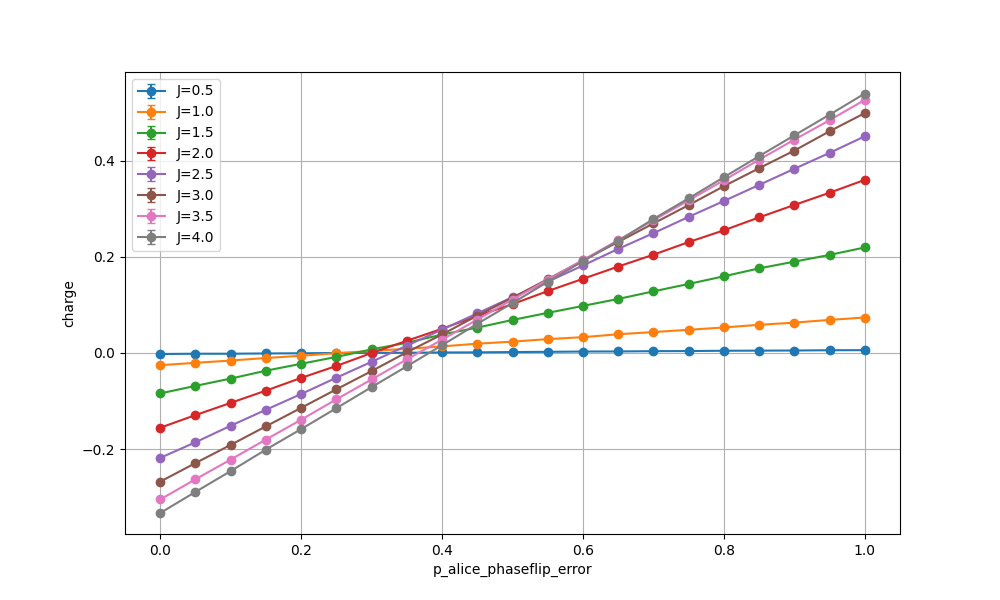}
    }

    \vspace{0.5em}
  
    \subfloat[Numerical, Bob's site]{
        \includegraphics[width=0.4\textwidth]{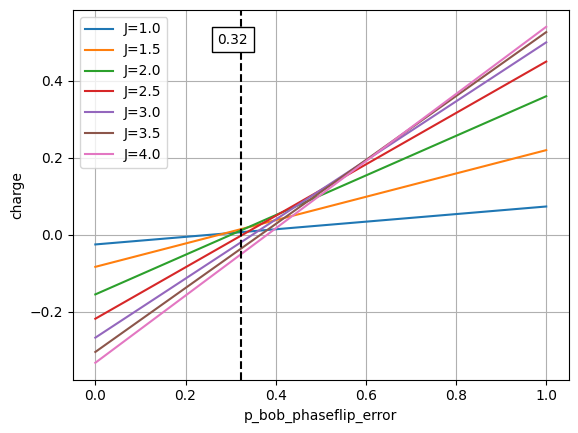}
    }
    \hfill
    \subfloat[Qiskit, Bob's site]{
        \includegraphics[width=0.55\textwidth]{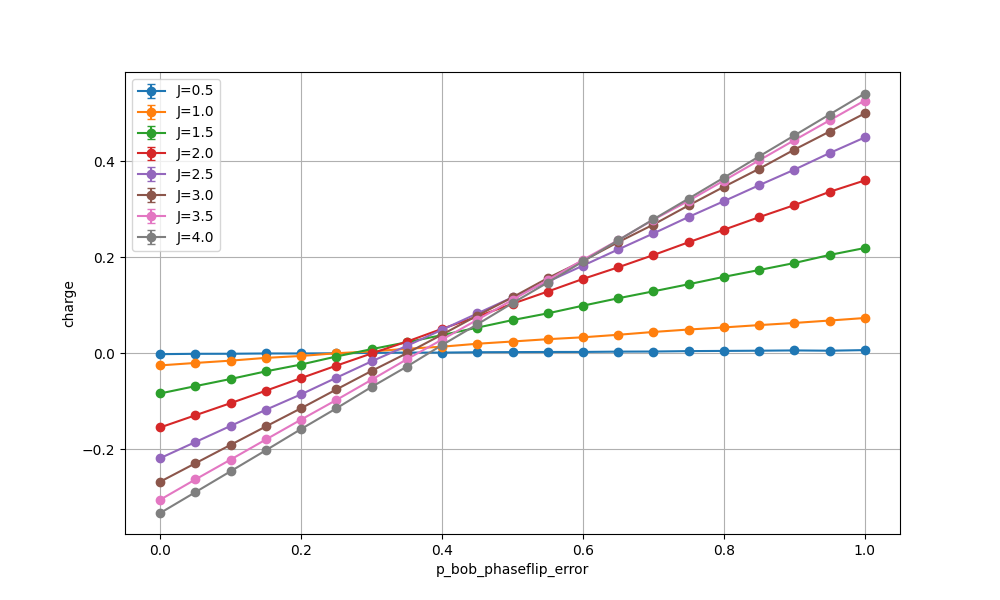}
    }

  \caption{Phase-flip error for the $H^{(2)}$ model (with $N=2, \sigma_A=X_0$). On Alice's site, the error attenuates the signals by decohering the measurement basis, but sign-flips are not prominent at low error probabilities. On Bob's site, the error is more detrimental, causing a rapid decay by inverting Bob's conditional rotation.}
  
  \label{fig:phaseflip-main}
\end{figure*}

\subsection{Superposition with an Excited State}
\label{subsec:superposition}

A coherent error creating a superposition $|\psi\rangle = \sqrt{1-p}|\psi_{gs}\rangle + e^{i\alpha}\sqrt{p}|\psi_{excited}\rangle$ introduces interference terms into the expectation values of $\xi$ and $\eta$. This leads to a non-linear dependence on the error probability $p$, as shown in Figure~\ref{fig:superposition-main}. Similar to the classical mixture, the energy protocol is more susceptible to sign-flips than the charge protocol, as shown in the energy data in Appendix Fig.~\ref{fig:nn_X_num_superposition_error_appendix} and Fig.~\ref{fig:nn_Y_num_superposition_error_appendix}. The phase sensitivity inherent in the superposition can destructively interfere with the commutator term $\eta$, which is the engine of teleportation, and this effect appears more pronounced for the energy observable. The Qiskit simulations again highlight the increased statistical variance, making it difficult to distinguish the true non-linear signal from finite-shot noise.

\subsection{Bit-Flip and Phase-Flip Errors}
\label{subsec:quantum-channel-errors}

We now consider local quantum channel noise acting on individual qubits.

\paragraph{Bit-Flip Error ($X_n$).} This error, modeled by $\rho \rightarrow (1-p)\rho + pX_n\rho X_n$, degrades the specific quantum correlations leveraged by the protocol. As shown in Figure~\ref{fig:bitflip-main}, both energy and charge signals decay smoothly with the error probability $p$. Sign-crossings are rare, indicating that the protocol is relatively robust against this error type. Energy generally maintains a larger absolute signal, which in principle makes it easier to measure, provided the noise level is low enough to prevent ambiguity (cf. the energy data in Appendix Figs.~\ref{fig:nn_X_num_bitflip_error_appendix} and~\ref{fig:nn_Y_num_bitflip_error_appendix}).

\paragraph{Phase-Flip Error ($Z_n$).} The impact of a phase-flip error depends critically on its location.

\begin{table*}[t]
\small
\centering
\caption{Qualitative comparison of teleported \textbf{energy} (E) vs.\ \textbf{charge} (Q) under various noise channels. The assessment is based on trends observed across different Hamiltonians and system sizes.}
\label{tab:noise-summary-improved}
\begin{tabularx}{\textwidth}{@{} l >{\hsize=0.8\hsize}X >{\hsize=0.8\hsize}X >{\hsize=1.4\hsize}X @{}}
\toprule
\textbf{Error Type} & \textbf{Signal Strength (E vs. Q)} & \textbf{Sign-Crossing (E vs. Q)} & \textbf{Dominant Physical Mechanism} \\
\midrule
Classical Comm. Error & E typically larger, but decays faster with N in $H^{(2)}$ & High for both at $p \approx 0.25-0.5$ & Linear mixing of the two classical ($a=0, 1$) branches, as predicted in \cite{QKDbyQET}. \\
\addlinespace
Mixture w/ Excited State & Both decay. E signal degrades faster & High risk for E; Very low for Q & E is highly sensitive to local energy offsets ($\xi_H$) of excited states. Q's sign (from $\eta_Q$) is more robust across low-energy manifolds. \\
\addlinespace
Superposition w/ Excited & Both decay non-linearly & High risk for E; Low for Q & Interference terms renormalize $\eta$ and $\xi$. This effect is more detrimental for the energy protocol. \\
\addlinespace
Bit-Flip Error (X) & E typically maintains a larger signal & Low for both & General degradation of the quantum correlations needed for teleportation. Requires high $p$ to cause sign-flip. \\
\addlinespace
Phase-Flip on Alice (Z) & E typically maintains a larger signal & Low for both & Decoheres Alice's measurement basis ($X_0/Y_0$), which attenuates the commutator term $\eta$ responsible for signal generation. \\
\addlinespace
Phase-Flip on Bob (Z) & Both decay rapidly & Moderate for E; Low for Q & Inverts the sign of Bob's rotation ($R_Y(\theta) \to R_Y(-\theta)$), which mimics a classical bit-flip and accelerates signal decay. \\
\bottomrule
\end{tabularx}
\end{table*}

\begin{itemize}
    \item \textbf{On Alice ($Z_0$):} A phase flip on Alice's qubit prior to her measurement effectively decoheres her measurement basis (since $\{Z_0, X_0\}=0$ and $\{Z_0, Y_0\}=0$). This directly suppresses the commutator term $\eta=i\langle[O_B, \sigma_B]\sigma_A\rangle$, thereby attenuating the teleported signal, a behavior observed for both charge (Fig.~\ref{fig:phaseflip-main}) and energy (Appendix Figs.~\ref{fig:nn_X_num_alice_phaseflip_error_appendix} and~\ref{fig:nn_Y_num_alice_phaseflip_error_appendix}).
    \item \textbf{On Bob ($Z_N$):} A phase flip on Bob's qubit is more detrimental. Since Bob's unitary operation is typically a rotation around the Y or X axis, the transformation $Z_N R_Y(\theta) Z_N^\dagger = R_Y(-\theta)$ shows that a phase flip inverts the sign of his rotation. This is equivalent to receiving a flipped classical bit from Alice and causes a rapid linear decay that accelerates the protocol's failure, an effect that is equally detrimental to the charge (Fig.~\ref{fig:phaseflip-main}) and energy (Appendix Figs.~\ref{fig:nn_X_num_bob_phaseflip_error_appendix} and~\ref{fig:nn_Y_num_bob_phaseflip_error_appendix}).
\end{itemize}

In both cases, energy and charge show similar qualitative degradation, although the protocol is clearly more sensitive to noise on Bob's side.

\subsection{Energy vs. Charge: A Consolidated View}
\label{subsec:consolidated}

Our analysis reveals a consistent trade-off between energy and charge teleportation. \textbf{Energy teleportation} can yield a stronger absolute signal, especially for the directly-coupled star Hamiltonian ($H^{(1)}$). However, this larger signal comes at the cost of fragility. Energy is fundamentally sensitive to errors that shift local energy offsets, such as contamination from excited states, which can induce premature sign-flips that are fatal to a QKD protocol. Moreover, its practical implementation is hampered by higher statistical noise arising from the need to measure non-commuting observables separately (Appendix~\ref{appendix:stats-analysis}).

In contrast, \textbf{charge teleportation} offers a more stable and robust signal. While its magnitude is often smaller, its resilience against sign-flips—particularly under excited-state mixture—is a decisive advantage. This robustness, combined with its perfect symmetry and superior statistical clarity from single-operator measurements, makes it a more reliable and scalable choice. These qualities are especially critical for the nearest-neighbor model, where the signal is inherently weaker and decays with distance, making the statistical stability of charge a key enabler. Table~\ref{tab:noise-summary-improved} summarizes these comparative findings.

Collectively, these results strongly suggest that the stability and resilience of charge teleportation make it a more promising candidate for the practical implementation of observable-based QKD on near-term quantum devices.

\medskip
\noindent\emph{Note.} Additional configurations, including extended \(N\) sweeps and further comparisons, are introduced in Appendix~\ref{appendix:noise-errors} for completeness.

\section{Cryptographic Security Analysis}
\label{sec:security_analysis}

The analysis in Section~\ref{sec:noise-errors} demonstrates the protocol's \textit{robustness}, identifying the physical error thresholds at which the teleported charge signal $\langle \Delta Q_B \rangle$ flips sign---a catastrophic failure. However, a full security proof requires a more rigorous, cryptographic framework to guarantee that a positive secret key rate $K$ exists, even in the presence of noise below this threshold. This section outlines this framework, transitioning from protocol-specific attacks to a quantitative security proof against general coherent attacks by an eavesdropper, Eve.

Our analysis assumes a trusted-device setting where the underlying Hamiltonian and ground state are trusted resources.

\subsection{Protocol-Specific Attack Models}
\label{sec:attack_models}

As established in \cite{QKDbyQET}, the protocol is inherently secure against several key attack models. In a Man-in-the-Middle (MITM) attack, Eve may know the Hamiltonian, intercept the classical communication (Alice's basis choice $\sigma_A$ and bit $a$), and attempt to forge Bob's result. This attack fails because Bob's outcome (the sign of $\langle \Delta Q_B \rangle$) depends not only on the classical bit $a$ but also on the shared quantum correlations. Eve cannot deterministically reproduce the correct sign without possessing the entangled state, and any attempt to do so would require post-selection capabilities, which are not physically realizable.

In an intercept-resend attack, where Eve establishes separate entangled states with Alice and Bob, the protocol also fails. Bob's measurement outcome will be uncorrelated with Alice's bit, resulting in a random key string that will be detected during parameter estimation.

Furthermore, as noted in Section~\ref{sec:numerical_sim}, the use of a random measurement basis by Alice (e.g., randomly choosing between $\sigma_A = X_0$ and $\sigma_A = Y_0$ for the $H^{(2)}$ model) protects against an adversary performing weak measurements to learn about state imperfections \cite{QKDbyQET}.

\subsection{Quantitative Security and Secret Key Rate}
\label{sec:key_rate_formula}

While robust against these specific attacks, a general security proof must bound the information Eve can gain from any physical interaction, including coherent attacks. The standard approach \cite{Scarani2009} is to compute the asymptotic secret key rate, $K_{\text{asym}}$, (in the limit of infinite signals) which is lower-bounded by the Devetak-Winter formula \cite{Devetak2005}:
\begin{equation}
    K_{\text{asym}} \ge 1 - h(e_{\text{bit}}) - h(e_{\text{ph}})
\end{equation}
where $h(x) = -x \log_2(x) - (1-x) \log_2(1-x)$ is the binary entropy function. Here, $1 - h(e_{\text{bit}})$ bounds the classical mutual information between Alice and Bob, $I(A:B)$, and $h(e_{\text{ph}})$ bounds the Holevo information $\chi(B:E)$, which is the upper bound on the information Eve can gain about Bob's key.

\subsection{Parameter Estimation}
\label{sec:parameter_estimation}

To calculate $K_{\text{asym}}$, Alice and Bob must publicly sacrifice a random subset of their measurement data to estimate two key parameters.

\begin{itemize}
    \item \textbf{$e_{\text{bit}}$ ($Q_{BER}$):} The Quantum Bit Error Rate, or $Q_{BER}$, is the error rate measured in the "key generation" basis (e.g., Alice $\sigma_A=X_0$). This is the observed probability that Bob measures the wrong sign (e.g., a positive charge, logical '0') when Alice intended the opposite (e.g., sent $a=0$, logical '1').

    \item \textbf{$e_{\text{ph}}$ (Phase Error):} The Phase Error Rate is the (unmeasured) error rate in the complementary "test" basis (e.g., Alice $\sigma_A=Y_0$).
\end{itemize}

The phase error rate $e_{\rm ph}$ is not measured directly. Instead, it is estimated by using the rounds where Alice and Bob randomly choose \textbf{different} bases. For example, in the $H^{(2)}$ model, Alice's random choice between $\sigma_A=X_0$ and $\sigma_A=Y_0$ is the mechanism for this. The rounds where they use the "mismatched" bases are sacrificed to test the quantum correlations (e.g., visibility). These correlations provide a verifiable upper bound on $e_{\rm ph}$ through entropic uncertainty relations \cite{Tomamichel2011, Koashi2009}.

\begin{figure}[htbp]
    \centering
    \subfloat[\( N = 2 \) (Secure)]{
        \includegraphics[width=0.32\textwidth]{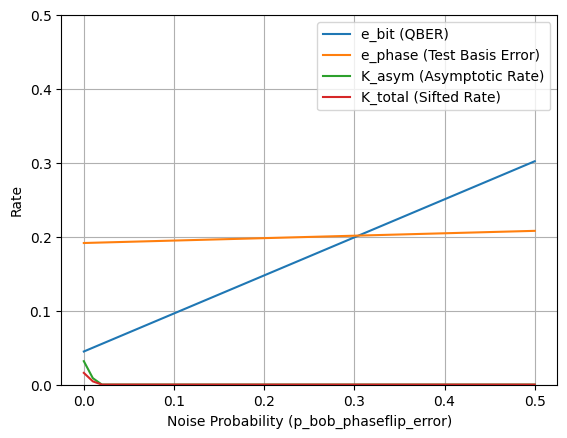}
        \label{fig:key_rates_n2}
    }
    \hfill
    \subfloat[\( N = 3 \) (Insecure)]{
        \includegraphics[width=0.32\textwidth]{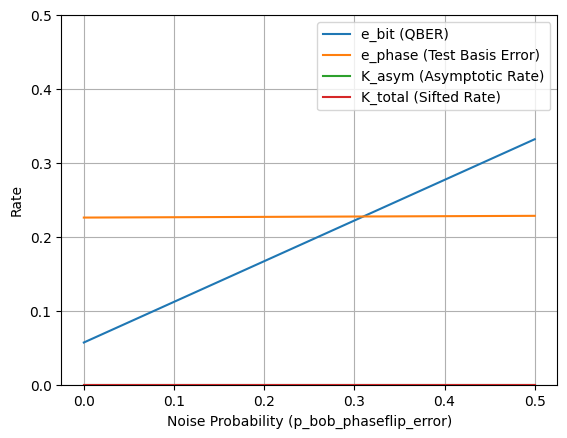}
        \label{fig:key_rates_n3}
    }
    \hfill
    \subfloat[\( N = 4 \) (Insecure)]{
        \includegraphics[width=0.32\textwidth]{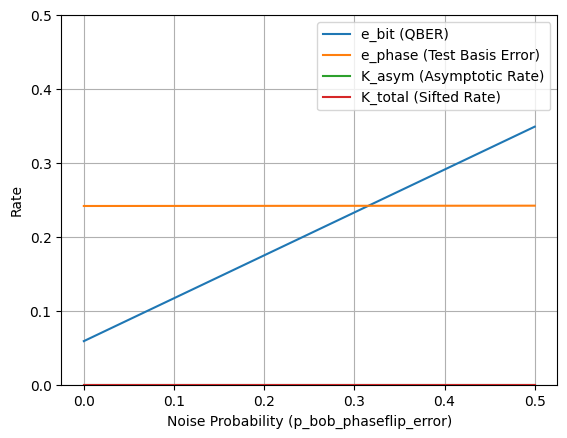}
        \label{fig:key_rates_n4}
    }

    \caption{Asymptotic key rate analysis for the $H^{(2)}$ model ($J=2, h=1$) under phase-flip noise on Bob's site. Plots show the $Q_{BER}$ ($e_{\text{bit}}$), the test basis error ($e_{\text{ph}}$), and the resulting asymptotic key rate ($K_{\text{asym}}$) as a function of noise probability $p$.}
    \label{fig:key_rates}
\end{figure}

\subsection{Simulation of Key Rate vs. Physical Noise}
\label{sec:simulation_key_rate}

To connect this framework to our physical model, we simulate the protocol under the "phase-flip at Bob's site" noise model ($p_{\text{bob\_phaseflip}}$) for the $H^{(2)}$ Hamiltonian. For each noise probability $p$, we run two parallel simulations to find $e_{\text{bit}}(p)$ and $e_{\text{ph}}(p)$. These error rates are calculated from the simulation as follows:

\begin{itemize}
    \item \textbf{Key Basis ($e_{\text{bit}}$):} We run the simulation in the key basis ($\sigma_A=X_0$). For a given $p$, we calculate the probabilities $P_{+}(p)$ (Bob measures $\Delta Q_B > 0$), $P_{-}(p)$ (Bob measures $\Delta Q_B < 0$), and $P_{0}(p)$ (Bob measures $\Delta Q_B = 0$), given that Alice sent $a=0$ (logical '1'). In this protocol, measurement outcomes of '0' are ambiguous and must be "sifted" (discarded). The final key is built only from the positive and negative outcomes. Due to the protocol's symmetry, an error is registered if Bob measures positive value. The $Q_{BER}$ (blue line) is the error probability normalized by the total probability of a \textbf{non-sifted} (kept) event: $e_{\text{bit}}(p) = P_{+}(p) / (P_{+}(p) + P_{-}(p))$.
    \item \textbf{Test Basis ($e_{\text{ph}}$):} We run a separate simulation in the test basis ($\sigma_A=Y_0$) with the same noise $p$. We calculate the analogous probabilities $P'_{+}(p)$, $P'_{-}(p)$, and $P'_{0}(p)$ for the $\Delta Q_B$ measurement. As with the key basis, the $P'_{0}(p)$ events are sifted out. The test-basis error $e_{\text{test}}(p) = P'_{+}(p) / (P'_{+}(p) + P'_{-}(p))$ (orange line) provides the upper bound for the phase error rate, so we set $e_{\text{ph}}(p) \approx e_{\text{test}}(p)$.
\end{itemize}
The asymptotic key rate $K_{\text{asym}}$ (green line) is then computed directly from these two error rates using $K_{\text{asym}}(p) = \max(0, 1 - h(e_{\text{bit}}(p)) - h(e_{\text{ph}}(p)))$.

The results for $N=2, 3,$ and $4$ (with $J=2, h=1$) are presented in Figure \ref{fig:key_rates}.

The simulation results in Fig. \ref{fig:key_rates} reveal a critical dependence on the system size $N$.
For the minimal chain $N=2$ (Fig. \ref{fig:key_rates_n2}), the protocol is secure. The asymptotic key rate $K_{\text{asym}}$ (green line) is positive for zero noise, $K_{\text{asym}}(p=0) > 0$, indicating a secure channel. This is achieved despite a high intrinsic phase error $e_{\text{ph}}(p=0) \approx 0.19$ because the bit error rate is sufficiently low, $e_{\text{bit}}(p=0) \approx 0.04$. The resulting key rate $K_{\text{asym}}(p=0) = 1 - h(0.04) - h(0.19) \approx 0.05$ is small but positive. The protocol can tolerate noise up to a threshold of $p_{\text{th}} \approx 0.02$, after which $K_{\text{asym}}$ drops to zero.

For longer chains, $N=3$ (Fig. \ref{fig:key_rates_n3}) and $N=4$ (Fig. \ref{fig:key_rates_n4}), the protocol is not secure for these Hamiltonian parameters ($J=2, h=1$). The key rate is zero for all noise levels, $K_{\text{asym}} = 0$. This failure is not a flaw, but a physical result: the intrinsic quantum correlations are too weak. As seen in the $N=3$ plot, the intrinsic phase error $e_{\text{ph}}(p=0) \approx 0.23$ is so high that the combined information leakage $h(e_{\text{bit}}) + h(e_{\text{ph}})$ exceeds 1, making $K_{\text{asym}}$ negative (and thus plotted as zero).

This result demonstrates that while the protocol is a valid proof-of-concept, its practical application requires careful optimization of the Hamiltonian parameters ($J, h$).

\subsection{Finite-Key Effects and Future Work}
\label{sec:finite_key}

The preceding analysis is in the asymptotic limit of infinite signals. For any practical implementation, one must account for the effects of realistic devices and finite statistics \cite{Xu2020, Renner2005}. For any practical implementation with a finite number of signals, $M$, the asymptotic rate $K_{\text{asym}}$ is an overestimation. A composable security proof must be used, which accounts for the statistical fluctuations in the estimation of $e_{\rm bit}$ and $e_{\rm ph}$ from a finite data set. The composable, finite-key security rate $\ell$ is given by more complex bounds \cite{Tomamichel2012, Lim2014}, which introduce correction terms that depend on $1/\sqrt{M}$ and the desired security parameter $\varepsilon_{\text{sec}}$.

Our analysis confirms the fundamental viability of the charge teleportation mechanism for QKD. The next steps involve a) optimizing the Hamiltonian parameters $J$ and $h$ to maximize the intrinsic key rate and secure distance $N$, and b) applying finite-key analysis to the optimized protocol to determine the practical key rates achievable with a realistic number of signals.

\section{Summary}
\label{sec:summary}

In this work, we have systematically analyzed the performance of charge observables within QET-based QKD protocols. Our primary contribution is the demonstration that charge teleportation offers distinct operational advantages over energy teleportation. By studying two distinct Transverse Field Ising Models—one with a central star-coupling interaction ($H^{(1)}$) and another with nearest-neighbor interactions ($H^{(2)}$)—we have systematically compared the performance of charge teleportation against energy teleportation. Our investigation, conducted through both numerical simulations and Qiskit-based quantum circuit implementations, confirms that charge teleportation offers significant advantages for cryptographic applications.

Our primary finding is the superior stability and robustness of the charge-based protocol. While energy teleportation can, in certain regimes, yield a larger absolute signal ($\langle \Delta H_B \rangle$), it is highly sensitive to statistical noise and exhibits an asymmetric response to Alice's classical bit choice. In contrast, the teleported charge expectation value ($\langle \Delta Q_B \rangle$) shows a perfect mirror symmetry, which is crucial for a reliable bit-to-key mapping. This symmetry, combined with a signal that remains stable and scalable as the system size $N$ increases, makes the charge protocol inherently more resilient to the finite-shot noise that characterizes near-term quantum hardware.

We also characterized the protocol's dependence on the system's physical parameters. The star-interaction Hamiltonian provides a stronger teleportation signal due to direct Alice-Bob coupling, whereas the signal in the more realistic nearest-neighbor model decays with distance. For the latter, the choice of Alice's measurement basis ($\sigma_A = X_0$ or $Y_0$) was shown to be a critical parameter influencing protocol efficiency. Critically, our protocol was validated on real quantum hardware for a minimal two-qubit system ($N=1$). Despite the presence of device noise, the experimental results successfully reproduced the characteristic separation of Bob's observable expectation values, confirming the fundamental viability of the QKD mechanism. Finally, our cryptographic security analysis quantified the precise trade-off between physical noise and secure key rate, demonstrating that the protocol is secure for $N=2$ (up to a noise threshold of $p_{\rm th} \approx 0.02$) but that longer chains ($N \ge 3$) would require further optimization of the Hamiltonian parameters to overcome high intrinsic error rates.

Looking forward, this work opens several avenues for future research. One promising direction is to systematically map other many-body Hamiltonians to their global symmetries to identify a wider set of conserved charges suitable for teleportation. While this paper demonstrated the utility of one such charge operator, a broader library could enable the tailoring of the protocol to specific physical systems for optimal performance. Furthermore, recent developments in \textit{Timelike} Quantum Energy Teleportation (TQET) offer an intriguing path to enhancing signal strength \cite{Ikeda2025_TQET, Pirandola2020, Lo2012, Acin2007}. TQET utilizes temporal correlations by introducing a waiting period before Bob's operation, which has been shown to boost energy transfer efficiency by over an order of magnitude. Since QKD protocols are often valued for security over speed, incorporating a timelike approach could dramatically improve the signal-to-noise ratio of charge teleportation, making it even more robust for real-world applications. The integration of these concepts could lead to a new generation of highly practical and secure QKD systems built upon the versatile framework of observable teleportation.

\begin{acknowledgments}
We would like to thank Kazuki Ikeda for valuable discussions. This work is supported in part by the Israeli Science Foundation Excellence Center, the US-Israel Binational Science Foundation, and the Israel Ministry of Science.
\end{acknowledgments}

\newpage
\appendix
\section*{Appendix}
\addcontentsline{toc}{section}{Appendix}
\section{Protocol Definition}
\subsection{Detailed Mathematical Derivation of Generic Observable Teleportation}
\label{appendix:detailed-protocol-analysis}

This appendix provides a detailed mathematical derivation of the generalized observable teleportation protocol, explicitly including the parameter $a$ representing Alice’s decision to communicate the true measurement outcome or the opposite bit.

\subsubsection{Protocol Definitions}
We start from the definitions established in the main text and in section 2.1 of the thesis document:
\begin{itemize}
\item Alice's measurement projection: 
\begin{equation}
P_A(b,\sigma_A)=\frac{1-(-1)^b\sigma_A}{2},\quad b\in\{0,1\}.
\end{equation}
\item Alice’s communicated bit: $c=b\oplus a$, with $a=0$ for the true bit and $a=1$ for the opposite bit.
\item Bob’s rotation:
\begin{equation}
U_B(c,\sigma_B)=e^{-i\theta(-1)^c\sigma_B}.
\end{equation}
\end{itemize}

\subsubsection{General Observable Expectation Value}

Expanding explicitly and using commutation properties $[P_A,O_B]=0$ and $[P_A,U_B]=0$, we derive the teleported observable expectation:
\begin{align}
&\langle\Delta O_B\rangle \nonumber\\
&= \frac{1}{2}\sum_bTr\left[\rho_{gs}(1-(-1)^b\sigma_A)(e^{-i\theta(-1)^{b\oplus a}\sigma_B}O_B e^{i\theta(-1)^{b\oplus a}\sigma_B}-O_B)\right] \nonumber\\
&=\frac{\xi}{2}(1-\cos 2\theta)-\frac{(-1)^a\eta}{2}\sin 2\theta,
\end{align}
where we defined:
\begin{equation}
\xi=\Tr[\rho_{gs}\sigma_B O_B\sigma_B]-\Tr[\rho_{gs}O_B],\quad\eta=i\Tr[\rho_{gs}[O_B,\sigma_B]].
\end{equation}

\subsubsection{Optimal Rotation Angle and Final Expression}

The optimal rotation angle $\theta$ maximizing $\langle\Delta O_B\rangle$ is chosen assuming $a=0$:
\begin{equation}
\tan(2\theta)=\frac{\eta}{\xi}.
\end{equation}

Thus, the maximal observable teleportation shift becomes:
\begin{equation}
\langle\Delta O_B\rangle=\frac{\xi}{2}-\frac{\sqrt{\xi^2+(-1)^a\eta^2}}{2}.
\end{equation}

This detailed derivation, explicitly including Alice’s choice of bit transmission, supports the implementation of secure quantum key distribution (QKD) protocols, ensuring robustness and optimality of the teleportation scheme.

\subsection{Detailed Analysis of Charge Teleportation}
\label{appendix:charge_teleportation}

We derive the expectation value shift for the charge operator \( Q_B = \frac{1}{2}(I + Z_N) \) using the general teleportation framework for commuting observables. The steps closely follow the procedure outlined.

\subsubsection{Charge Operator and Teleportation Protocol}

The charge operator acts as a projector onto the \(|0\rangle\) state and satisfies:
\[
Q_B |0\rangle = |0\rangle, \qquad Q_B |1\rangle = 0.
\]

Following the teleportation protocol with Alice measuring \( \sigma_A \in \{X_0, Y_0\} \) and Bob applying a rotation \( U_B(b\oplus a, \sigma_B) = e^{-i\theta(-1)^{b\oplus a}\sigma_B} \), we analyze two operator pairings:
\begin{align*}
(\sigma_A, \sigma_B) = (X_0, Y_N), \quad (\sigma_A, \sigma_B) = (Y_0, X_N).
\end{align*}

We denote \( a \in \{0,1\} \) as Alice’s classical decision bit controlling whether she sends the true measurement outcome or its flipped value.

\subsubsection{Computing \texorpdfstring{$\xi$}{xi} and \texorpdfstring{$\eta$}{eta}}

We define:
\begin{align}
\xi &= \Tr[\rho_{gs} \sigma_B Q_B \sigma_B] - \Tr[\rho_{gs} Q_B], \\
\eta &= i \Tr[\rho_{gs} \sigma_A [Q_B, \sigma_B]].
\end{align}

For \( Q_B = \frac{1}{2}(I + Z_N) \), using \( \sigma_B^2 = I \) and the identity \( \sigma_B Z_N \sigma_B = -Z_N \) for \( \sigma_B \in \{X_N, Y_N\} \), we find:
\begin{equation}
\sigma_B Q_B \sigma_B = \frac{1}{2}(I - Z_N) \quad \Rightarrow \quad \xi = -\langle Z_N \rangle.
\end{equation}

For the commutator term:
\[
[Z_N, \sigma_B] =
\begin{cases}
2i Y_N & \text{if } \sigma_B = X_N, \\
-2i X_N & \text{if } \sigma_B = Y_N,
\end{cases}
\]
we obtain:
\begin{align}
\text{If } (\sigma_A, \sigma_B) = (X_0, Y_N) &\Rightarrow \eta = 2 \langle X_0 X_N \rangle, \\
\text{If } (\sigma_A, \sigma_B) = (Y_0, X_N) &\Rightarrow \eta = -2 \langle Y_0 Y_N \rangle.
\end{align}

\subsubsection{Final Expression and Interpretation}

Inserting into the general expression:
\begin{equation}
\langle \Delta Q_B \rangle = \frac{1}{2} \xi - \frac{1}{2} \sqrt{\xi^2 + (-1)^a \eta^2},
\end{equation}
we obtain:
\begin{equation}
\langle \Delta Q_B \rangle = \frac{1}{2} \langle Z_N \rangle - \frac{1}{2} \sqrt{ \langle Z_N \rangle^2 + 4 \langle O_A O_B \rangle^2 },
\end{equation}
with \( O_A O_B = X_0 X_N \) or \( Y_0 Y_N \) depending on the measurement basis.

This expression encapsulates the full analytical result for the teleportation of the charge observable, and underlies the mechanism by which logical bits can be extracted from sign changes in the expectation value. The classical bit $a$ determines the logical encoding convention and enforces the QKD protocol’s security.

\section{Detailed Analytical Derivations for Quantum Observable Teleportation}

\subsection{Energy Extraction in the Star-Coupled Hamiltonian \texorpdfstring{$H^{(1)}$}{H1}}

\subsubsection{System Definition and Local Hamiltonian}

The star-coupled Hamiltonian is defined as:
\begin{equation}
H^{(1)} = J \sum_{k=1}^{N} X_0 X_k + h \sum_{k=0}^{N} Z_k.
\end{equation}

The localized Hamiltonian at Bob's site (site \( N \)) is obtained by isolating the terms acting on site \( N \):
\begin{equation}
H_B^{(1)} = h Z_N + J X_0 X_N.
\end{equation}

\subsubsection{Protocol and Teleported Energy}

In this protocol, Alice measures \( \sigma_A = X_0 \), with projection:
\begin{equation}
P_A(b) = \frac{1}{2} \left( I - (-1)^b X_0 \right),
\end{equation}
and Bob applies:
\begin{equation}
U_B(b) = \exp(-i \theta (-1)^b Y_N).
\end{equation}

The extracted energy after applying the operation is derived via:
\begin{equation}
\langle \Delta H_B \rangle = \text{Tr}[\rho' H_B] - \text{Tr}[\rho H_B],
\end{equation}
where \( \rho \) is the ground state and \( \rho' \) is the post-rotation state.

Define:
\begin{align}
\xi &= -2 \langle H_B^{(1)} \rangle = -2h \langle Z_N \rangle - 2J \langle X_0 X_N \rangle, \\
\eta &= 2h \langle X_0 X_N \rangle - 2J \langle Z_N \rangle.
\end{align}

Then the energy extracted at Bob, maximized over \( \theta \), is:
\begin{equation}
\langle \Delta H_B^{(1)} \rangle = -\frac{\xi}{2} - \frac{1}{2} \sqrt{\xi^2 + \eta^2}.
\end{equation}

This leads to the explicit form:
\begin{align}
\langle \Delta H_B^{(1)} \rangle ={}& -h \langle Z_N \rangle - J \langle X_0 X_N \rangle \nonumber \\
& - \sqrt{(h^2 + J^2)(\langle Z_N \rangle^2 + \langle X_0 X_N \rangle^2)}.
\end{align}

\subsection{Energy Extraction in the Nearest-Neighbor Hamiltonian \texorpdfstring{$H^{(2)}$}{H2}}

\subsubsection{System Definition and Local Hamiltonian}

The nearest-neighbor Hamiltonian is:
\begin{equation}
H^{(2)} = J \sum_{k=1}^{N} X_{k-1} X_k + h \sum_{k=0}^{N} Z_k,
\end{equation}
and the corresponding local Hamiltonian on Bob's site is:
\begin{equation}
H_B^{(2)} = h Z_N + J X_{N-1} X_N.
\end{equation}

We analyze two cases separately.

\subsubsection{Case I: \texorpdfstring{$\sigma_A = X_0$}{sigmaAX0}}

\paragraph{Measurement and Rotation.}
Alice projects using:
\begin{equation}
P_A^{(X)}(b) = \frac{1}{2} \left( I - (-1)^b X_0 \right),
\end{equation}
and Bob rotates with:
\begin{equation}
U_B^{(X)}(b) = \exp(-i \theta (-1)^b Y_N).
\end{equation}

\paragraph{Energy Extraction Parameters.}
The relevant quantities for calculating energy change are:
\begin{align}
\xi &= -2h \langle Z_N \rangle - 2J \langle X_{N-1} X_N \rangle, \\
\eta &= 2h \langle X_0 X_N \rangle - 2J \langle X_0 X_{N-1} Z_N \rangle.
\end{align}

Hence, the extracted energy is:
\begin{equation}
\langle \Delta H_B^{(2)} \rangle = -\frac{\xi}{2} - \frac{1}{2} \sqrt{\xi^2 + \eta^2}.
\end{equation}

This expression is structurally identical to the \( H^{(1)} \) case but involves longer-range correlations (e.g., \( \langle X_0 X_{N-1} Z_N \rangle \)) due to the indirect connectivity.

\subsubsection{Case II: \texorpdfstring{$\sigma_A = Y_0$}{sigmaAY0}}

\paragraph{Measurement and Rotation.}
Alice now measures:
\begin{equation}
P_A^{(Y)}(b) = \frac{1}{2} \left( I - (-1)^b Y_0 \right),
\end{equation}
and Bob again uses:
\begin{equation}
U_B^{(Y)}(b) = \exp(-i \theta (-1)^b Y_N).
\end{equation}

\paragraph{Energy Extraction Parameters.}
The expressions simplify because the correlator \( \langle X_{N-1} X_N \rangle \) vanishes under this measurement. Define:
\begin{align}
\xi &= -2h \langle Z_N \rangle, \\
\eta &= -2h \langle Y_0 X_N \rangle.
\end{align}

Therefore, for \( a = 0 \), the teleported energy becomes:
\begin{align}
\langle \Delta H_B^{(2)} \rangle &= -h\langle Z_N \rangle - h\frac{\langle Z_N \rangle^2 + \langle Y_0Y_N \rangle^2}{\sqrt{\langle Z_N \rangle^2 + \langle Y_0Y_N \rangle^2}} \nonumber \\
&= -\frac{\xi}{2} - \frac{1}{2} \sqrt{\xi^2 + \eta^2},
\end{align}
which again matches the general QET structure.

\subsection{Detailed Analytical Derivation for the \texorpdfstring{$N=1$}{N=1} Case}

\subsubsection{System Hamiltonian and Matrix Representation}

We consider the minimal model of the quantum teleportation protocol for a two-qubit system, \( N = 1 \), with Alice assigned to site 0 and Bob to site 1. In this case, both interaction Hamiltonians coincide:
\begin{equation}
    H^{(1)} = H^{(2)} = J X_0 X_1 + h(Z_0 + Z_1),
\end{equation}
where \( J \) is the coupling strength between the qubits and \( h \) is the strength of the local transverse field.

In the computational basis \( \{ \ket{00}, \ket{01}, \ket{10}, \ket{11} \} \), the Pauli operators are given by:
\begin{equation*}
Z_0 + Z_1 = 
\begin{pmatrix}
2 & 0 & 0 & 0 \\
0 & 0 & 0 & 0 \\
0 & 0 & 0 & 0 \\
0 & 0 & 0 & -2
\end{pmatrix}, \quad
X_0 X_1 =
\begin{pmatrix}
0 & 0 & 0 & 1 \\
0 & 0 & 1 & 0 \\
0 & 1 & 0 & 0 \\
1 & 0 & 0 & 0
\end{pmatrix}.
\end{equation*}

The total Hamiltonian becomes:
\begin{equation}
H =
\begin{pmatrix}
2h & 0 & 0 & J \\
0 & 0 & J & 0 \\
0 & J & 0 & 0 \\
J & 0 & 0 & -2h
\end{pmatrix}.
\end{equation}

This matrix can be decomposed into two independent blocks based on parity symmetry:
\[
H =
\begin{pmatrix}
\text{Block}_{\text{even}} & 0 \\
0 & \text{Block}_{\text{odd}}
\end{pmatrix},
\]
where the even-parity subspace is spanned by \( \{\ket{00}, \ket{11}\} \) and the odd-parity subspace by \( \{\ket{01}, \ket{10}\} \). The Hamiltonian thus becomes:
\begin{align}
H_{\text{even}} &= 
\begin{pmatrix}
2h & J \\
J & -2h
\end{pmatrix}, \quad
H_{\text{odd}} = 
\begin{pmatrix}
0 & J \\
J & 0
\end{pmatrix}.
\end{align}

\subsubsection{Ground State Derivation}

The ground state resides in the even-parity subspace since it contains the lowest eigenvalue of the total Hamiltonian. The eigenvalues of \( H_{\text{even}} \) are:
\begin{equation}
E_{\pm} = \pm \sqrt{4h^2 + J^2}.
\end{equation}

We define \( J = 2k \), such that \( E_0 = \sqrt{h^2 + k^2} \), and the ground state energy becomes \( E_{gs} = -2E_0 \).

The normalized ground state is:
\begin{equation}
\ket{gs} = \frac{1}{\sqrt{2E_0}} \left( -\sqrt{E_0 - h} \ket{00} + \sqrt{E_0 + h} \ket{11} \right).
\end{equation}

\subsubsection{Expectation Values and Correlations}

Let us define the ratio:
\begin{equation}
r = \frac{k}{E_0} = \frac{J/2}{\sqrt{h^2 + (J/2)^2}}.
\end{equation}

We now compute the relevant correlation functions in the ground state:
\begin{align}
\langle Z_0 \rangle &= \langle gs | Z_0 | gs \rangle = \frac{h}{E_0}, \\
\langle Z_1 \rangle &= \langle gs | Z_1 | gs \rangle = \frac{h}{E_0}, \\
\langle X_0 X_1 \rangle &= \langle gs | X_0 X_1 | gs \rangle = -\frac{k}{E_0}, \\
\langle X_0 \rangle &= \langle X_1 \rangle = \langle Y_0 \rangle = \langle Y_1 \rangle = 0.
\end{align}

The charge operator is defined as:
\begin{equation}
Q_B = \frac{1 - Z_1}{2},
\end{equation}
and thus its expectation value becomes:
\begin{equation}
\langle Q_B \rangle = \frac{1 - \langle Z_1 \rangle}{2} = \frac{1}{2} \left(1 - \frac{h}{E_0} \right).
\end{equation}

\subsubsection{Analytical Expressions for Teleported Energy and Charge}

Following the teleportation protocol where Bob receives the classical bit \( a \in \{0, 1\} \), and assuming the optimal rotation angle is chosen, the expectation values of the teleported observables are:

\paragraph{Energy:}
\begin{equation}
\langle \Delta H_B \rangle = h + Jr - \frac{ \left( h^2 + (-1)^a J^2 \right) \left( 1 + (-1)^a r^2 \right) }{ \sqrt{ (h^2 + J^2)(1 + r^2) } }
\end{equation}

\paragraph{Charge:}
\begin{equation}
\langle \Delta Q_B \rangle = \frac{1}{2} \left( 1 - \frac{ 1 + (-1)^a r^2 }{ \sqrt{1 + r^2} } \right)
\end{equation}

These expressions make explicit the dependence on the classical bit \( a \), as well as on the ratio \( r = k/E_0 \), which encapsulates the coupling-to-field ratio \( J/h \). Both expressions are smooth and analytic in the parameter \( r \), and provide a clear demonstration of how bit-dependent observable differences arise from the underlying quantum correlations in the ground state.

\subsubsection{Limit Behavior and Interpretation}

Let us consider the limiting cases:

\paragraph{Weak Coupling (\( J \ll h \))}:
\[
r \to 0 \Rightarrow \langle \Delta H_B \rangle \to 0, \quad \langle \Delta Q_B \rangle \to 0.
\]

\paragraph{Strong Coupling (\( J \gg h \))}:
\[
r \to 1 \Rightarrow \langle \Delta H_B \rangle \to 0, \quad \langle \Delta Q_B \rangle \to -\frac{1}{2} (-1)^a.
\]

Hence, energy teleportation vanishes in both limits, while charge teleportation asymptotically converges to a maximal bit-dependent value. This reinforces the interpretation that energy teleportation is optimal in intermediate coupling regimes, while charge teleportation retains discriminative power even in asymptotic regimes, which is crucial for its application in secure quantum key distribution (QKD).

\subsubsection{Conclusion}

This detailed analytical derivation of the two-qubit system illustrates the entire teleportation protocol explicitly, including the dependence of teleported observables on the bit \( a \), the system parameters \( h, J \), and the resulting physical correlations. These results validate the structure of the QET protocol, justify the expressions used in the main text, and serve as a precise benchmark for simulations conducted in larger systems.

\section{Numerical Simulation}
\subsection{Detailed Simulation Results for Star Interaction Hamiltonian \texorpdfstring{\( H^{(1)} \)}{H1}}
\label{appendix:numerical-simulation}

In this appendix, we present a more comprehensive study of the teleportation protocols under the star-type Hamiltonian \( H^{(1)} \), focusing on the dependence of the results on the transverse field \( h \), while keeping the coupling strength \( J = 1 \) fixed.

\subsubsection{Teleportation Fidelity vs External Field \texorpdfstring{\( h \)}{h}}

\begin{figure*}[t!]
    \centering
    \subfloat[Energy, \( N = 1 \), \( h \)]{
        \includegraphics[width=0.225\textwidth]{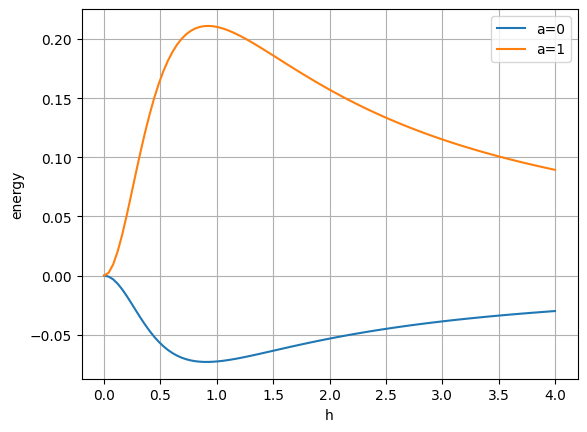}
    }
    \hfill
    \subfloat[Energy, \( N = 2 \), \( h \)]{
        \includegraphics[width=0.225\textwidth]{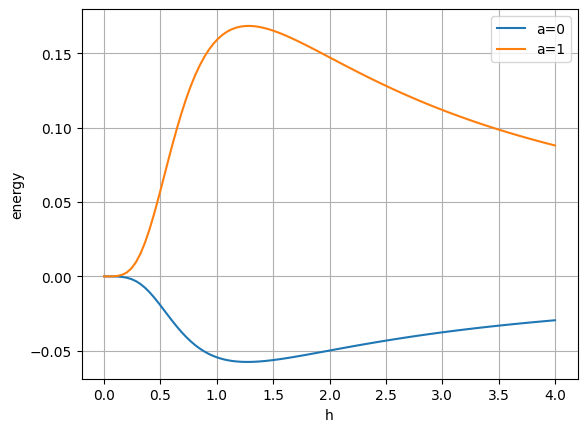}
    }
    \hfill
    \subfloat[Energy, \( N = 3 \), \( h \)]{
        \includegraphics[width=0.225\textwidth]{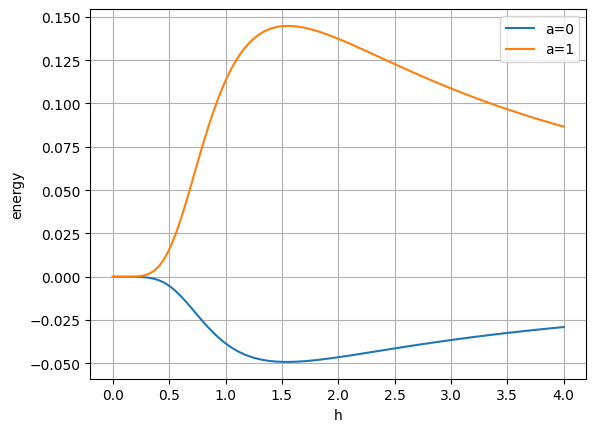}
    }
    \hfill
    \subfloat[Energy, \( N = 4 \), \( h \)]{
        \includegraphics[width=0.225\textwidth]{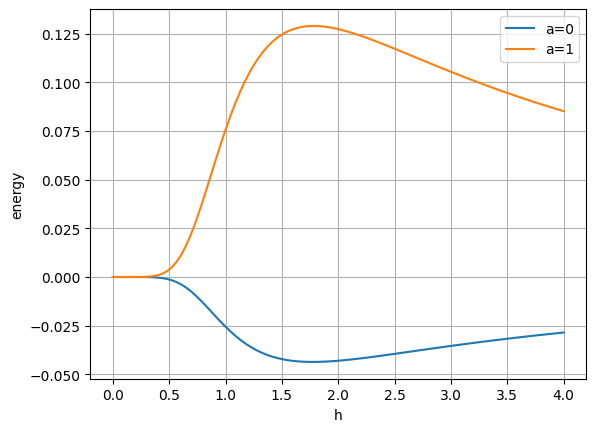}
    }

    \vspace{0.5em}

    \subfloat[Charge, \( N = 1 \), \( h \)]{
        \includegraphics[width=0.225\textwidth]{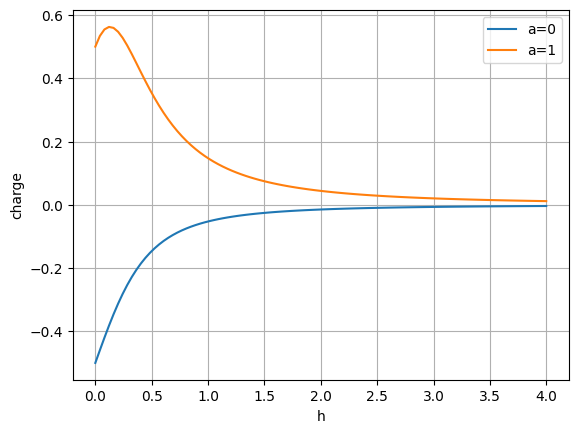}
    }
    \hfill
    \subfloat[Charge, \( N = 2 \), \( h \)]{
        \includegraphics[width=0.225\textwidth]{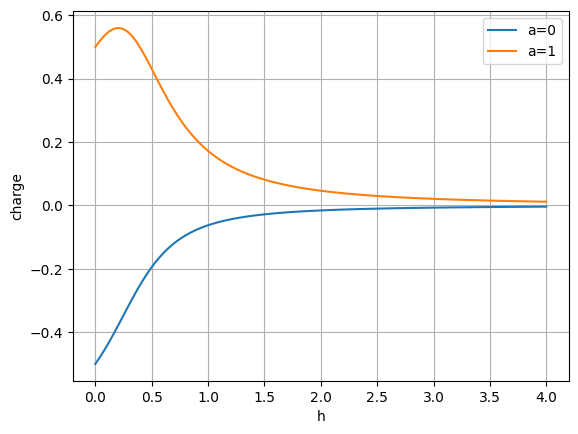}
    }
    \hfill
    \subfloat[Charge, \( N = 3 \), \( h \)]{
        \includegraphics[width=0.225\textwidth]{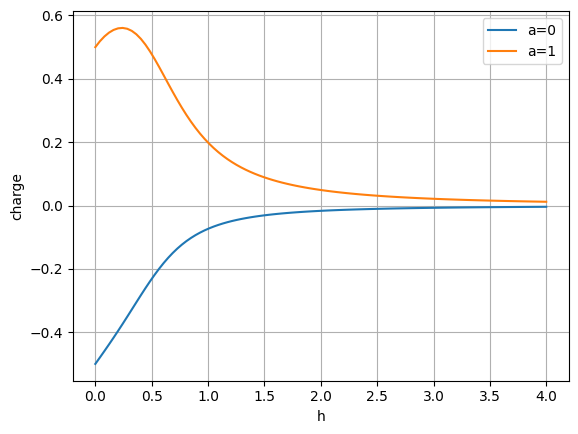}
    }
    \hfill
    \subfloat[Charge, \( N = 4 \), \( h \)]{
        \includegraphics[width=0.225\textwidth]{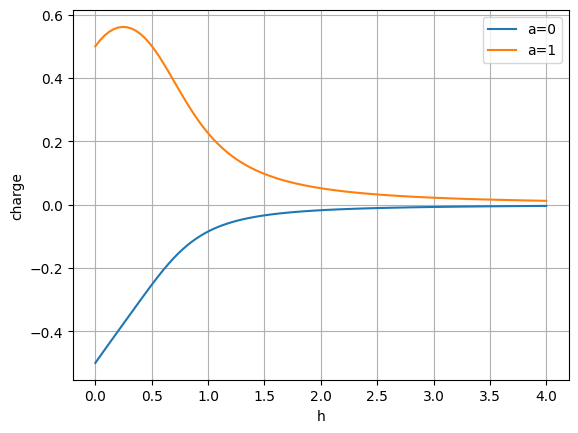}
    }
    
    \vspace{0.5em}

    \subfloat[Energy, \( N = 1 \), \( J \)]{
        \includegraphics[width=0.225\textwidth]{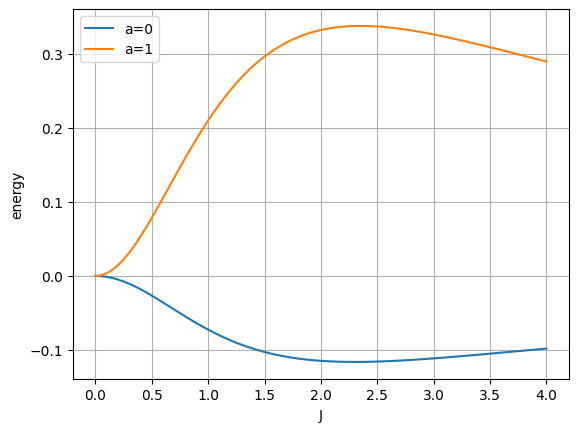}
    }
    \hfill
    \subfloat[Energy, \( N = 2 \), \( J \)]{
        \includegraphics[width=0.225\textwidth]{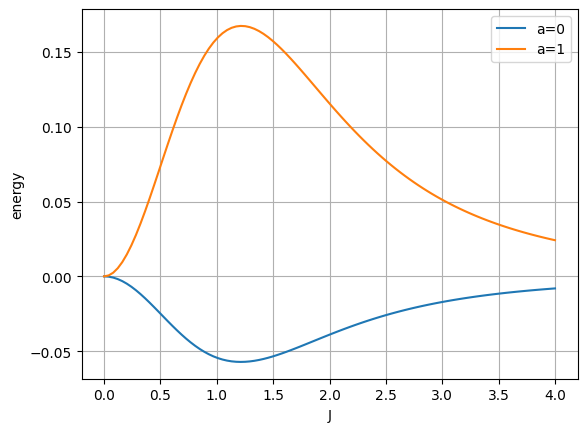}
    }
    \hfill
    \subfloat[Energy, \( N = 3 \), \( J \)]{
        \includegraphics[width=0.225\textwidth]{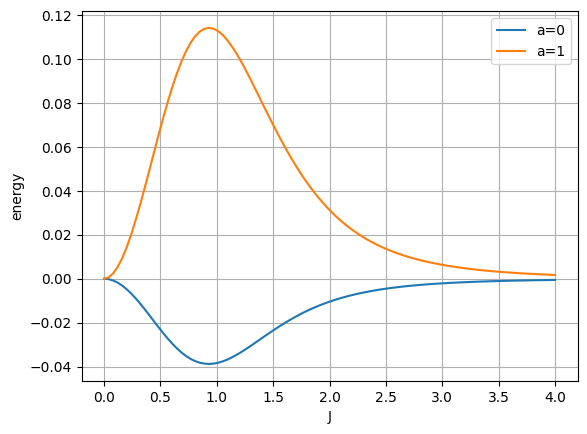}
    }
    \hfill
    \subfloat[Energy, \( N = 4 \), \( J \)]{
        \includegraphics[width=0.225\textwidth]{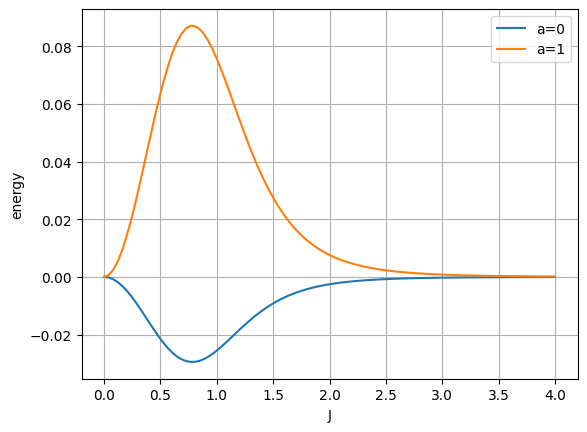}
    }

    \caption{Extracted value at Bob's site, under Star-Interaction Hamiltonian for Energy and Charge.}
    \label{fig:alice_vs_hJ}
\end{figure*}

The plots in Figure~\ref{fig:alice_vs_hJ} show the behavior of Bob's expectation values across varying values of \( h \) and \( J \) as completion to Figure~\ref{fig:alice_x_vs_j} from main text. We observe:

\begin{itemize}
    \item For \textbf{energy teleportation}, the fidelity degrades as \( h \) increases, due to the weakening of \( X_0 X_k \) correlations dominating the interaction term.
    \item For \textbf{charge teleportation}, a similar trend is observed, but the degradation is smoother, indicating better resilience to increasing transverse field strengths.
\end{itemize}

\subsubsection{Protocol Optimization and Design Implications}

From the simulations, we find that optimal teleportation occurs at intermediate field strengths and coupling ratios, typically where entanglement entropy is near its peak. This aligns with previous findings in~\cite{Ikeda2023, IkedaTeleportingCharge}. These results suggest:

\begin{itemize}
    \item Energy teleportation is more sensitive to parameter tuning, requiring careful Hamiltonian engineering.
    \item Charge teleportation, while delivering slightly smaller absolute observable values, is more stable and symmetric, offering better QKD performance under noise or imperfect conditions.
\end{itemize}

\subsubsection{Robustness to Bit Flipping}

As shown in both main text and simulations, the expectation values under charge teleportation exhibit perfect mirror symmetry under the bit-flipping of Alice’s result \( b \to b \oplus 1 \). This provides a useful check for correctness and a natural framework for encoding binary key bits, as emphasized in~\cite{QKDbyQET}.

These findings strengthen the case for preferring charge-based QKD implementations in practical systems, especially where channel errors or imperfect hardware may break energy teleportation asymmetries.

\subsection{Additional Results for Nearest-Neighbor Interaction}
\label{appendix:nn_detailed}

This appendix includes further numerical results for the nearest-neighbor Hamiltonian, focusing on the dependence of teleportation fidelity on the external field \( h \) for different measurement bases.

\subsubsection{Observable Expectation Values vs. \texorpdfstring{\( h \)}{h}}

\begin{figure*}[t!]
    \centering
    \subfloat[Energy, \( N = 2 \)]{
        \includegraphics[width=0.3\textwidth]{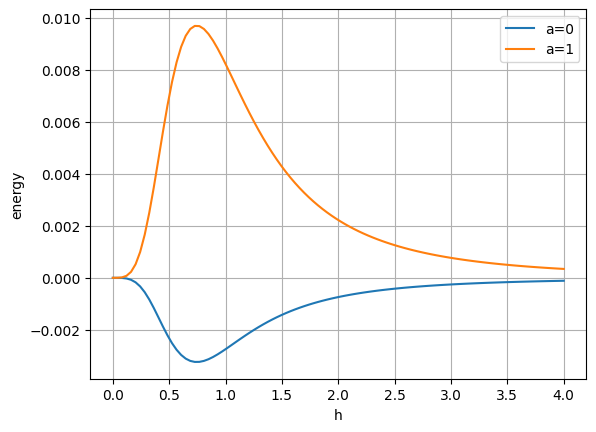}
    }
    \hfill
    \subfloat[Energy, \( N = 3 \)]{
        \includegraphics[width=0.3\textwidth]{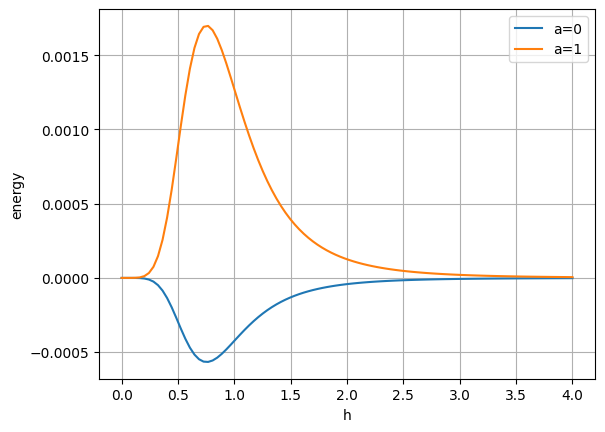}
    }
    \hfill
    \subfloat[Energy, \( N = 4 \)]{
        \includegraphics[width=0.3\textwidth]{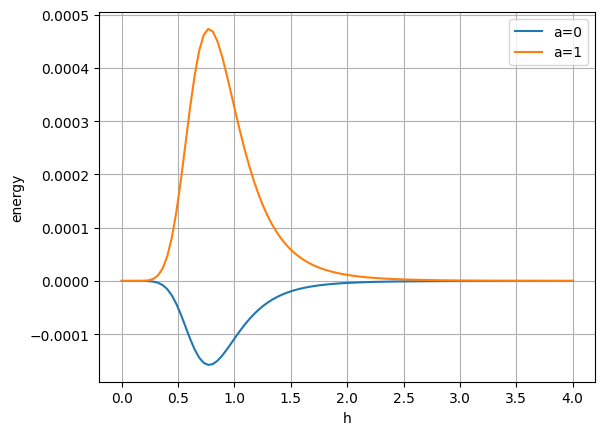}
    }

    \vspace{0.5em}

    \subfloat[Charge, \( N = 2 \)]{
        \includegraphics[width=0.3\textwidth]{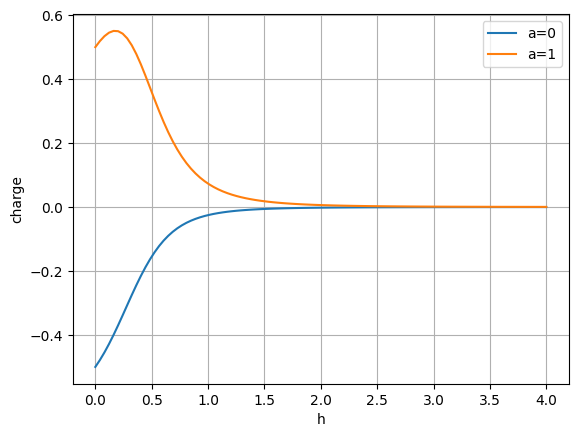}
    }
    \hfill
    \subfloat[Charge, \( N = 3 \)]{
        \includegraphics[width=0.3\textwidth]{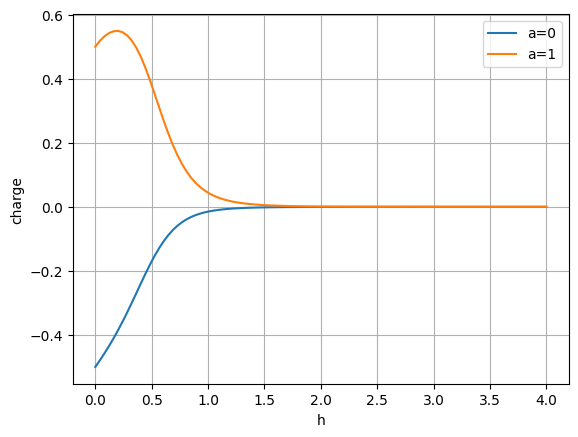}
    }
    \hfill
    \subfloat[Charge, \( N = 4 \)]{
        \includegraphics[width=0.3\textwidth]{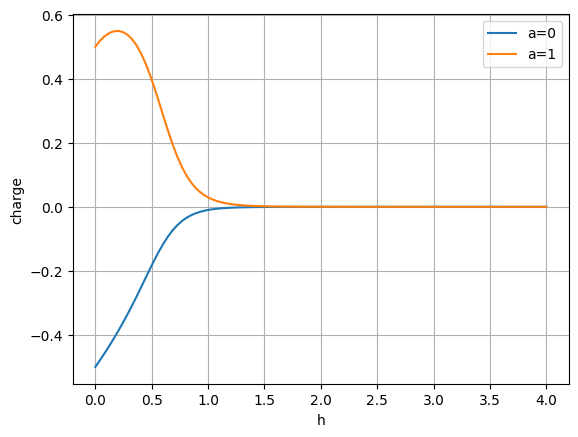}
    }

    \caption{Extracted value at Bob's site vs. \( h \), under nearest-neighbor Hamiltonian for Energy and Charge for \( \sigma_A = X_0 \).}
    \label{fig:nn_vs_h_X}
\end{figure*}

\begin{figure*}[t!]
    \centering
    \subfloat[Energy, \( N = 2 \)]{
        \includegraphics[width=0.3\textwidth]{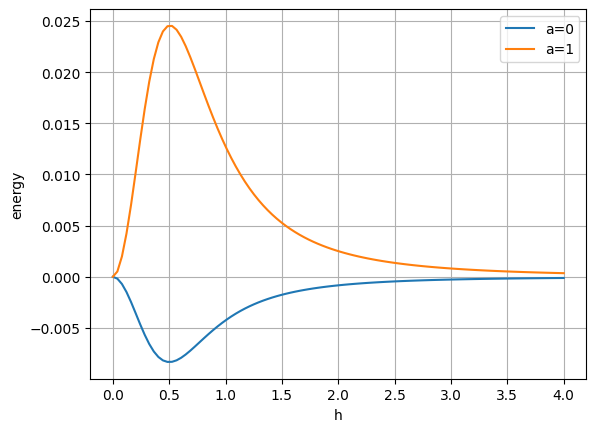}
    }
    \hfill
    \subfloat[Energy, \( N = 3 \)]{
        \includegraphics[width=0.3\textwidth]{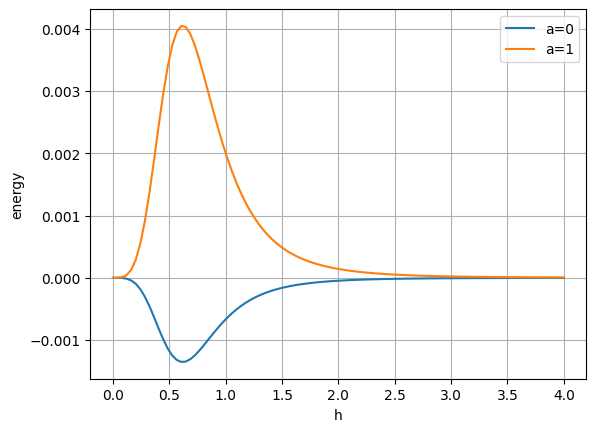}
    }
    \hfill
    \subfloat[Energy, \( N = 4 \)]{
        \includegraphics[width=0.3\textwidth]{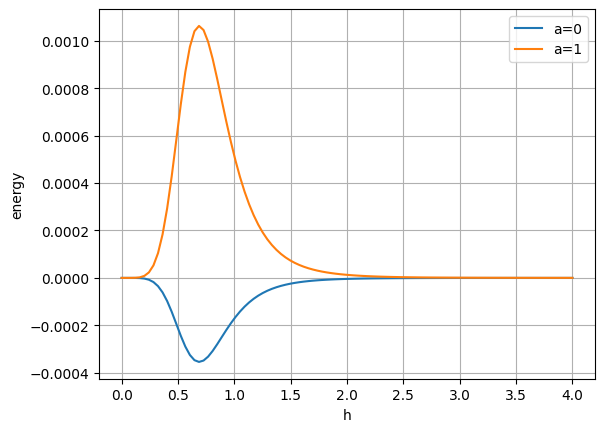}
    }
    \vspace{0.5em}

    \subfloat[Charge, \( N = 2 \)]{
        \includegraphics[width=0.3\textwidth]{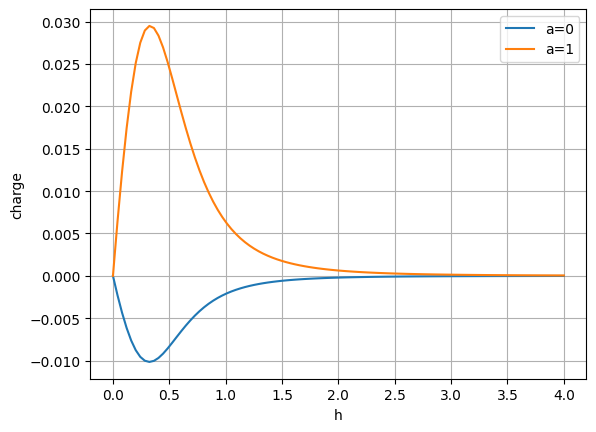}
    }
    \hfill
    \subfloat[Charge, \( N = 3 \)]{
        \includegraphics[width=0.3\textwidth]{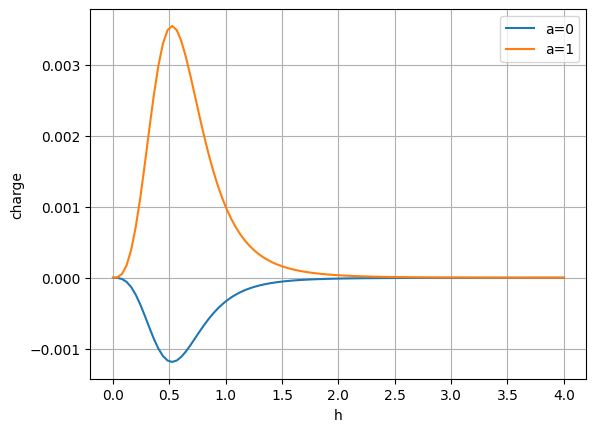}
    }
    \hfill
    \subfloat[Charge, \( N = 4 \)]{
        \includegraphics[width=0.3\textwidth]{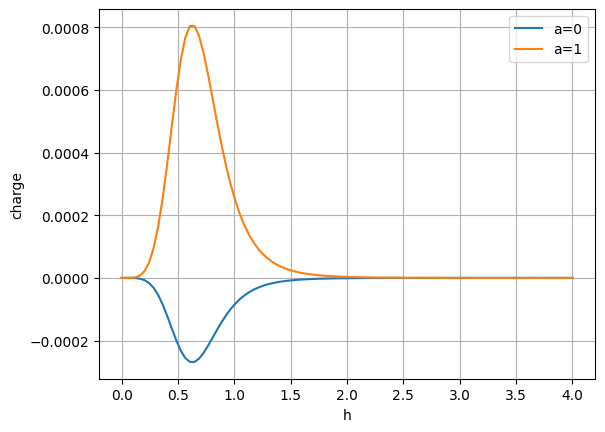}
    }
    
    \caption{Extracted value at Bob's site vs. \( h \), under nearest-neighbor Hamiltonian for Energy and Charge for \( \sigma_A = Y_0 \).}
    \label{fig:nn_vs_h_Y}
\end{figure*}

\begin{figure*}[t!]
    \centering
    \subfloat[Energy, \( N = 2 \)]{
        \includegraphics[width=0.3\textwidth]{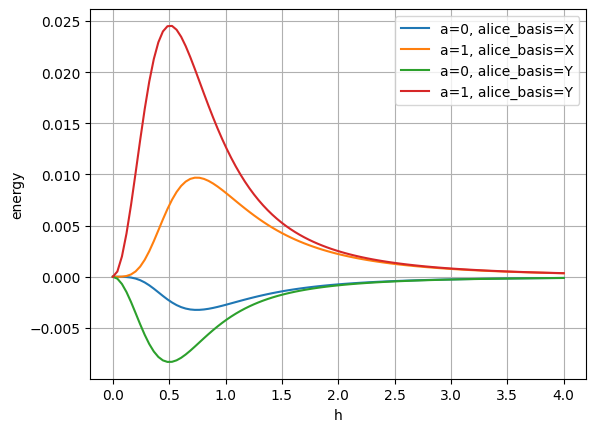}
    }
    \hfill
    \subfloat[Energy, \( N = 3 \)]{
        \includegraphics[width=0.3\textwidth]{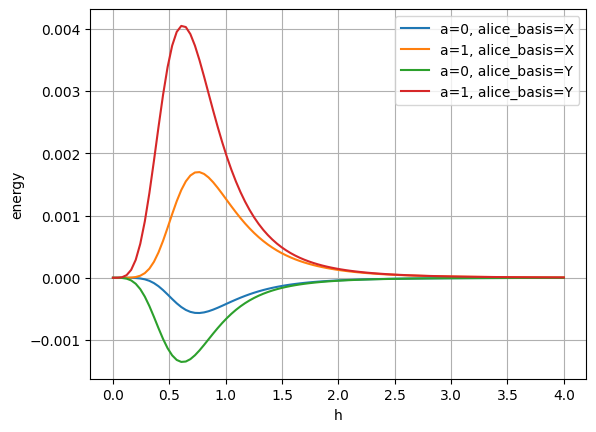}
    }
    \hfill
    \subfloat[Energy, \( N = 4 \)]{
        \includegraphics[width=0.3\textwidth]{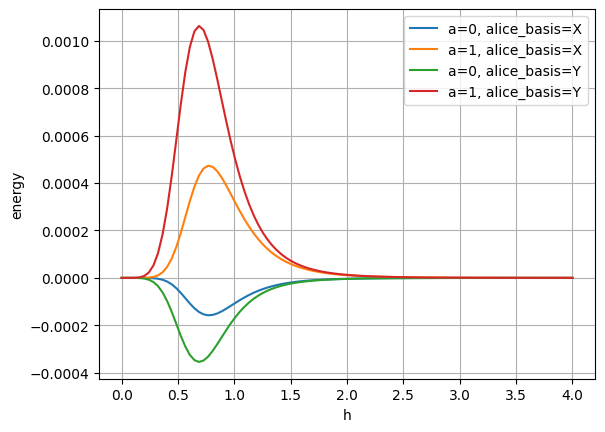}
    }

    \vspace{0.5em}

    \subfloat[Charge, \( N = 2 \)]{
        \includegraphics[width=0.3\textwidth]{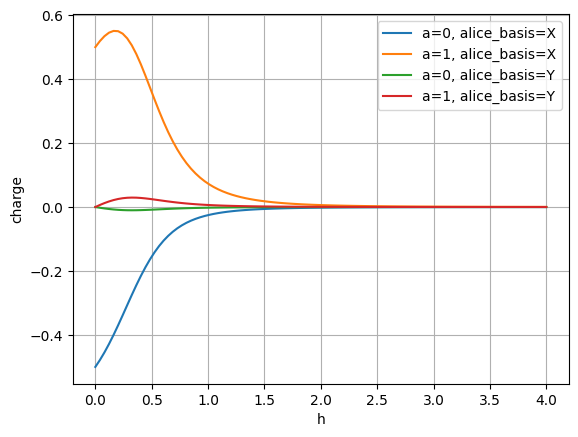}
    }
    \hfill
    \subfloat[Charge, \( N = 3 \)]{
        \includegraphics[width=0.3\textwidth]{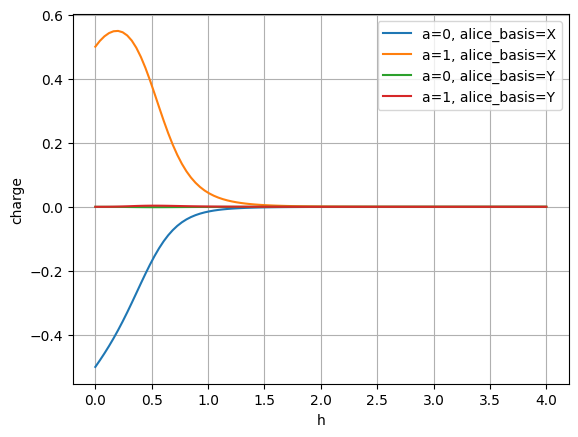}
    }
    \hfill
    \subfloat[Charge, \( N = 4 \)]{
        \includegraphics[width=0.3\textwidth]{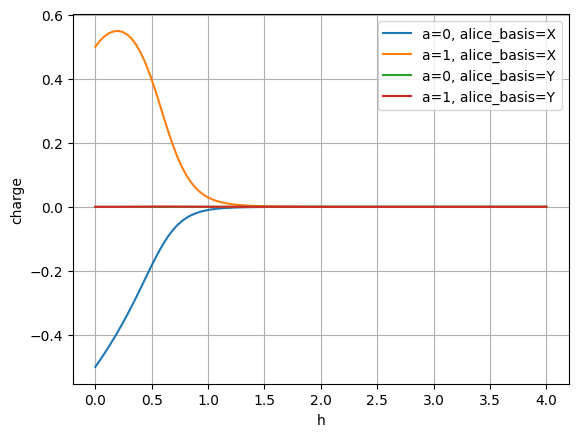}
    }
    
    \caption{Extracted value at Bob's site vs. \( h \), under nearest-neighbor Hamiltonian for Energy and Charge for both Alice Bases.}
    \label{fig:nn_vs_h_both}
\end{figure*}

Figures~\ref{fig:nn_vs_h_X},~\ref{fig:nn_vs_h_Y}, and~\ref{fig:nn_vs_h_both} present the behavior of teleported observables as a function of \( h \), with fixed coupling \( J=1 \). The trends are qualitatively similar to the \( J \)-dependence, but with more rapid decay in observable magnitude for large \( h \), due to the dominance of the local \( Z \)-field.

\begin{figure*}[t!]
    \centering
    \subfloat[\( \sigma_A = X_0 \), \( N = 2 \)]{
        \includegraphics[width=0.3\textwidth]{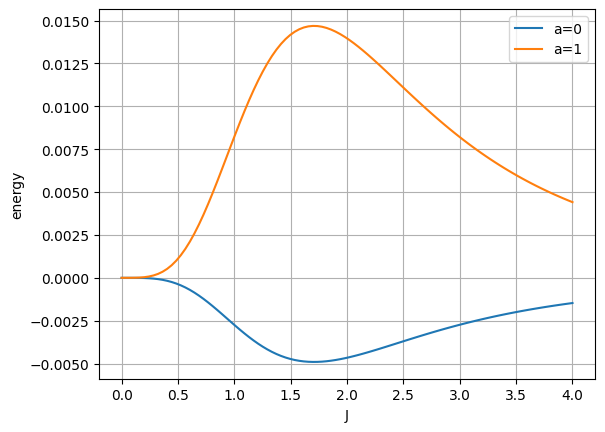}
    }
    \hfill
    \subfloat[\( \sigma_A = X_0 \), \( N = 3 \)]{
        \includegraphics[width=0.3\textwidth]{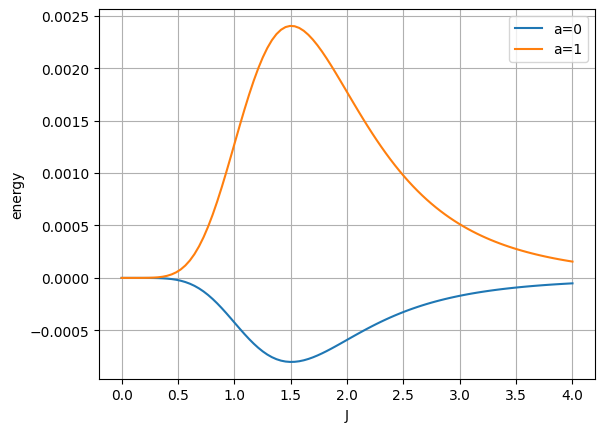}
    }
    \hfill
    \subfloat[\( \sigma_A = X_0 \), \( N = 4 \)]{
        \includegraphics[width=0.3\textwidth]{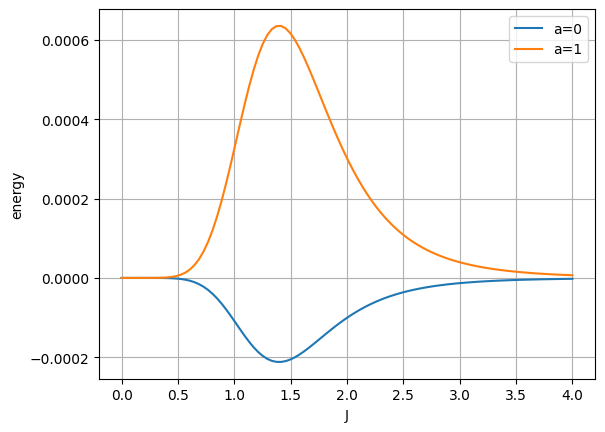}
    }

    \vspace{0.5em}

    \subfloat[\( \sigma_A = Y_0 \), \( N = 2 \)]{
        \includegraphics[width=0.3\textwidth]{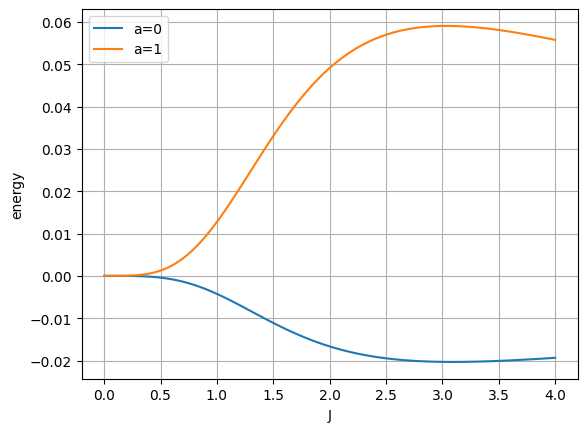}
    }
    \hfill
    \subfloat[\( \sigma_A = Y_0 \), \( N = 3 \)]{
        \includegraphics[width=0.3\textwidth]{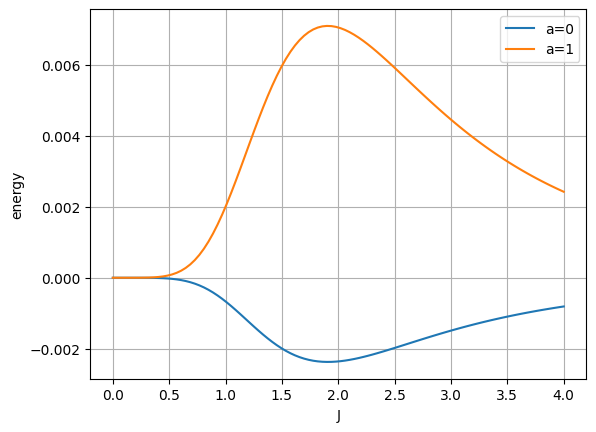}
    }
    \hfill
    \subfloat[\( \sigma_A = Y_0 \), \( N = 4 \)]{
        \includegraphics[width=0.3\textwidth]{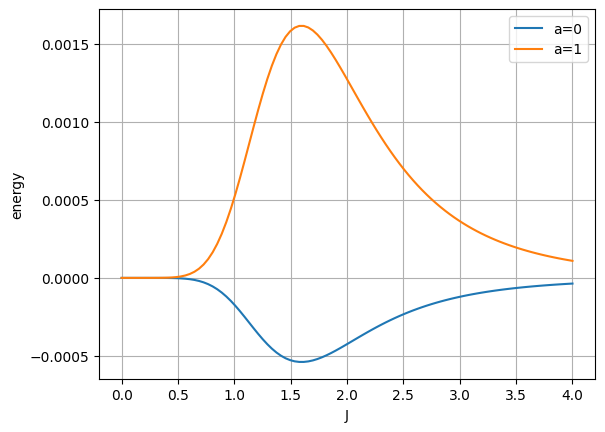}
    }

    \vspace{0.5em}

    \subfloat[\( N = 2 \)]{
        \includegraphics[width=0.3\textwidth]{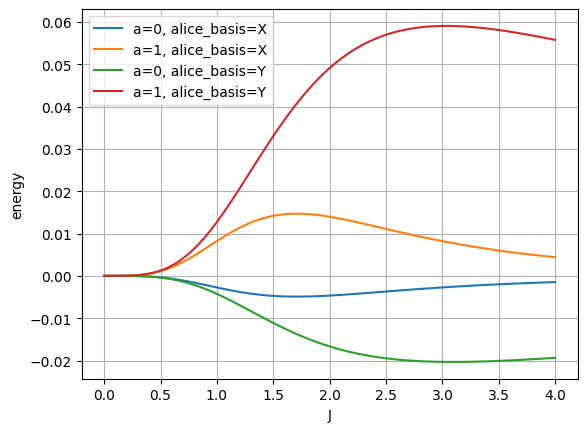}
    }
    \hfill
    \subfloat[\( N = 3 \)]{
        \includegraphics[width=0.3\textwidth]{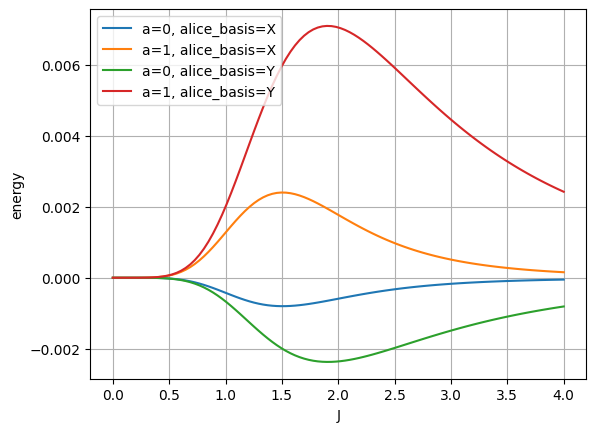}
    }
    \hfill
    \subfloat[\( N = 4 \)]{
        \includegraphics[width=0.3\textwidth]{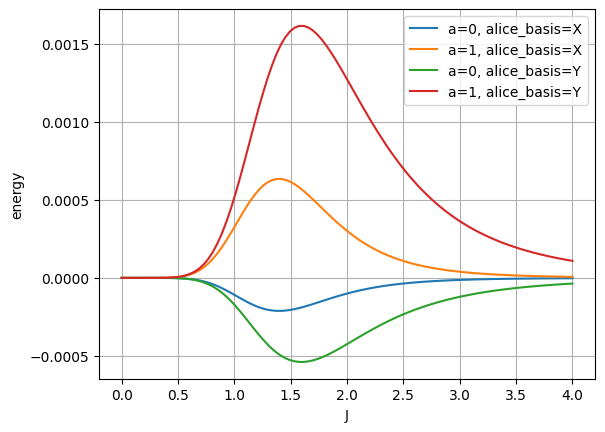}
    }

    \caption{Extracted Energy value at Bob's site vs. \( J \), under nearest-neighbor Hamiltonian.}
    \label{fig:nn_vs_J}
\end{figure*}

\subsubsection{Summary of Observations}

\begin{itemize}
    \item The charge protocol remains strongly dependent on the basis choice, with the difference between X and Y observables increasing with \( N \).
    \item Energy teleportation results are more uniform across bases, indicating a \textbf{broader operational regime}.
    \item The field strength \( h \) controls the overall scale of extracted observables, but does not invert qualitative trends.
\end{itemize}

These findings reinforce the interpretation that energy teleportation offers \textbf{greater robustness to control imperfections}, while charge teleportation may provide \textbf{stronger signals} under optimized conditions.

\section{Qiskit Simulation}
\label{appendix:qiskit-simulation}

\begin{figure}[ht!]
    \centering
    \includegraphics[width=0.45\textwidth]{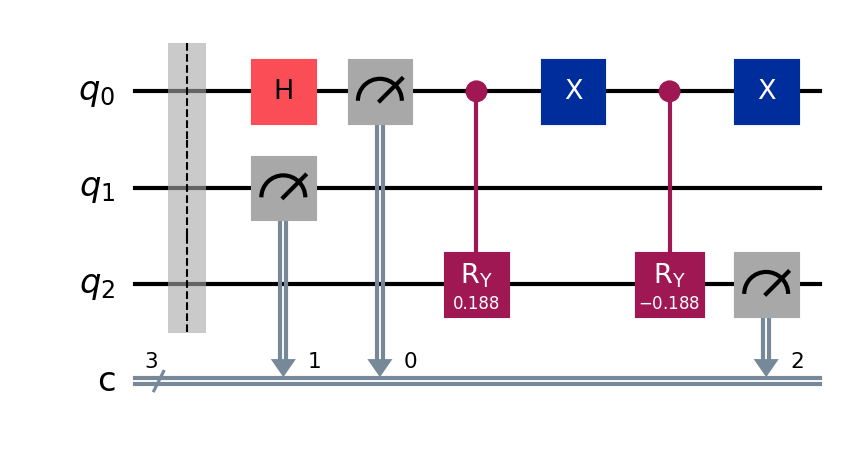}
    \caption{Quantum circuit measuring the local term \( H_b \) in the energy teleportation protocol.}
    \label{fig:H1_qc}
\end{figure}

\begin{figure}[ht!]
    \centering
    \includegraphics[width=0.45\textwidth]{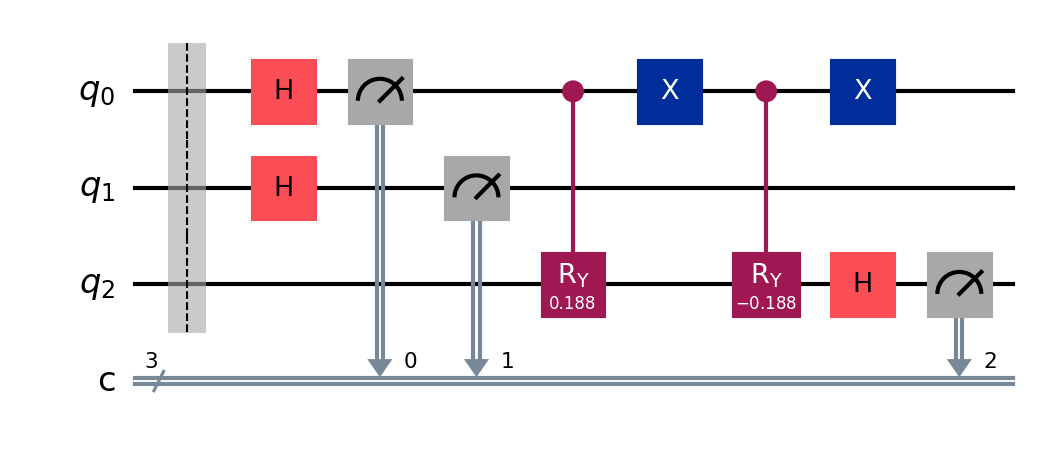}
    \caption{Quantum circuit measuring the interaction term \( V_{ab} \) in the energy teleportation protocol.}
    \label{fig:V_qc}
\end{figure}

This appendix presents the quantum circuits used throughout this work. We include the energy-teleportation components \(H_1\) and \(V\), each measured at Bob’s site. As discussed in Section~\ref{sec:complex-sum-constructed-operator}, \([H_1,V]\neq 0\), so \(H_1\) and \(V\) must be evaluated in two separate circuits. The circuit designs closely follow Ikeda’s implementation of quantum energy teleportation on superconducting hardware~\cite{Ikeda2023}.

\section{Statistical Framework for Quantum Circuit Simulations}
\label{appendix:stats-analysis}

The results presented in Section~\ref{sec:qiskit_sim} of the main paper are derived from simulations of quantum circuits executed on Qiskit's `aer\_simulator` and on real quantum hardware. Each execution, or "shot," is a probabilistic measurement of the system's $(N+1)$ qubits in the computational basis. This appendix details the statistical methodology used to calculate the expectation values of observables and their associated uncertainties.

\subsection{Core Assumptions}
Our analysis is based on the following standard assumptions for quantum circuit simulations:
\begin{itemize}
    \item \textbf{Independent and Identically Distributed (i.i.d.) Trials:} Each shot is an independent measurement drawn from the same underlying probability distribution defined by the final quantum state. The outcome of one shot does not influence any other.
    \item \textbf{Multinomial Distribution:} For a total of $n_{\text{shots}}$ trials, the set of counts $\{c_0, c_1, \dots, c_{2^{N+1}-1}\}$ for each of the $2^{N+1}$ basis states follows a multinomial distribution.
    \item \textbf{Central Limit Theorem (CLT):} The expectation value of an observable $\langle O \rangle$ is a sample mean calculated from a large number of shots. According to the CLT, for a sufficiently large $n_{\text{shots}}$, the sampling distribution of this mean is well-approximated by a Normal (Gaussian) distribution. This justifies the use of standard error as a measure of statistical uncertainty.
\end{itemize}

\subsection{Calculating Expectation Values and Statistical Uncertainty}

\subsubsection{Expectation Value}
An observable $O$ is measured by post-processing the counts from the simulation. The expectation value $\langle O \rangle$ is computed as:
\begin{equation}
    \langle O \rangle = \sum_{i=0}^{2^{N+1}-1} p_i \lambda_i
\end{equation}
where $p_i = c_i / n_{\text{shots}}$ is the estimated probability of measuring the basis state $|i\rangle$, and $\lambda_i = \langle i | O | i \rangle$ is the corresponding eigenvalue of the observable $O$.

\subsubsection{Statistical Uncertainty (Standard Error of the Mean)}
The statistical uncertainty of the estimated expectation value is given by the Standard Error of the Mean (SEM), denoted $\sigma_{\langle O \rangle}$. This is the quantity represented by the error bars in the plots. The SEM is calculated as:
\begin{equation}
    \sigma_{\langle O \rangle} = \frac{\sigma_O}{\sqrt{n_{\text{shots}}}}
\end{equation}
where $\sigma_O$ is the standard deviation of a \textit{single} measurement of the observable $O$. The variance, $\sigma_O^2$, is calculated directly from the simulation results using the formula:
\begin{equation}
    \begin{aligned}
        \sigma_O^2 &= \text{Var}(O) = \langle O^2 \rangle - \langle O \rangle^2 \\
        &= \left( \sum_{i} p_i \lambda_i^2 \right) - \left( \sum_{i} p_i \lambda_i \right)^2
    \end{aligned}
\end{equation}
This provides a direct method to quantify the statistical noise inherent in the Monte Carlo nature of the simulation.

\subsection{Application to Teleportation Observables}

We now apply this framework to the specific observables measured in the charge and energy teleportation protocols.

\subsubsection{Charge Teleportation (\texorpdfstring{$Q_B$}{QB})}
The charge at Bob's site is given by the operator $Q_B = \frac{1}{2}(I + Z_N)$. Since $Q_B$ is a projector, its eigenvalues are 0 and 1. The variance of a single measurement is:
\begin{equation}
    \text{Var}(Q_B) = \langle Q_B^2 \rangle - \langle Q_B \rangle^2 = \langle Q_B \rangle - \langle Q_B \rangle^2
\end{equation}
Using the relation $\langle Q_B \rangle = \frac{1}{2}(1 + \langle Z_N \rangle)$, we find:
\begin{equation}
    \begin{aligned}
        \text{Var}(Q_B) &= \frac{1}{2}(1 + \langle Z_N \rangle) - \frac{1}{4}(1 + \langle Z_N \rangle)^2 \\
        &= \frac{1}{4}(1 - \langle Z_N \rangle^2)
    \end{aligned}
\end{equation}
Thus, the SEM for the charge measurement is:
\begin{equation}
    \sigma_{\langle Q_B \rangle} = \frac{1}{2\sqrt{n_{\text{shots}}}} \sqrt{1 - \langle Z_N \rangle^2}
\end{equation}
Since this calculation relies on a single circuit execution (measuring in the Z-basis), the statistical analysis is straightforward.

\subsubsection{Energy Teleportation (\texorpdfstring{$H_B$}{HB})}
Bob's local Hamiltonian, for instance in the nearest-neighbor model ($H^{(2)}$), is $H_B = hZ_N + JX_{N-1}X_N$. The two terms, $Z_N$ and $X_{N-1}X_N$, do not commute and must be measured in separate circuit executions. This necessitates the use of error propagation to find the total uncertainty.

The expectation value is $\langle H_B \rangle = h\langle Z_N \rangle + J\langle X_{N-1}X_N \rangle$. Since the measurements for $\langle Z_N \rangle$ and $\langle X_{N-1}X_N \rangle$ are from independent sets of shots, their statistical errors are uncorrelated. The total variance of the mean is the sum of the individual variances:
\begin{equation}
    \begin{aligned}
        \sigma_{\langle H_B \rangle}^2 &= \text{Var}(h\langle Z_N \rangle) + \text{Var}(J\langle X_{N-1}X_N \rangle) \\
        &= h^2 \sigma_{\langle Z_N \rangle}^2 + J^2 \sigma_{\langle X_{N-1}X_N \rangle}^2
    \end{aligned}
\end{equation}
The individual SEMs are calculated as:
\begin{align}
    \sigma_{\langle Z_N \rangle}^2 &= \frac{1}{n_{\text{shots,Z}}} \left( \langle Z_N^2 \rangle - \langle Z_N \rangle^2 \right) \nonumber\\
    &= \frac{1}{n_{\text{shots,Z}}} \left( 1 - \langle Z_N \rangle^2 \right) \\
    \sigma_{\langle X_{N-1}X_N \rangle}^2 &= \frac{1}{n_{\text{shots,X}}} \left( 1 - \langle X_{N-1}X_N \rangle^2 \right)
\end{align}
The final expression for the uncertainty in the energy measurement is:
\begin{equation}
    \sigma_{\langle H_B \rangle} = \sqrt{
        \begin{aligned}    
        &\frac{h^2}{n_{\text{shots,Z}}} (1 - \langle Z_N \rangle^2) \\
        &+ \frac{J^2}{n_{\text{shots,X}}} (1 - \langle X_{N-1}X_N \rangle^2)
        \end{aligned}
    }
\end{equation}
This explicitly shows that the statistical uncertainty for the energy is a combination of noise from two independent experiments, quantitatively explaining the increased fluctuations observed in the Qiskit energy plots compared to the charge plots.

\subsection{Summary and Conclusions}

This analysis provides a rigorous basis for the statistical error bars and the discussion of noise in the main text. Key conclusions include:
\begin{itemize}
    \item The increased statistical noise in energy measurements is a direct consequence of error propagation from measuring non-commuting observables in separate experiments.
    \item The "meaningful statistical noise" observed for larger systems (e.g., $N=3, 4$ in the nearest-neighbor model) is a result of a decreasing Signal-to-Noise Ratio (SNR). As $N$ increases, the teleported signal $|\langle O \rangle|$ diminishes due to weaker correlations, while the single-shot variance $\sigma_O^2$ remains approximately constant (approaching 1). Consequently, the SEM $\sigma_{\langle O \rangle}$ becomes comparable to or larger than the signal itself, requiring a significantly larger $n_{\text{shots}}$ to resolve.
    \item The stability of the charge teleportation protocol is attributed to its reliance on a single measurement basis and a consistently strong signal across different coupling regimes.
\end{itemize}

\section{Extended Results: Noise and Errors}
\label{appendix:noise-errors}

This appendix provides an extended analysis of the QKD protocol's performance under various noise models, complementing the discussion in Section~\ref{sec:noise-errors}. We present a comprehensive set of numerical simulations for both the star-interaction Hamiltonian ($H^{(1)}$) and the nearest-neighbor Hamiltonian ($H^{(2)}$), examining the resilience of both charge and energy teleportation.

\begin{table*}[ht!]
\centering
\begin{tabular}{|l|p{6cm}|p{6cm}|}
\hline
\textbf{Error Model} & \textbf{Star-Coupled ($H^{(1)}$) Trend} & \textbf{Nearest-Neighbor ($H^{(2)}$) Trend} \\ \hline
Classical Comm. Error & Linear decay; Sign flip at $p \approx 0.25$. Energy signal stronger but equally fragile. & Linear decay. Charge signal is smaller but maintains sign integrity longer for large $N$. \\ \hline
Excited State Mixture & Energy signal flips sign rapidly ($p < 0.2$). Charge signal decays but remains negative (secure). & Same trend. Energy fails almost immediately; Charge is highly robust ($p > 0.4$). \\ \hline
Bit-Flip ($X$) & Gradual decay for both. No sharp failure threshold. & Smooth decay. Charge signal remains distinguishable up to high error rates. \\ \hline
Phase-Flip ($Z$) on Bob & Rapid linear decay. Fails at $p \approx 0.33$. & Rapid decay for both. Dominant failure mode for Bob's side errors. \\ \hline
\end{tabular}
\caption{Summary of noise resilience trends. The Star-Coupled ($H^{(1)}$) results (omitted for brevity) exhibit qualitatively similar behavior to the Nearest-Neighbor ($H^{(2)}$) results shown below, though with generally higher absolute signal magnitudes due to direct connectivity.}
\label{tab:noise_summary}
\end{table*}

\subsection{Classical Communication Error}

\begin{figure*}[tb!]
    \centering
    \subfloat[Star-Coupled interaction \( H^{(1)} \)]{
        \includegraphics[width=0.3\textwidth]{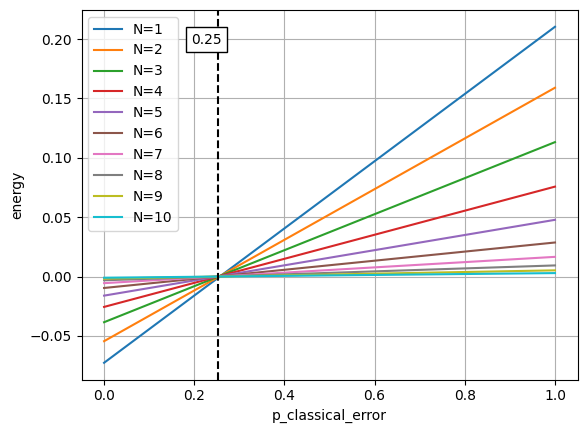}
    }
    \hfill
    \subfloat[Nearest-Neighbors interaction \( H^{(2)} \), \( \sigma_A = X_0 \)]{
        \includegraphics[width=0.3\textwidth]{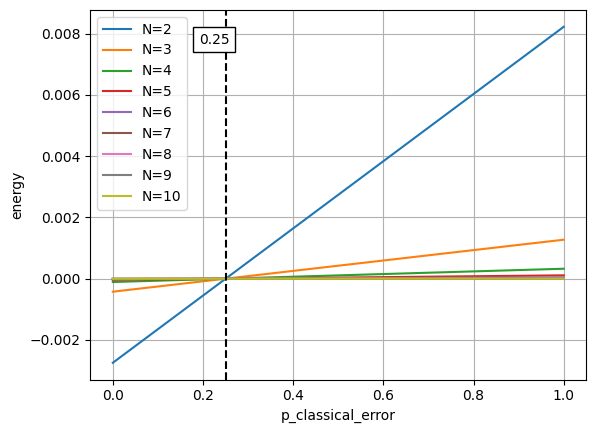}
    }
    \hfill
    \subfloat[Nearest-Neighbors interaction \( H^{(2)} \), \( \sigma_A = Y_0 \)]{
        \includegraphics[width=0.3\textwidth]{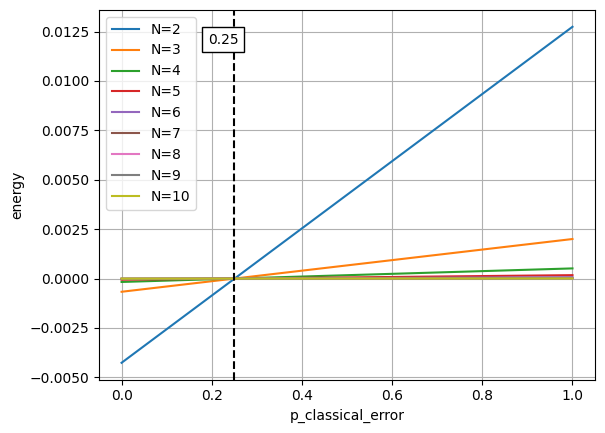}
    }

    \caption{Energy vs. Classical communication error for different \(N\) values. The linear degradation of the signal is evident across all models, with a consistent failure threshold where the expectation value crosses zero. The signal magnitude for the $H^{(2)}$ model visibly diminishes with increasing system size $N$.}
    \label{fig:energy_vs_classical_err_for_n}
\end{figure*}

\begin{figure*}[tb!]
    \centering
    
    \subfloat[Energy, \( N = 2 \)]{
        \includegraphics[width=0.3\textwidth]{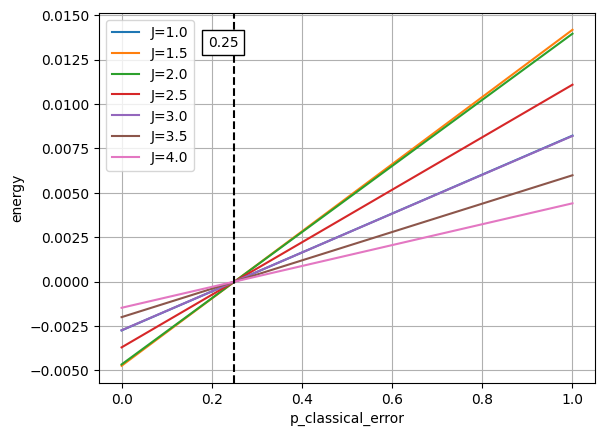}
    }
    \hfill
    \subfloat[Energy, \( N = 3 \)]{
        \includegraphics[width=0.3\textwidth]{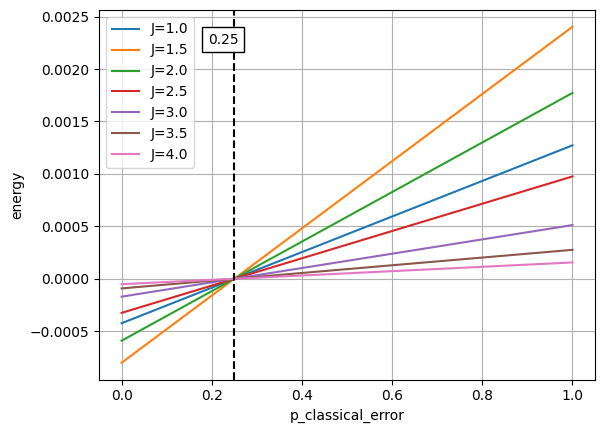}
    }
    \hfill
    \subfloat[Energy, \( N = 4 \)]{
        \includegraphics[width=0.3\textwidth]{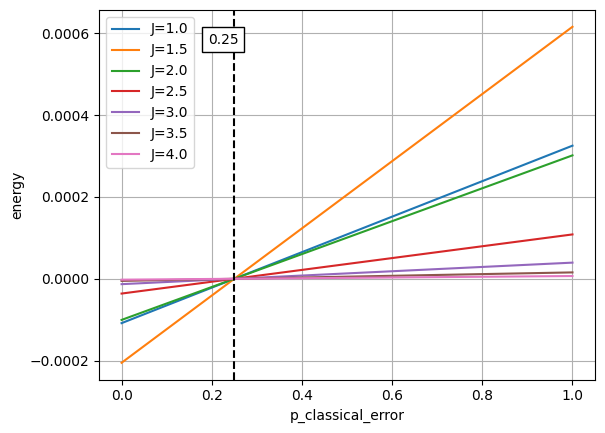}
    }

    \vspace{0.5em}

    \subfloat[Charge, \( N = 2 \)]{
        \includegraphics[width=0.3\textwidth]{plots/errors/numerical/N2_charge_vs_p_classical_error.png}
    }
    \hfill
    \subfloat[Charge, \( N = 3 \)]{
        \includegraphics[width=0.3\textwidth]{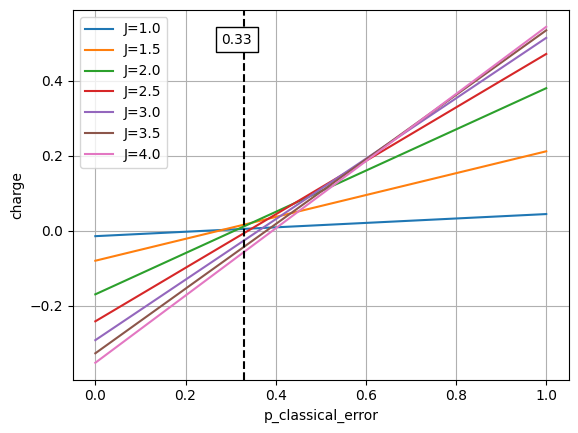}
    }
    \hfill
    \subfloat[Charge, \( N = 4 \)]{
        \includegraphics[width=0.3\textwidth]{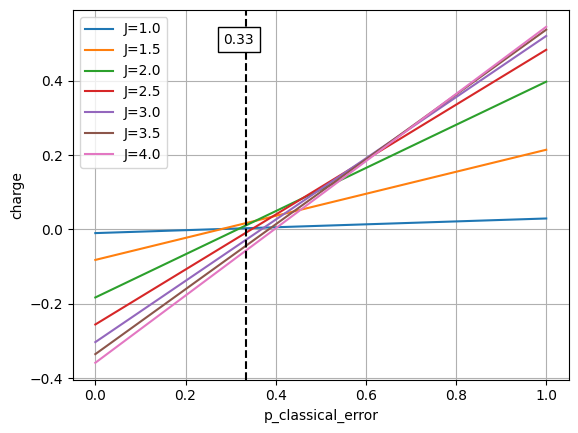}
    }

    \caption{Classical communication error vs the coupling \(J\) for the nearest-neighbor Hamiltonian $H^{(2)}$ with Alice Base \(\sigma_A = X_0\).}
    \label{fig:classical_err_nn_ham_x}
\end{figure*}

\begin{figure*}[tb!]
    \centering
    
    \subfloat[Energy, \( N = 2 \)]{
        \includegraphics[width=0.3\textwidth]{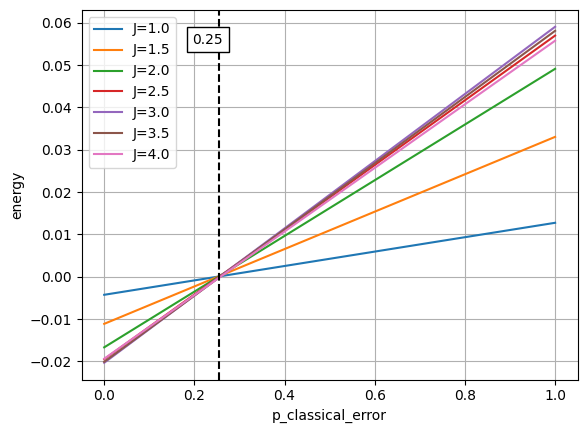}
    }
    \hfill
    \subfloat[Energy, \( N = 3 \)]{
        \includegraphics[width=0.3\textwidth]{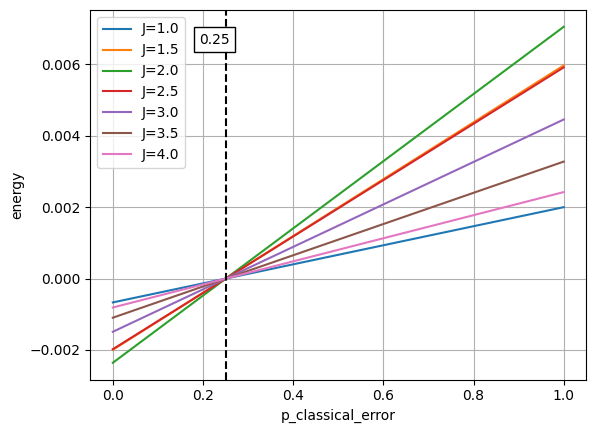}
    }
    \hfill
    \subfloat[Energy, \( N = 4 \)]{
        \includegraphics[width=0.3\textwidth]{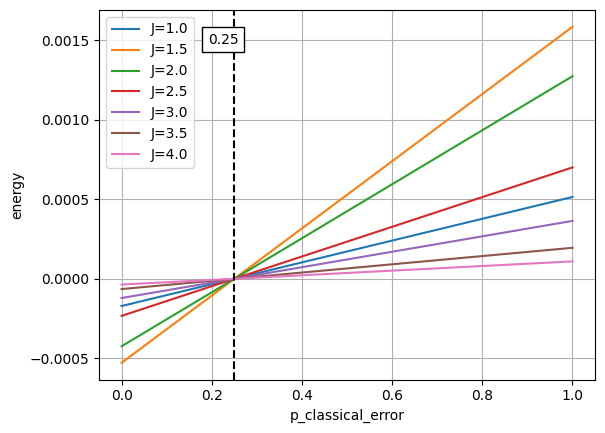}
    }

    \vspace{0.5em}

    \subfloat[Charge, \( N = 2 \)]{
        \includegraphics[width=0.3\textwidth]{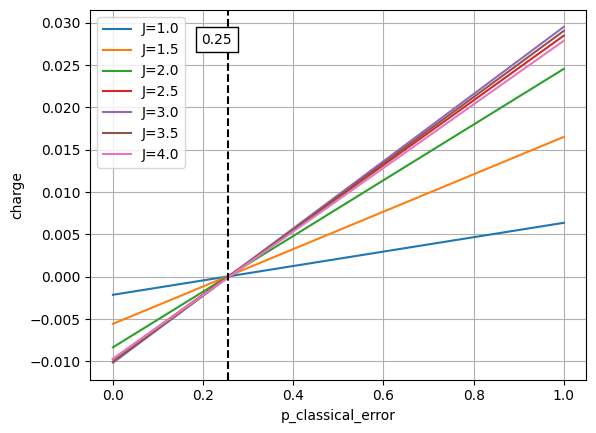}
    }
    \hfill
    \subfloat[Charge, \( N = 3 \)]{
        \includegraphics[width=0.3\textwidth]{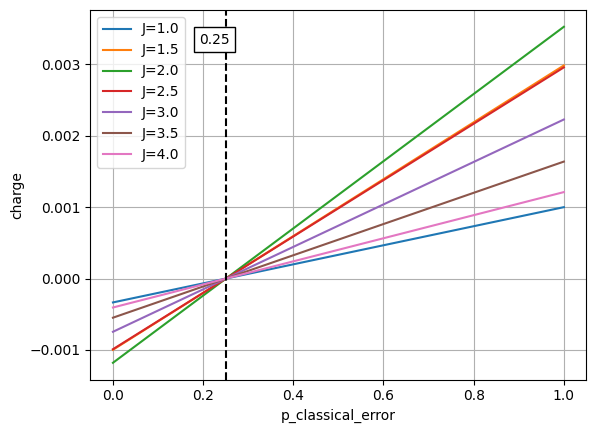}
    }
    \hfill
    \subfloat[Charge, \( N = 4 \)]{
        \includegraphics[width=0.3\textwidth]{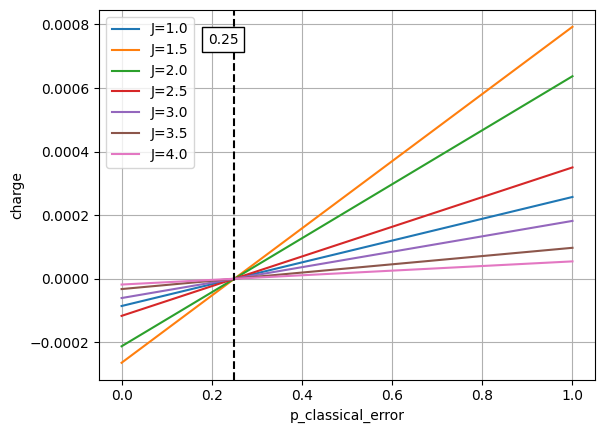}
    }

    \caption{Classical communication error vs the coupling \(J\) for the nearest-neighbor Hamiltonian $H^{(2)}$ with Alice Base \(\sigma_A = Y_0\).}
    \label{fig:classical_err_nn_ham_y}
\end{figure*}

An error in the classical communication channel represents a scenario where the measurement outcome bit sent by Alice is flipped with a probability $p$ before it reaches Bob. This type of error directly impacts Bob's conditional operation, causing him to apply the incorrect unitary rotation. Consequently, the final state at Bob's side becomes a statistical mixture of the state corresponding to the intended outcome (transmitted with probability $1-p$) and the state corresponding to the flipped outcome (transmitted with probability $p$) \cite{QKDbyQET}. The resulting density matrix at Bob's site is given by:

\begin{equation*}
\begin{aligned}
\rho_{B,p} &= (1-p)\sum_{b}U_{B}(b)P_{A}(b)\rho_{gs}P_{A}(b)U_{B}^{\dagger}(b) \\
&+ p\sum_{b'=b\oplus1}U_{B}(b')P_{A}(b)\rho_{gs}P_{A}(b)U_{B}^{\dagger}(b')
\end{aligned}
\end{equation*}

This leads to a linear degradation of the expected signal, which can be expressed as \cite{QKDbyQET}:

\begin{equation*}
\langle\Delta O_{B}\rangle_{p} = (1-p)\langle\Delta O_{B}\rangle_{a=0} + p\langle\Delta O_{B}\rangle_{a=1}
\end{equation*}

Since the protocol is designed such that the intended signal $\langle\Delta O_{B}\rangle_{a=0}$ is negative and the flipped signal $\langle\Delta O_{B}\rangle_{a=1}$ is positive and of similar magnitude, the overall expectation value is attenuated linearly as $p$ increases, moving from a negative value towards a positive one. The protocol fails when this value crosses zero, as this constitutes a bit-error in the generated key, rendering it insecure.

The numerical results presented in Figure \ref{fig:energy_vs_classical_err_for_n} and Figures \ref{fig:classical_err_nn_ham_x}-\ref{fig:classical_err_nn_ham_y} comprehensively illustrate this behavior across all simulated configurations.

Figure \ref{fig:energy_vs_classical_err_for_n} shows the teleported energy as a function of the error probability $p$ for various system sizes $N$. For all Hamiltonians, the expected linear decay is observed. A critical failure threshold, where the sign of the teleported energy flips, consistently appears around $p \approx 0.25$, corroborating the analysis in \cite{QKDbyQET}. Notably, for the nearest-neighbor model ($H^{(2)}$), the magnitude of the teleported energy signal weakens significantly as the system size $N$ increases, which makes resolving the signal from noise more challenging in larger systems.

Figures \ref{fig:classical_err_nn_ham_x} and \ref{fig:classical_err_nn_ham_y} provide a more detailed view, showing the behavior for different coupling strengths $J$.
\textbf{Nearest-Neighbor Model ($H^{(2)}$):} The results for this model, depicted in Figures \ref{fig:classical_err_nn_ham_x} ($\sigma_A = X_0$) and \ref{fig:classical_err_nn_ham_y} ($\sigma_A = Y_0$), confirm a similar linear degradation. However, a key distinction emerges: the charge signal's magnitude remains more robust against increases in system size $N$ compared to the energy signal. While the fundamental error threshold remains in the same $p \approx 0.25-0.32$ range, the rapid decay of the energy signal's absolute value means that for larger distances between Alice and Bob, the charge-based protocol provides a more resilient and discernible signal. This highlights a significant practical advantage of charge teleportation for implementing QKD in larger, more realistic quantum systems.

In summary, while the protocol is fundamentally vulnerable to classical communication errors with a consistent failure threshold, the charge teleportation protocol demonstrates superior stability and resilience, especially in the nearest-neighbor model, where its signal magnitude degrades less rapidly with system size compared to energy teleportation. This makes it a more robust choice for practical implementations.

\subsection{Mixture with an Excited State}

\begin{figure*}[tb!]
    \centering
    \subfloat[Energy, \( N = 2 \)]{
        \includegraphics[width=0.3\textwidth]{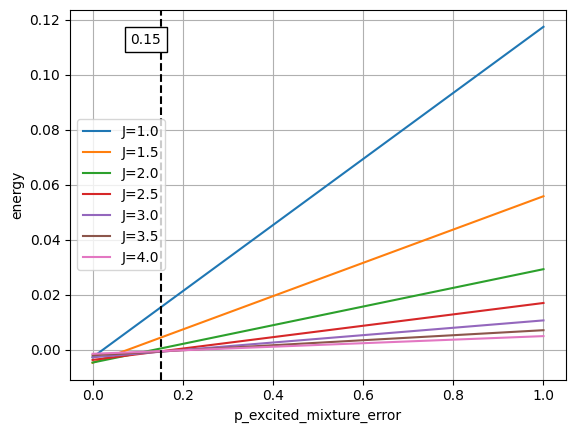}
    }
    \hfill
    \subfloat[Energy, \( N = 3 \)]{
        \includegraphics[width=0.3\textwidth]{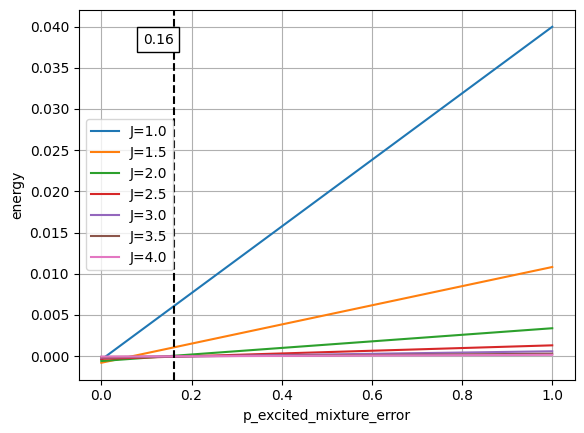}
    }
    \hfill
    \subfloat[Energy, \( N = 4 \)]{
        \includegraphics[width=0.3\textwidth]{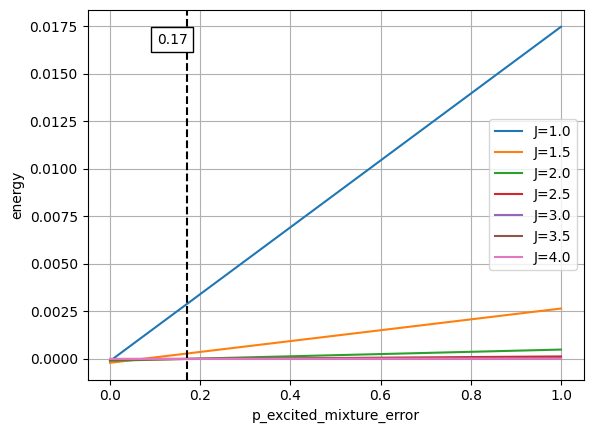}
    }

    \vspace{0.5em}

    \subfloat[Charge, \( N = 2 \)]{
        \includegraphics[width=0.3\textwidth]{plots/errors/numerical/N2_charge_vs_p_excited_mixture_error.png}
    }
    \hfill
    \subfloat[Charge, \( N = 3 \)]{
        \includegraphics[width=0.3\textwidth]{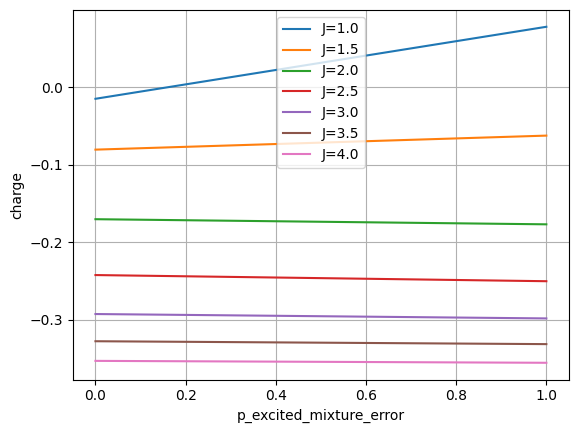}
    }
    \hfill
    \subfloat[Charge, \( N = 4 \)]{
        \includegraphics[width=0.3\textwidth]{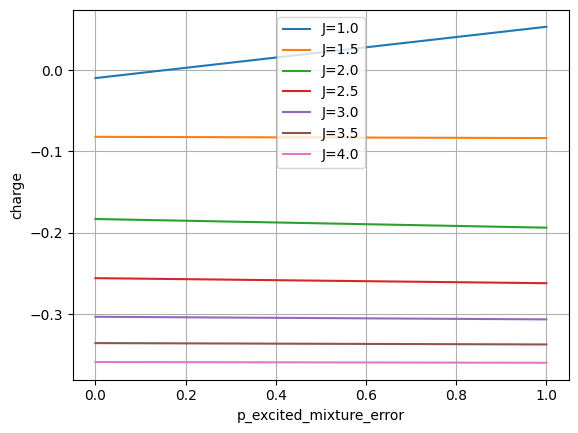}
    }
    
    \caption{Numerical simulation results for teleported expectation value vs. the probability $p$ of mixture with the first excited state for Nearest Neighbors Hamiltonian \(H^{(2)}\) with Alice base \(\sigma_A = X_0\).}
    \label{fig:nn_X_num_mixture_error_appendix}
\end{figure*}

\begin{figure*}[tb!]
    \centering
    \subfloat[Energy, \( N = 2 \)]{
        \includegraphics[width=0.3\textwidth]{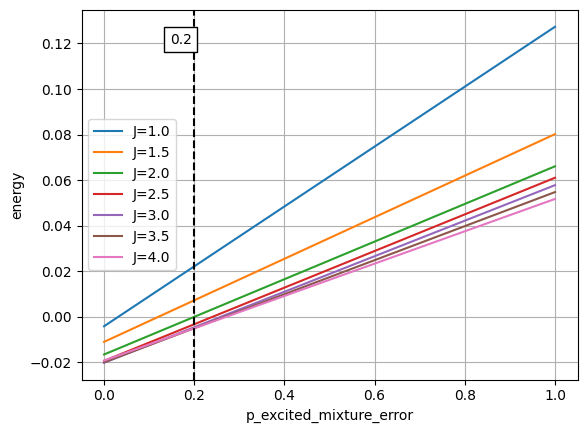}
    }
    \hfill
    \subfloat[Energy, \( N = 3 \)]{
        \includegraphics[width=0.3\textwidth]{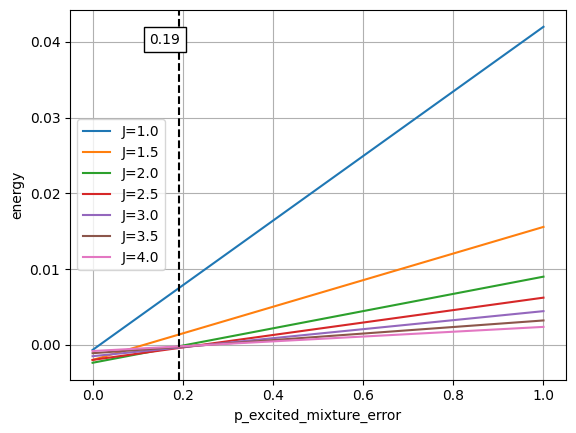}
    }
    \hfill
    \subfloat[Energy, \( N = 4 \)]{
        \includegraphics[width=0.3\textwidth]{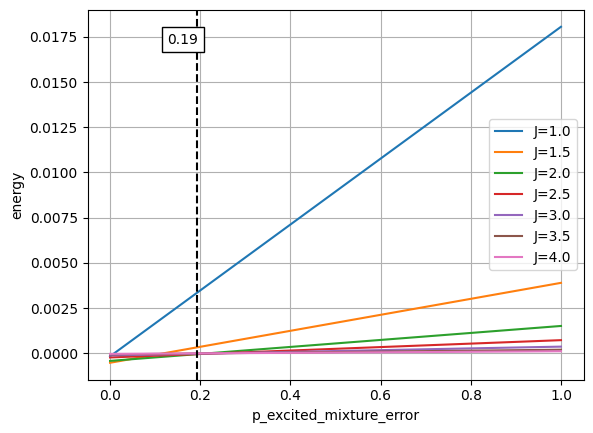}
    }

    \vspace{0.5em}

    \subfloat[Charge, \( N = 2 \)]{
        \includegraphics[width=0.3\textwidth]{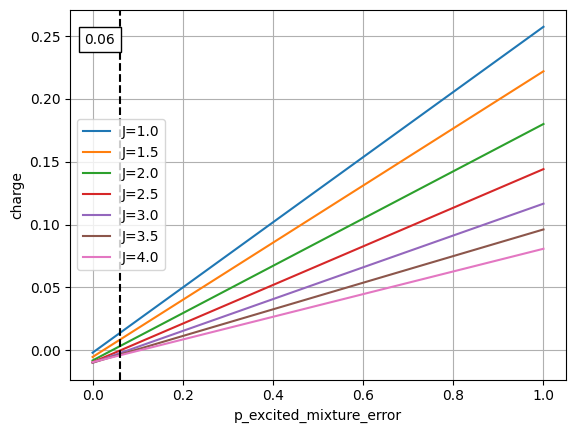}
    }
    \hfill
    \subfloat[Charge, \( N = 3 \)]{
        \includegraphics[width=0.3\textwidth]{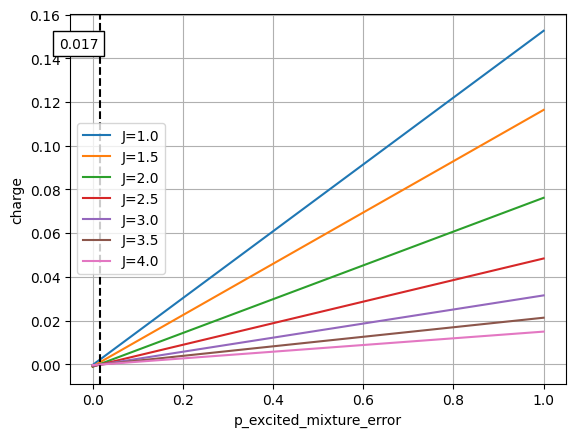}
    }
    \hfill
    \subfloat[Charge, \( N = 4 \)]{
        \includegraphics[width=0.3\textwidth]{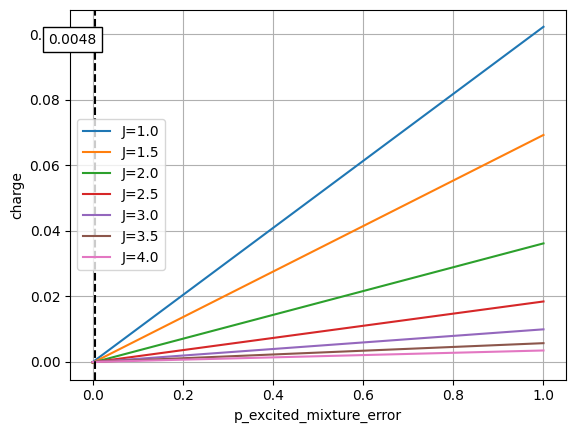}
    }
    
    \caption{Numerical simulation results for teleported expectation value vs. the probability $p$ of mixture with the first excited state for Nearest Neighbors Hamiltonian \(H^{(2)}\) with Alice base \(\sigma_A = Y_0\).}
    \label{fig:nn_Y_num_mixture_error_appendix}
\end{figure*}

\begin{figure*}[tb!]
    \centering
    \subfloat[Energy, \( N = 2 \), \( \sigma_A = X_0 \)]{
        \includegraphics[width=0.225\textwidth]{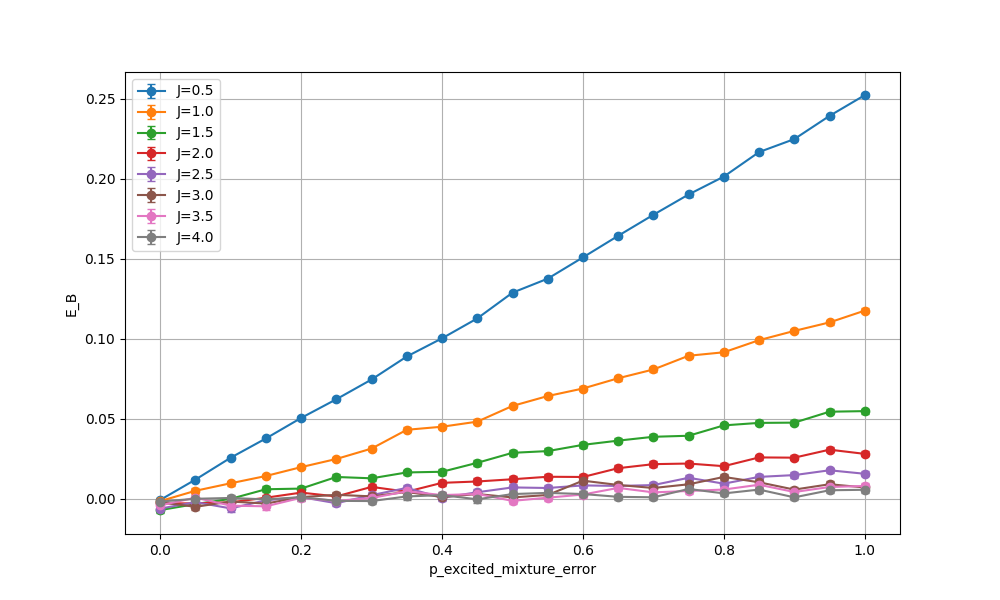}
    }
    \hfill
    \subfloat[Energy, \( N = 3 \), \( \sigma_A = X_0 \)]{
        \includegraphics[width=0.225\textwidth]{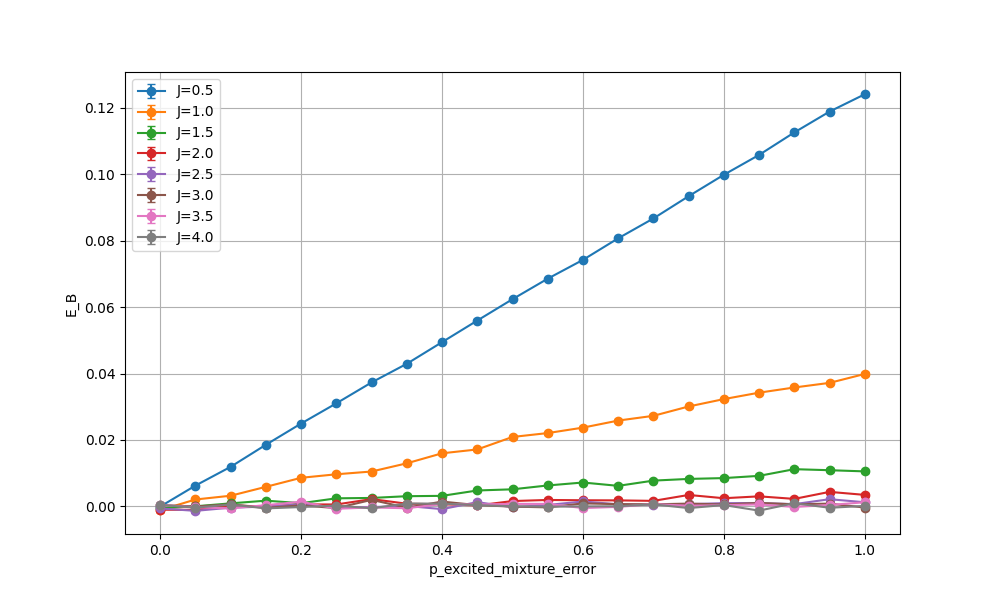}
    }
    \hfill
    \subfloat[Charge, \( N = 2 \), \( \sigma_A = X_0 \)]{
        \includegraphics[width=0.225\textwidth]{plots/errors/qiskit/N2_charge_vs_p_excited_mixture_error.png}
    }
    \hfill
    \subfloat[Charge, \( N = 3 \), \( \sigma_A = X_0 \)]{
        \includegraphics[width=0.225\textwidth]{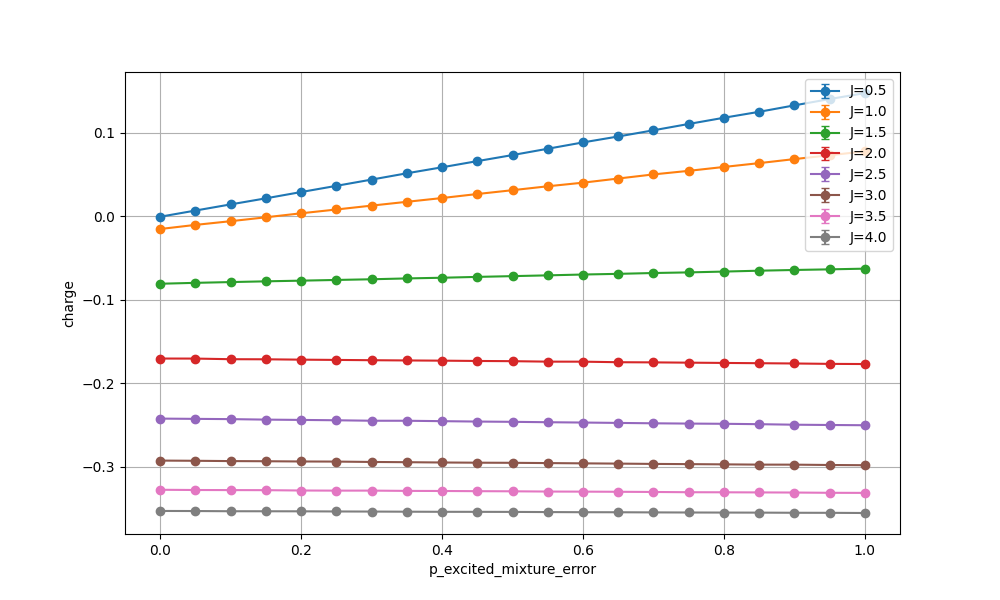}
    }

    \vspace{0.5em}

    \subfloat[Energy, \( N = 2 \), \( \sigma_A = Y_0 \)]{
        \includegraphics[width=0.225\textwidth]{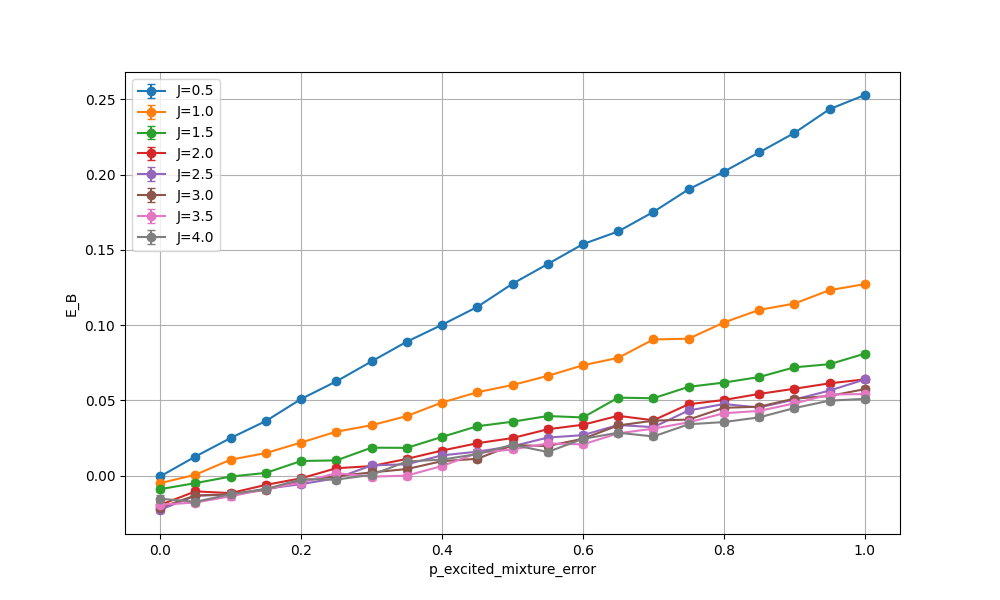}
    }
    \hfill
    \subfloat[Energy, \( N = 3 \), \( \sigma_A = Y_0 \)]{
        \includegraphics[width=0.225\textwidth]{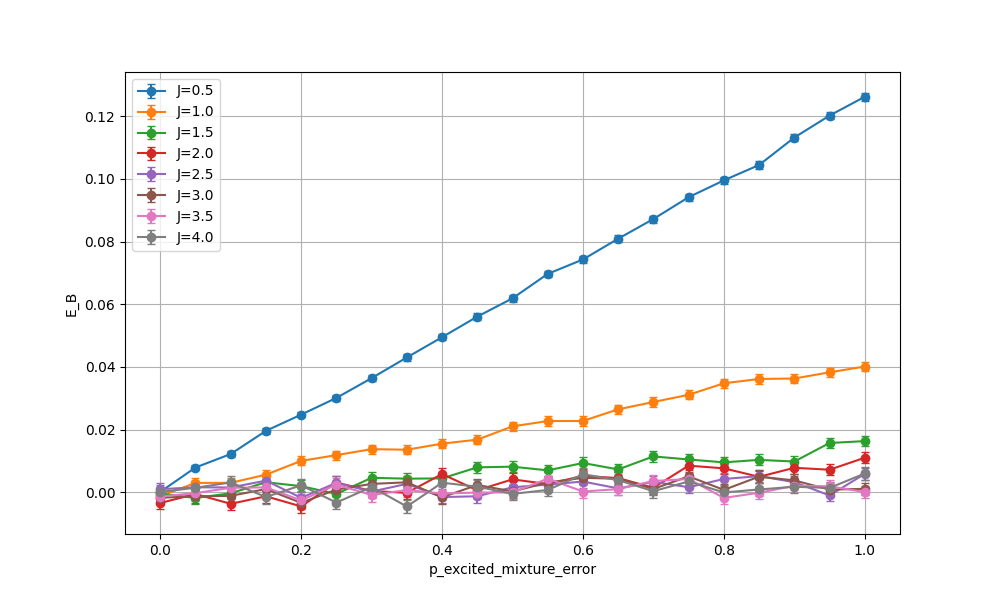}
    }
    \hfill
    \subfloat[Charge, \( N = 2 \), \( \sigma_A = Y_0 \)]{
        \includegraphics[width=0.225\textwidth]{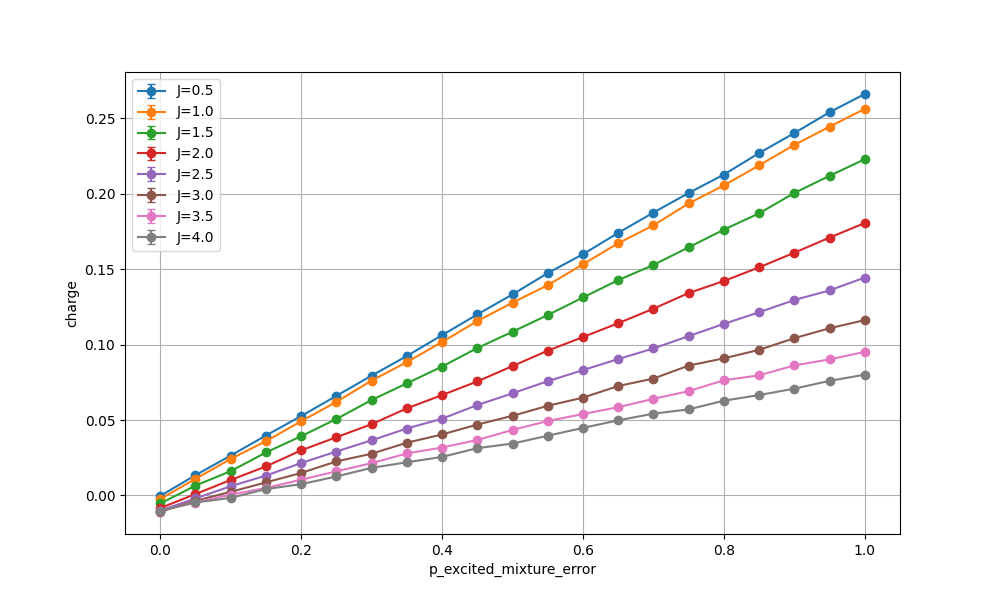}
    }
    \hfill
    \subfloat[Charge, \( N = 3 \), \( \sigma_A = Y_0 \)]{
        \includegraphics[width=0.225\textwidth]{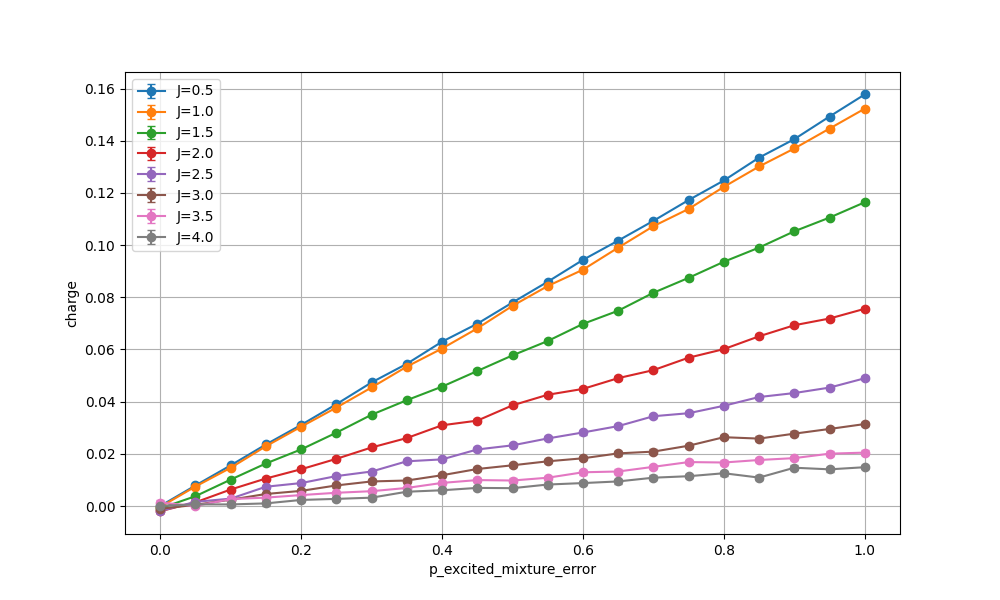}
    }
    
    \caption{Qiskit simulation results for teleported expectation value vs. the probability $p$ of mixture with the first excited state for Nearest Neighbors Hamiltonian \(H^{(2)}\) for both Alice bases.}
    \label{fig:nn_qiskit_mixture_error_appendix}
\end{figure*}

Perfect ground state preparation is experimentally challenging; a common imperfection is a residual population of excited states. This section analyzes the protocol's performance when the initial resource state is not the pure ground state $\rho_{gs} = |gs\rangle\langle gs|$, but rather a statistical mixture with the first excited state, $\rho_1$. This error model is particularly relevant for systems with a small energy gap, where distinguishing the ground state from low-lying excitations is difficult \cite{QKDbyQET}. The state is modeled as:
\begin{equation*}
\rho = (1-p)\rho_{gs} + p\rho_1
\end{equation*}
where $p$ is the probability of the system being in the excited state. The teleported signal becomes a weighted average of the contributions from each state, and the change in Bob's local energy is given by \cite{QKDbyQET}:
\begin{equation*}
\langle\Delta H_B\rangle_p = (1-p)\text{Tr}[(\rho_B - \rho_{gs})H_B] + p\,\text{Tr}[(\sigma_B - \sigma)H_B]
\end{equation*}
where $\rho_B$ and $\sigma_B$ are the final states at Bob's side originating from the ground and excited states, respectively.

Our analysis, presented in Figure \ref{fig:nn_qiskit_mixture_error_appendix}, reveals a critical performance difference between the energy and charge teleportation protocols under this noise model.

A crucial insight is that energy teleportation is highly vulnerable to this error. Low-lying excitations significantly alter the local energy expectation value $\langle H_B \rangle$, often contributing a large positive offset that overwhelms the negative teleported signal from the ground state component \cite{QKDbyQET}. This causes the total signal to rapidly cross zero and become positive, even for moderate error probabilities ($p \approx 0.15-0.29$), as seen in the top rows of Figures \ref{fig:nn_X_num_mixture_error_appendix} and \ref{fig:nn_Y_num_mixture_error_appendix}. Such a sign-flip is fatal to the QKD protocol.

In contrast, the charge protocol is far more robust. The charge observable $Q_B$, being related to local spin parity rather than total energy, is less sensitive to such energy shifts. The quantum correlations responsible for charge teleportation often retain their sign and structure across low-energy states. As a result, the contribution from the excited state does not introduce a large positive offset. The charge signal therefore remains negative across a wide range of mixture probabilities, merely decreasing in magnitude as $p$ increases. This remarkable stability is evident in the bottom rows of all numerical simulation figures, demonstrating a decisive advantage for practical QKD.

The Qiskit simulations, shown in Figure \ref{fig:nn_qiskit_mixture_error_appendix}, corroborate these numerical findings. They also highlight the practical measurement challenges: the energy plots exhibit significant statistical variance, which is a direct consequence of both the weaker signal and the error propagation from measuring non-commuting terms in separate circuits (see Appendix F). The charge plots, in comparison, are markedly smoother, underscoring not only its theoretical robustness against this error model but also its superior statistical stability in a measurement context. This trend holds across all tested Hamiltonians and system sizes.

\subsection{Superposition with an Excited State}

\begin{figure*}[tb!]
    \centering
    \subfloat[Energy, \( N = 2 \)]{
        \includegraphics[width=0.3\textwidth]{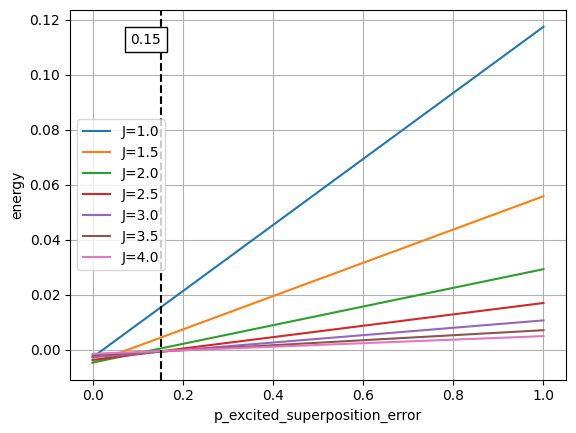}
    }
    \hfill
    \subfloat[Energy, \( N = 3 \)]{
        \includegraphics[width=0.3\textwidth]{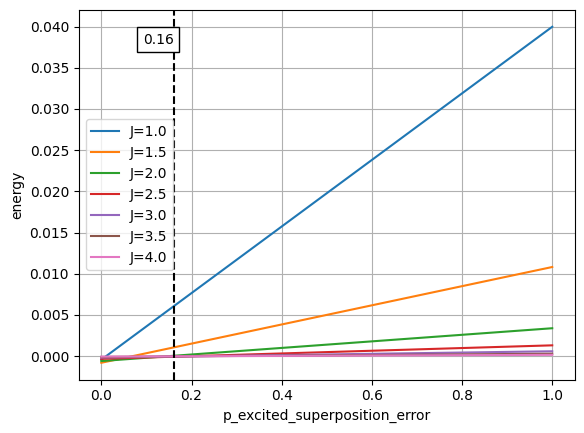}
    }
    \hfill
    \subfloat[Energy, \( N = 4 \)]{
        \includegraphics[width=0.3\textwidth]{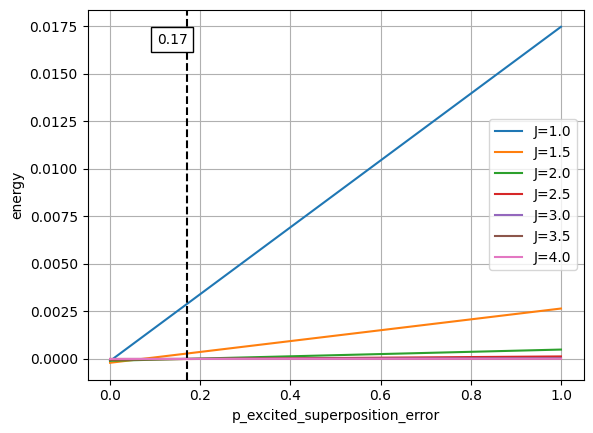}
    }

    \vspace{0.5em}

    \subfloat[Charge, \( N = 2 \)]{
        \includegraphics[width=0.3\textwidth]{plots/errors/numerical/N2_charge_vs_p_excited_superposition_error.png}
    }
    \hfill
    \subfloat[Charge, \( N = 3 \)]{
        \includegraphics[width=0.3\textwidth]{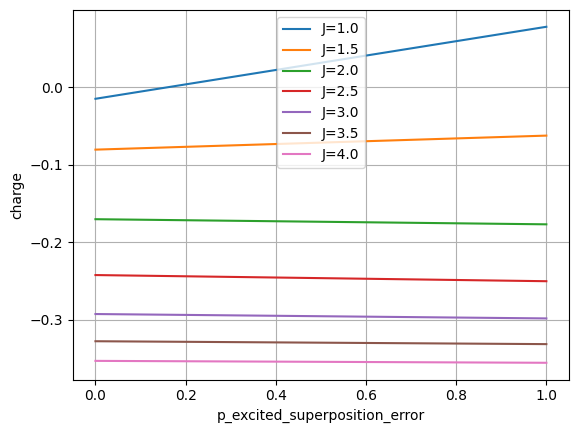}
    }
    \hfill
    \subfloat[Charge, \( N = 4 \)]{
        \includegraphics[width=0.3\textwidth]{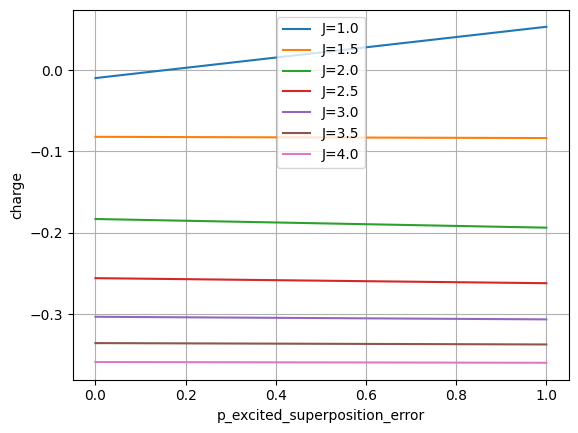}
    }
    
    \caption{Numerical simulation results for teleported expectation value vs. the probability $p$ of superposition with the first excited state for Nearest Neighbors Hamiltonian \(H^{(2)}\) with Alice base \(\sigma_A = X_0\).}
    \label{fig:nn_X_num_superposition_error_appendix}
\end{figure*}

\begin{figure*}[tb!]
    \centering
    \subfloat[Energy, \( N = 2 \)]{
        \includegraphics[width=0.3\textwidth]{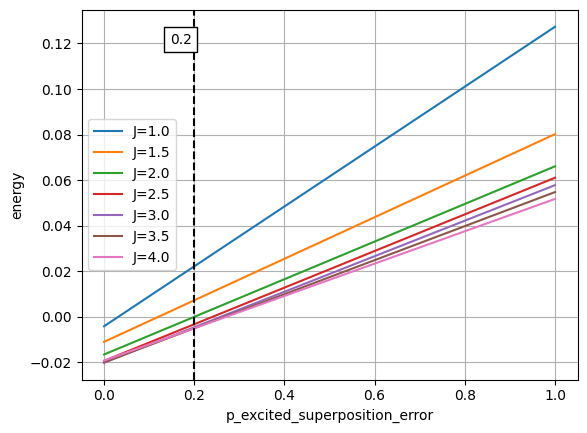}
    }
    \hfill
    \subfloat[Energy, \( N = 3 \)]{
        \includegraphics[width=0.3\textwidth]{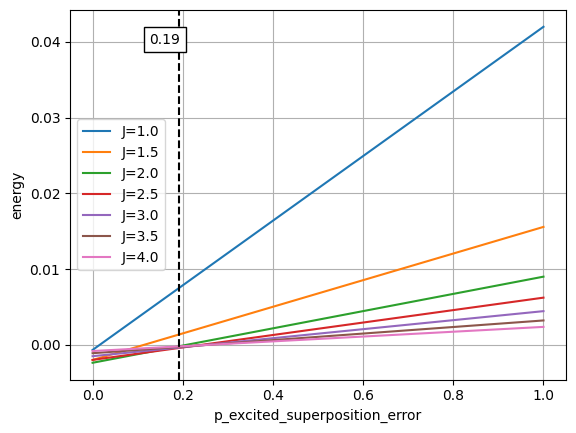}
    }
    \hfill
    \subfloat[Energy, \( N = 4 \)]{
        \includegraphics[width=0.3\textwidth]{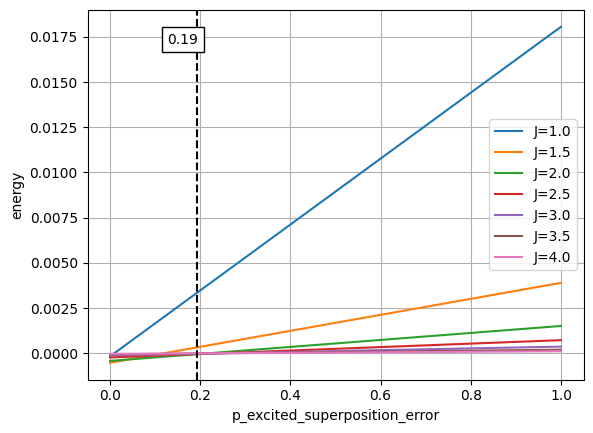}
    }

    \vspace{0.5em}

    \subfloat[Charge, \( N = 2 \)]{
        \includegraphics[width=0.3\textwidth]{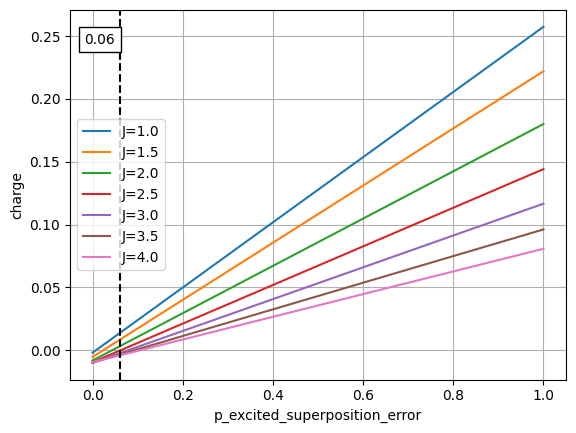}
    }
    \hfill
    \subfloat[Charge, \( N = 3 \)]{
        \includegraphics[width=0.3\textwidth]{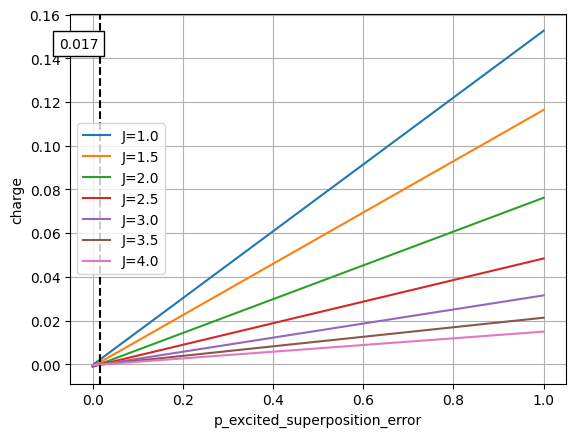}
    }
    \hfill
    \subfloat[Charge, \( N = 4 \)]{
        \includegraphics[width=0.3\textwidth]{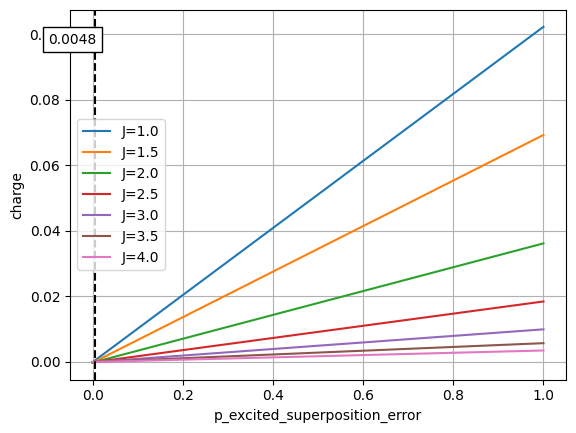}
    }
    
    \caption{Numerical simulation results for teleported expectation value vs. the probability $p$ of superposition with the first excited state for Nearest Neighbors Hamiltonian \(H^{(2)}\) with Alice base \(\sigma_A = Y_0\).}
    \label{fig:nn_Y_num_superposition_error_appendix}
\end{figure*}

\begin{figure*}[tb!]
    \centering
    \subfloat[Energy, \( N = 2 \), \( \sigma_A = X_0 \)]{
        \includegraphics[width=0.225\textwidth]{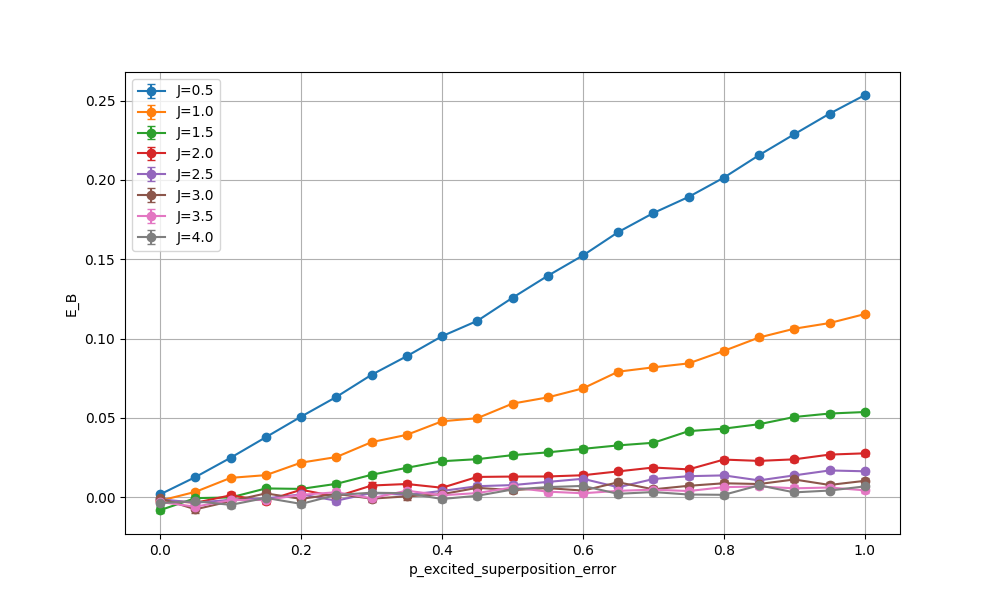}
    }
    \hfill
    \subfloat[Energy, \( N = 3 \), \( \sigma_A = X_0 \)]{
        \includegraphics[width=0.225\textwidth]{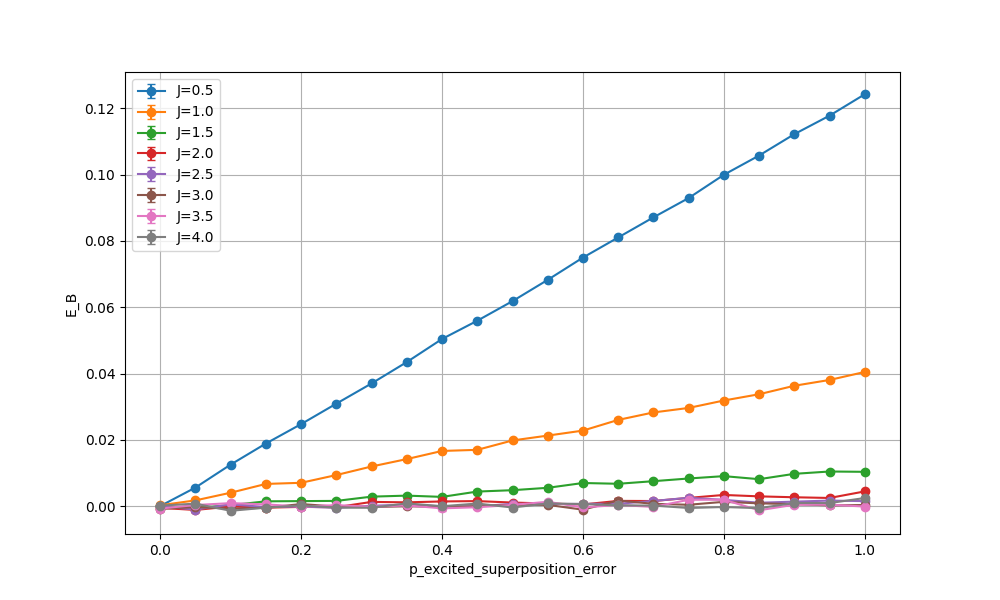}
    }
    \hfill
    \subfloat[Charge, \( N = 2 \), \( \sigma_A = X_0 \)]{
        \includegraphics[width=0.225\textwidth]{plots/errors/qiskit/N2_charge_vs_p_excited_superposition_error.png}
    }
    \hfill
    \subfloat[Charge, \( N = 3 \), \( \sigma_A = X_0 \)]{
        \includegraphics[width=0.225\textwidth]{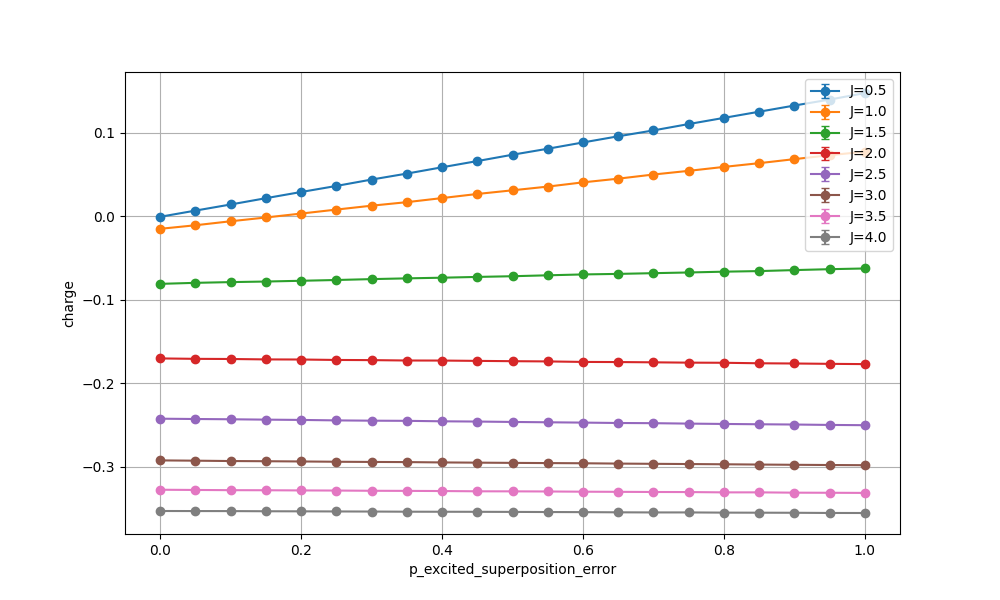}
    }

    \vspace{0.5em}

    \subfloat[Energy, \( N = 2 \), \( \sigma_A = Y_0 \)]{
        \includegraphics[width=0.225\textwidth]{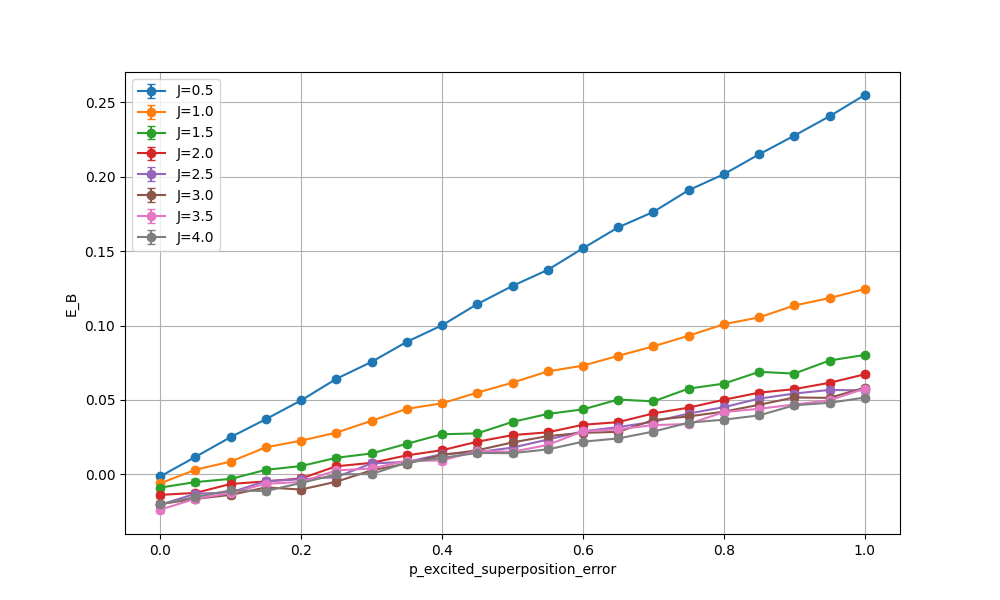}
    }
    \hfill
    \subfloat[Energy, \( N = 3 \), \( \sigma_A = Y_0 \)]{
        \includegraphics[width=0.225\textwidth]{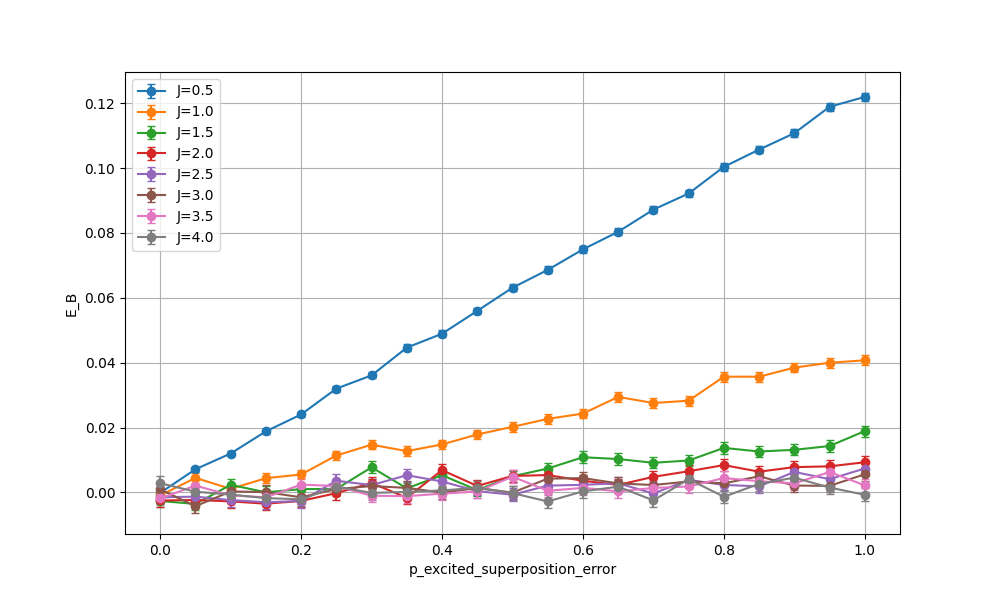}
    }
    \hfill
    \subfloat[Charge, \( N = 2 \), \( \sigma_A = Y_0 \)]{
        \includegraphics[width=0.225\textwidth]{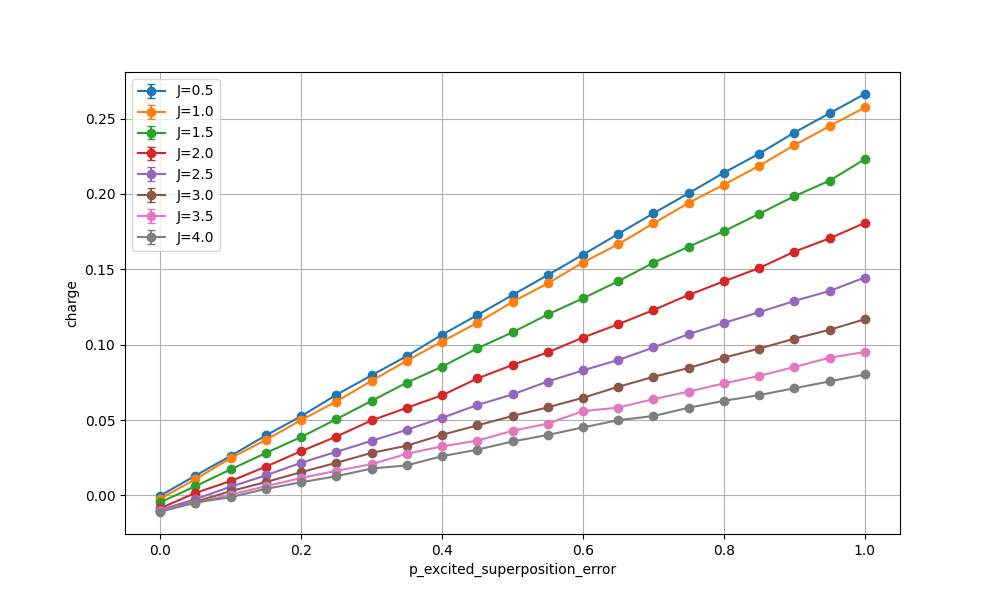}
    }
    \hfill
    \subfloat[Charge, \( N = 3 \), \( \sigma_A = Y_0 \)]{
        \includegraphics[width=0.225\textwidth]{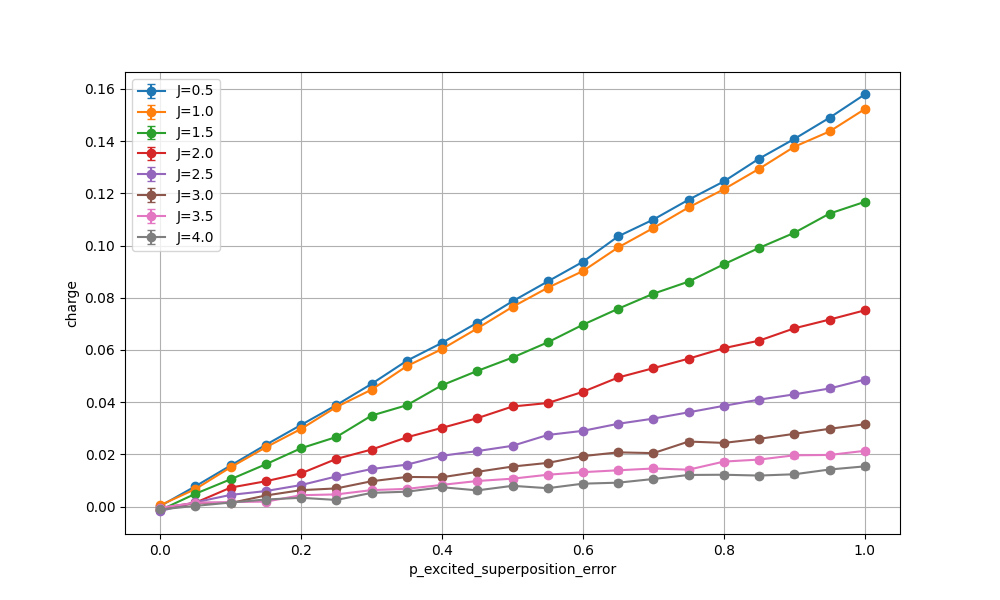}
    }
    
    \caption{Qiskit simulation results for teleported expectation value vs. the probability $p$ of superposition with the first excited state for Nearest Neighbors Hamiltonian \(H^{(2)}\) for both Alice bases.}
    \label{fig:nn_qiskit_superposition_error_appendix}
\end{figure*}

A coherent error in state preparation can result in a superposition of the ground state $\ket{\psi_{gs}}$ and the first excited state $\ket{\psi_{1}}$. Unlike a statistical mixture, this creates a pure state with contributions from both energy levels, modeled as:
\begin{equation*}
\ket{\psi} = \sqrt{1-p}\ket{\psi_{gs}} + e^{i\alpha}\sqrt{p}\ket{\psi_{1}}
\end{equation*}
where $p$ is the probability of the excited state amplitude and $\alpha$ is a relative phase. As shown in \cite{QKDbyQET}, the effect of the phase $\alpha$ is generally negligible, so the analysis focuses on the impact of the probability $p$.

The key distinction of this coherent error is the introduction of interference terms (e.g., cross-terms involving $\bra{\psi_{gs}}...\ket{\psi_{1}}$) into the expectation values of the parameters $\xi$ and $\eta$ that govern the teleportation protocol. This leads to a non-linear degradation of the signal as a function of $p$, which is a distinct signature of a coherent error. The phase sensitivity inherent in the superposition can cause destructive interference that is particularly detrimental to the commutator term $\eta$, which is the primary driver of the teleportation process.

The trend, shown across Figure \ref{fig:nn_qiskit_superposition_error_appendix}, is qualitatively similar to the statistical mixture case but quantitatively distinct due to the non-linear signal degradation. Once again, the energy protocol proves highly susceptible to this error, with the signal quickly flipping to positive at low error probabilities. The charge protocol, however, demonstrates significant resilience. The correlations enabling charge teleportation are less affected by this coherent mixing, allowing the signal to maintain its negative sign and thus preserve the integrity of the key bit across a much wider range of $p$.

The Qiskit simulations presented in Figure \ref{fig:nn_qiskit_superposition_error_appendix} confirm these non-linear trends. The increased statistical variance in the plots, particularly for the energy observable in the nearest-neighbor model, makes it challenging to precisely resolve the non-linear signal from finite-shot noise. This observation reinforces that the charge protocol is not only more robust to the underlying physical error but also more statistically stable in a realistic measurement scenario.

\subsection{Bit-Flip Error}

\begin{figure*}[tb!]
    \centering
    \subfloat[Energy, \( N = 2 \)]{
        \includegraphics[width=0.3\textwidth]{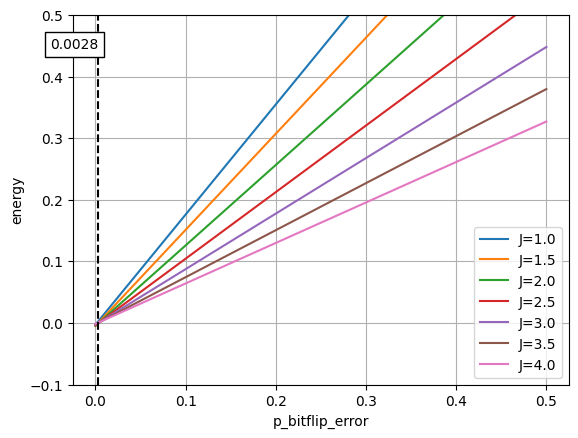}
    }
    \hfill
    \subfloat[Energy, \( N = 3 \)]{
        \includegraphics[width=0.3\textwidth]{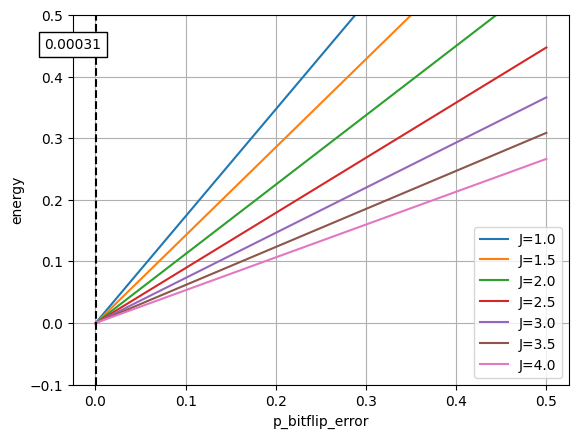}
    }
    \hfill
    \subfloat[Energy, \( N = 4 \)]{
        \includegraphics[width=0.3\textwidth]{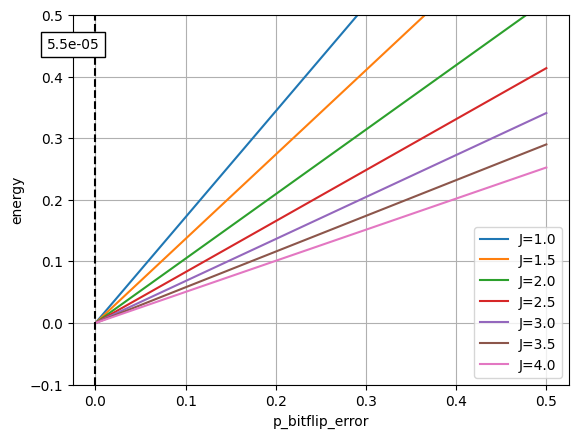}
    }

    \vspace{0.5em}

    \subfloat[Charge, \( N = 2 \)]{
        \includegraphics[width=0.3\textwidth]{plots/errors/numerical/N2_charge_vs_p_bitflip_error.png}
    }
    \hfill
    \subfloat[Charge, \( N = 3 \)]{
        \includegraphics[width=0.3\textwidth]{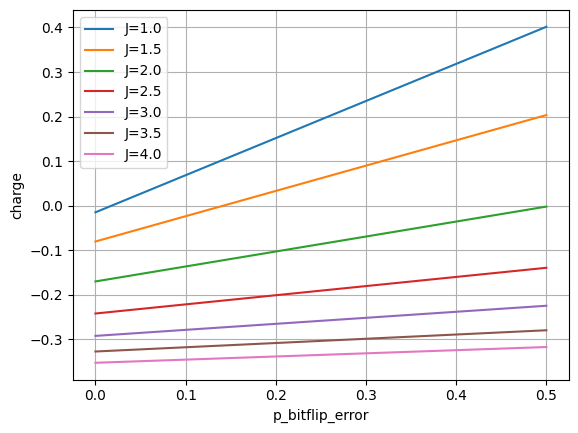}
    }
    \hfill
    \subfloat[Charge, \( N = 4 \)]{
        \includegraphics[width=0.3\textwidth]{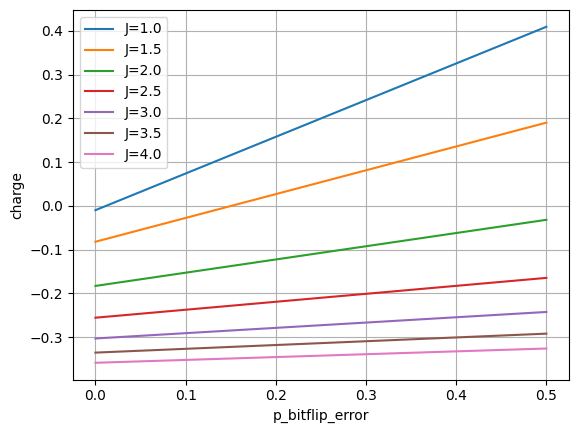}
    }
    
    \caption{Numerical simulation results for teleported expectation value vs. the probability $p$ of bitflip at Bob's site for Nearest Neighbors Hamiltonian \(H^{(2)}\) with Alice base \(\sigma_A = X_0\).}
    \label{fig:nn_X_num_bitflip_error_appendix}
\end{figure*}

\begin{figure*}[tb!]
    \centering
    \subfloat[Energy, \( N = 2 \)]{
        \includegraphics[width=0.3\textwidth]{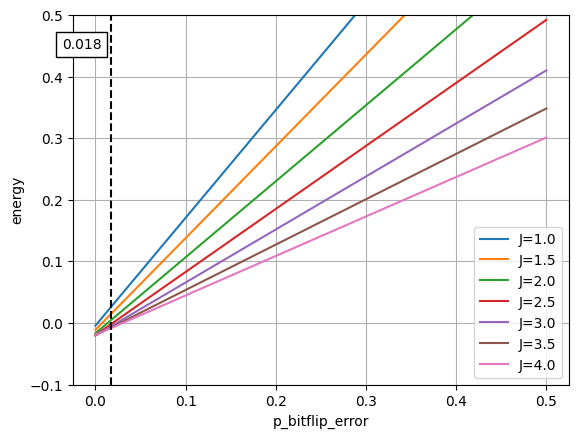}
    }
    \hfill
    \subfloat[Energy, \( N = 3 \)]{
        \includegraphics[width=0.3\textwidth]{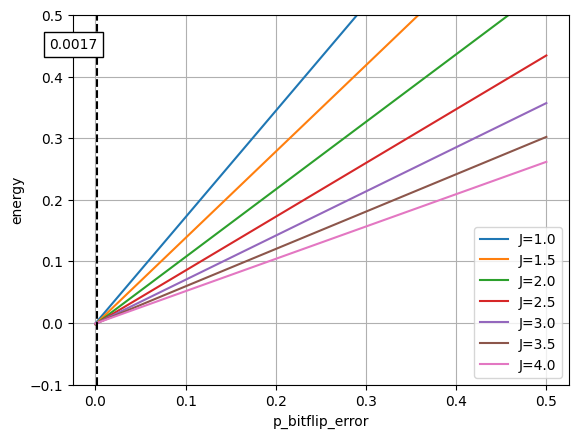}
    }
    \hfill
    \subfloat[Energy, \( N = 4 \)]{
        \includegraphics[width=0.3\textwidth]{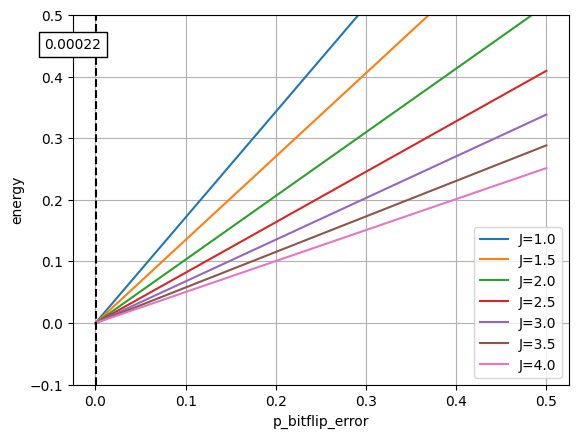}
    }

    \vspace{0.5em}

    \subfloat[Charge, \( N = 2 \)]{
        \includegraphics[width=0.3\textwidth]{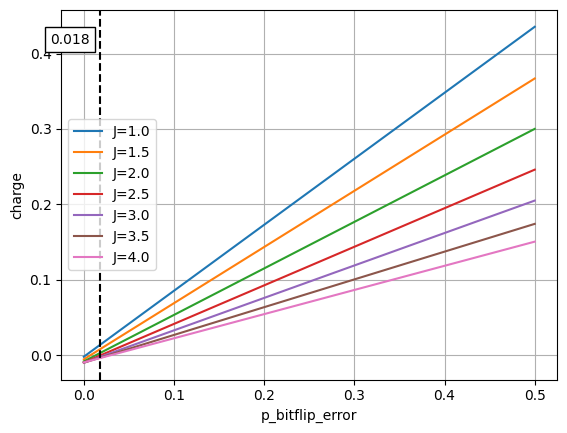}
    }
    \hfill
    \subfloat[Charge, \( N = 3 \)]{
        \includegraphics[width=0.3\textwidth]{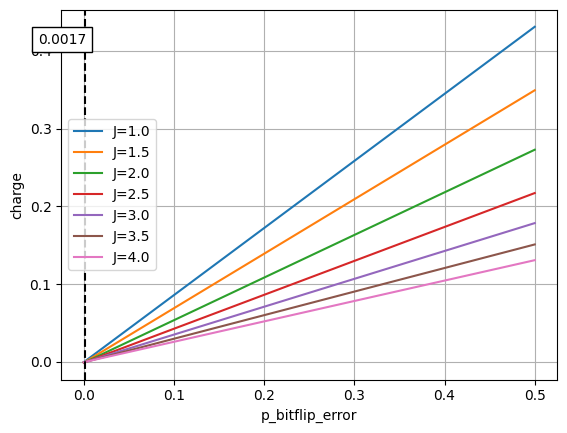}
    }
    \hfill
    \subfloat[Charge, \( N = 4 \)]{
        \includegraphics[width=0.3\textwidth]{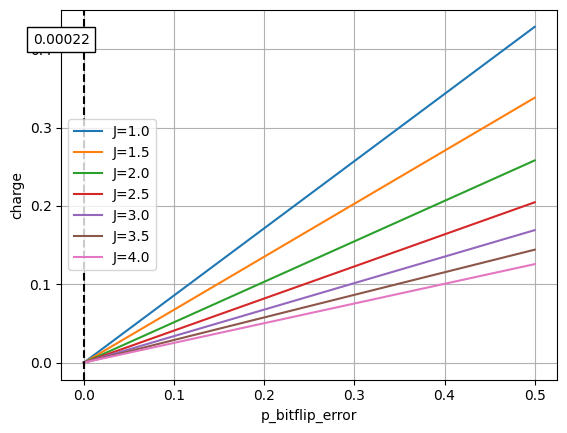}
    }
    
    \caption{Numerical simulation results for teleported expectation value vs. the probability $p$ of bitflip at Bob's site for Nearest Neighbors Hamiltonian \(H^{(2)}\) with Alice base \(\sigma_A = Y_0\).}
    \label{fig:nn_Y_num_bitflip_error_appendix}
\end{figure*}

\begin{figure*}[tb!]
    \centering
    \subfloat[Energy, \( N = 2 \), \( \sigma_A = X_0 \)]{
        \includegraphics[width=0.225\textwidth]{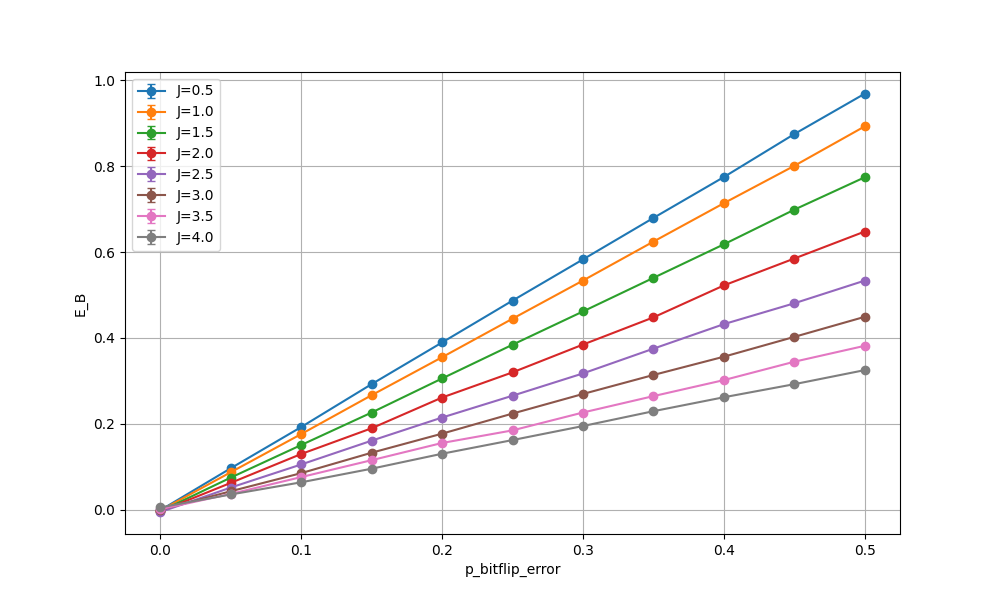}
    }
    \hfill
    \subfloat[Energy, \( N = 3 \), \( \sigma_A = X_0 \)]{
        \includegraphics[width=0.225\textwidth]{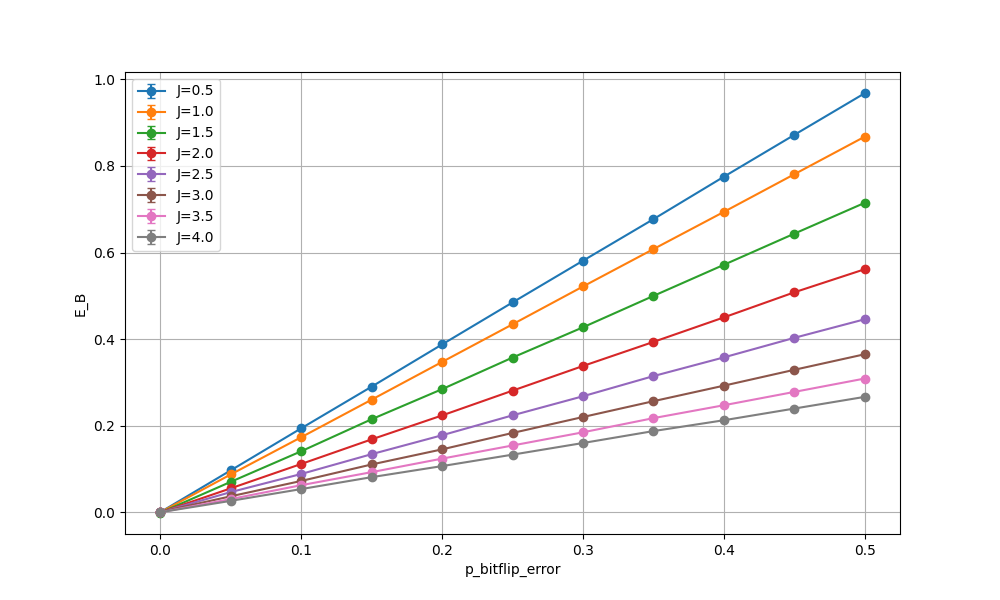}
    }
    \hfill
    \subfloat[Charge, \( N = 2 \), \( \sigma_A = X_0 \)]{
        \includegraphics[width=0.225\textwidth]{plots/errors/qiskit/N2_charge_vs_p_bitflip_error.png}
    }
    \hfill
    \subfloat[Energy, \( N = 3 \), \( \sigma_A = X_0 \)]{
        \includegraphics[width=0.225\textwidth]{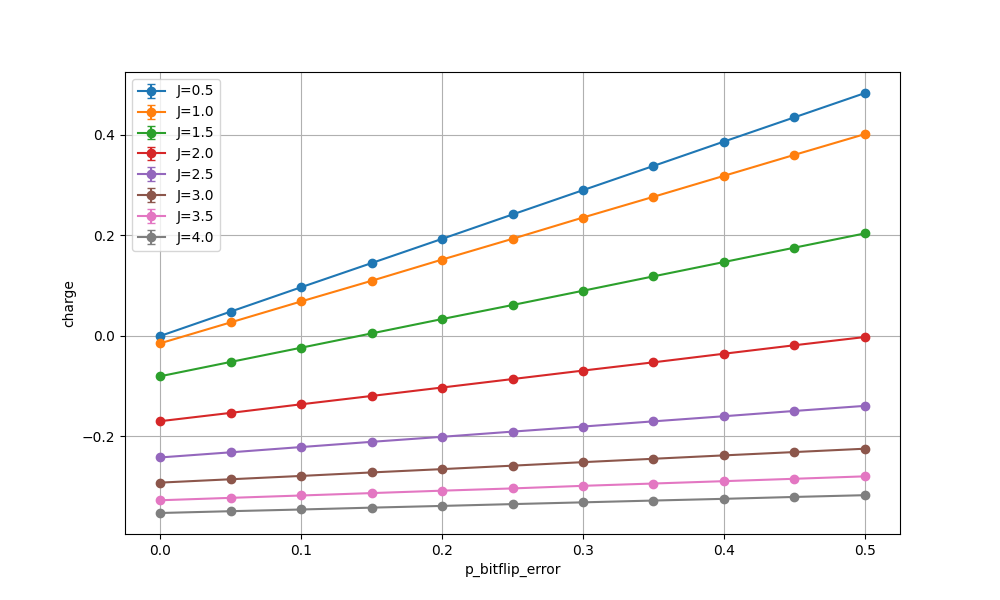}
    }

    \vspace{0.5em}

    \subfloat[Energy, \( N = 2 \), \( \sigma_A = Y_0 \)]{
        \includegraphics[width=0.225\textwidth]{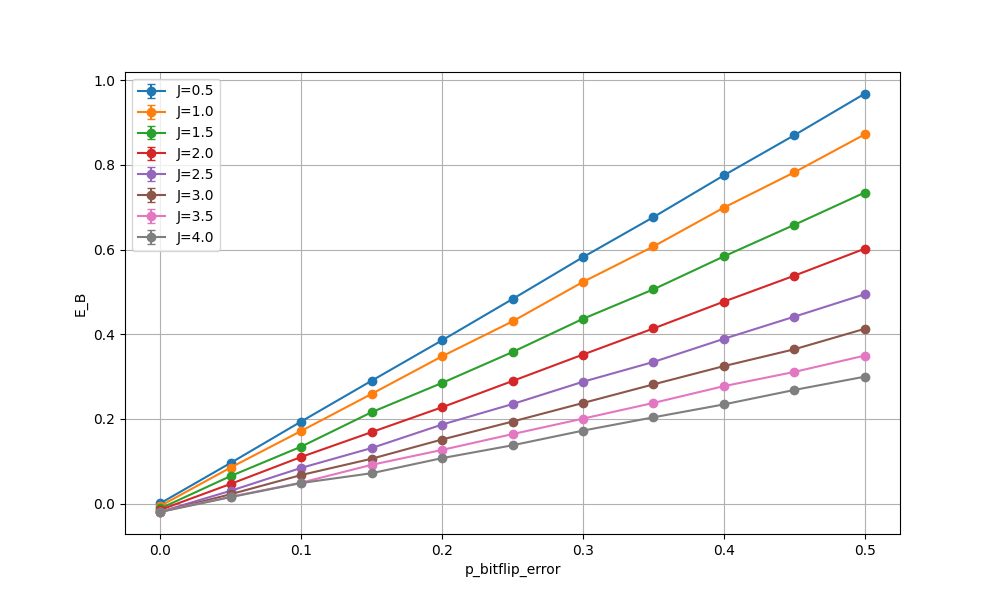}
    }
    \hfill
    \subfloat[Energy, \( N = 3 \), \( \sigma_A = Y_0 \)]{
        \includegraphics[width=0.225\textwidth]{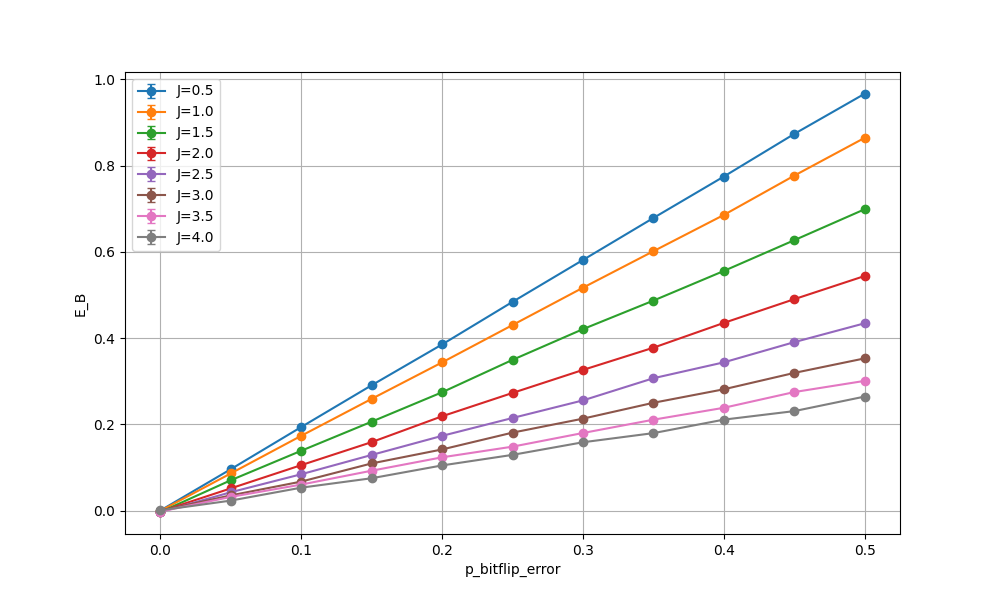}
    }
    \hfill
    \subfloat[Charge, \( N = 2 \), \( \sigma_A = Y_0 \)]{
        \includegraphics[width=0.225\textwidth]{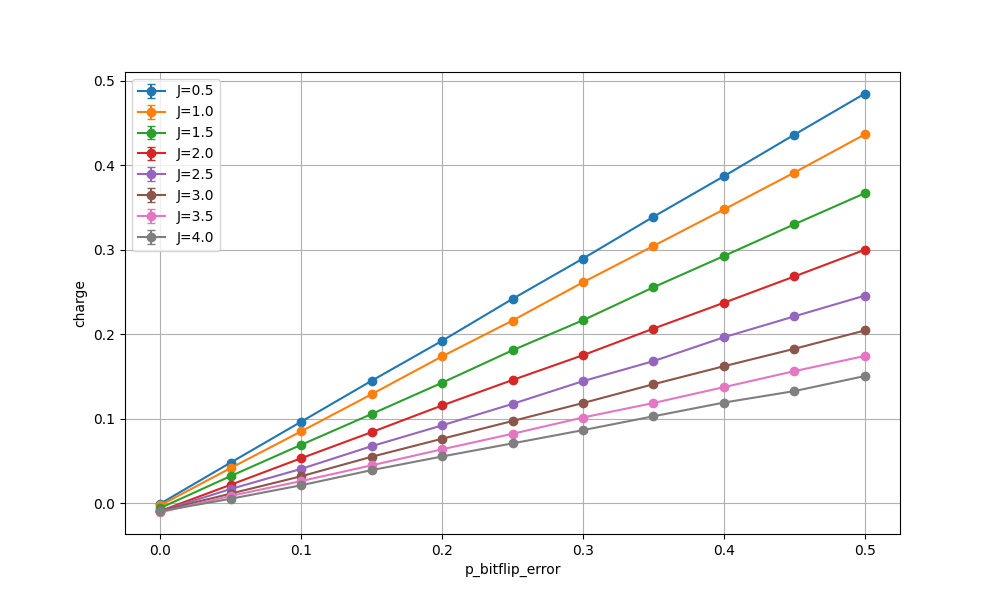}
    }
    \hfill
    \subfloat[Charge, \( N = 3 \), \( \sigma_A = Y_0 \)]{
        \includegraphics[width=0.225\textwidth]{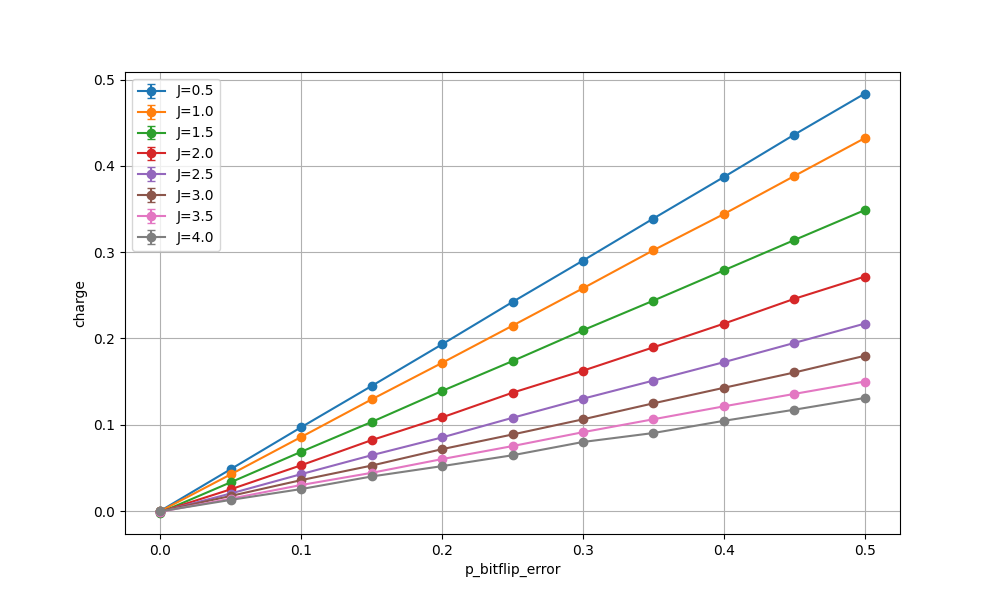}
    }
    
    \caption{Qiskit simulation results for teleported expectation value vs. the probability $p$ of bitflip at Bob's site for Nearest Neighbors Hamiltonian \(H^{(2)}\) for both Alice bases.}
    \label{fig:nn_qiskit_bitflip_error_appendix}
\end{figure*}

We now consider the effect of a local bit-flip error channel on the shared resource state. This error is modeled by the transformation $\rho \rightarrow (1-p)\rho + pX_{n}\rho X_{n}$, which applies a Pauli-\(X\) gate to the qubit at site $n$ with probability $p$. Due to the local nature of the protocol's key operators, it is sufficient to analyze the impact of such errors occurring at Alice's site ($n=0$) and Bob's site ($n=N$) \cite{QKDbyQET}.

A remarkable feature of the protocol is its inherent robustness to bit-flip errors on Alice's qubit when she measures in the corresponding basis. If a bit-flip error ($X_0$) occurs on Alice's qubit in the resource state before her measurement in the $\sigma_A = X_0$ basis, it has no effect on Bob's final expectation value. This is because the measurement operator $P_A(b)$ projects the state onto an eigenstate of $X_0$. Since the error operator $X_0$ commutes with the measurement projectors, the term corresponding to the error state, $P_A(b)X_0\rho_{gs}X_0P_A(b)$, reduces to $P_A(b)\rho_{gs}P_A(b)$, making the error undetectable and harmless in this specific configuration.

The more impactful scenario is a bit-flip error on Bob's qubit, which directly corrupts the entangled correlations he needs to perform his operation. This error attenuates the signal by degrading the non-local correlations in the ground state. As shown in the numerical simulations in Figures \ref{fig:nn_X_num_bitflip_error_appendix} and \ref{fig:nn_Y_num_bitflip_error_appendix}, both the energy and charge signals decay smoothly as the error probability $p$ increases. Unlike the state preparation errors, bit-flips do not introduce a large offsetting term that causes a premature sign-flip. Instead, the failure is gradual, manifesting as a reduced signal-to-noise ratio. While the energy protocol often yields a larger absolute signal, both observables exhibit a similar resilience characterized by this smooth degradation.

The Qiskit simulations presented in Figure \ref{fig:nn_qiskit_bitflip_error_appendix} validate these numerical findings. The results show a stable decay, confirming that the protocol's relative robustness to bit-flip errors on Bob's qubit is well-represented at the quantum circuit level \cite{Lo2005, Wang2005}.

\subsection{Alice site Phase-Flip Error}

\begin{figure*}[tb!]
    \centering
    \subfloat[Energy, \( N = 2 \)]{
        \includegraphics[width=0.3\textwidth]{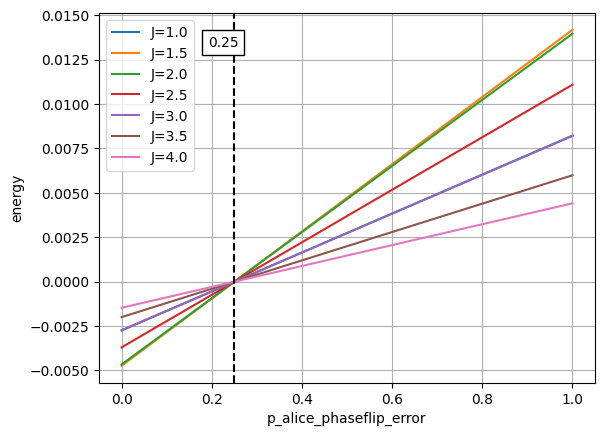}
    }
    \hfill
    \subfloat[Energy, \( N = 3 \)]{
        \includegraphics[width=0.3\textwidth]{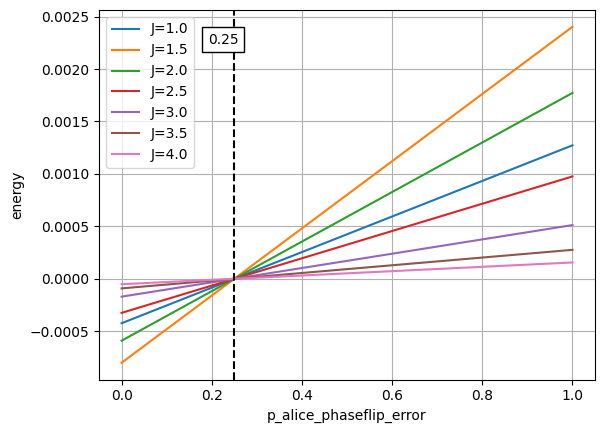}
    }
    \hfill
    \subfloat[Energy, \( N = 4 \)]{
        \includegraphics[width=0.3\textwidth]{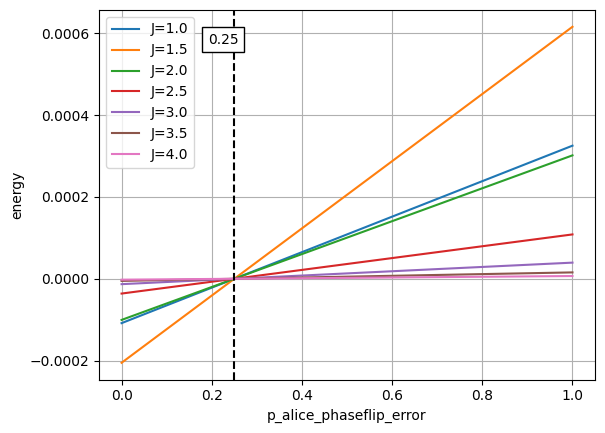}
    }

    \vspace{0.5em}

    \subfloat[Charge, \( N = 2 \)]{
        \includegraphics[width=0.3\textwidth]{plots/errors/numerical/N2_charge_vs_p_alice_phaseflip_error.png}
    }
    \hfill
    \subfloat[Charge, \( N = 3 \)]{
        \includegraphics[width=0.3\textwidth]{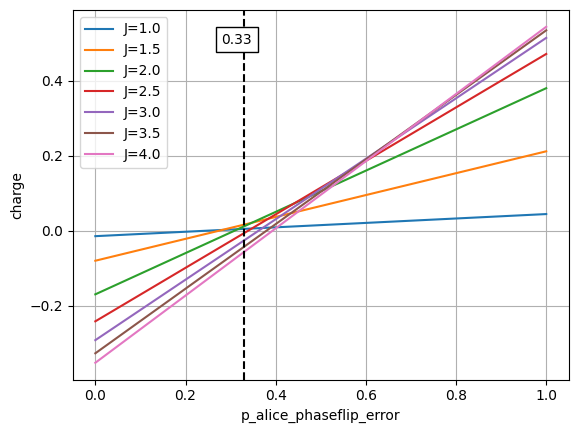}
    }
    \hfill
    \subfloat[Charge, \( N = 4 \)]{
        \includegraphics[width=0.3\textwidth]{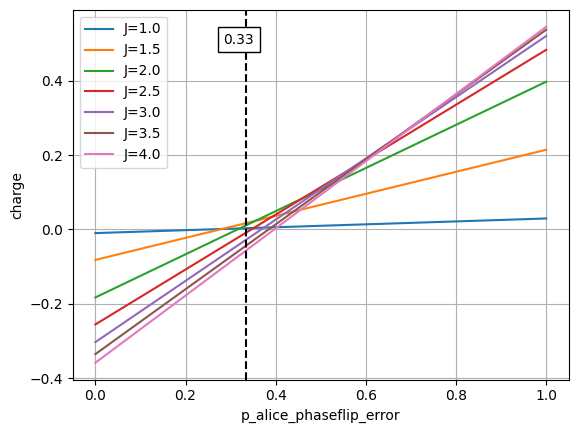}
    }
    
    \caption{Numerical simulation results for teleported expectation value vs. the probability $p$ of phaseflip at Alice's site for Nearest Neighbors Hamiltonian \(H^{(2)}\) with Alice base \(\sigma_A = X_0\).}
    \label{fig:nn_X_num_alice_phaseflip_error_appendix}
\end{figure*}

\begin{figure*}[tb!]
    \centering
    \subfloat[Energy, \( N = 2 \)]{
        \includegraphics[width=0.3\textwidth]{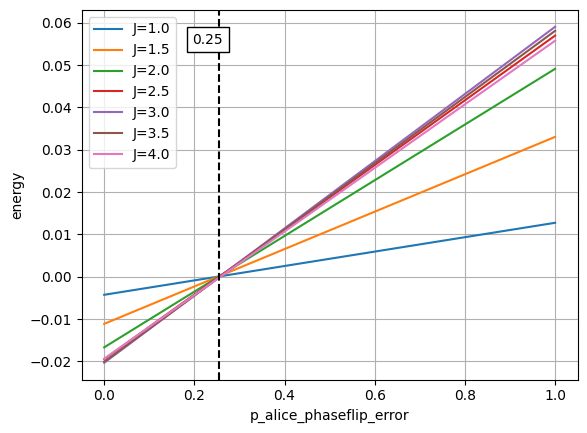}
    }
    \hfill
    \subfloat[Energy, \( N = 3 \)]{
        \includegraphics[width=0.3\textwidth]{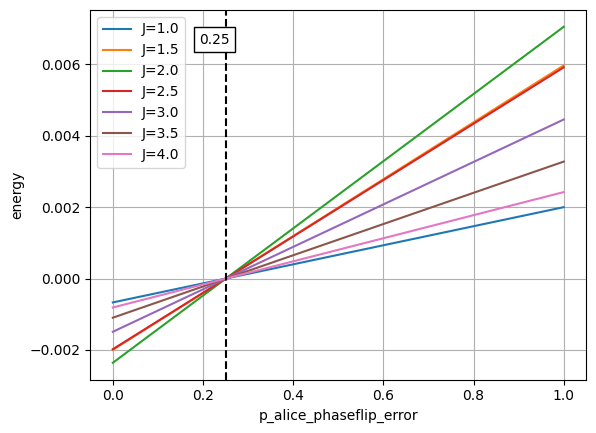}
    }
    \hfill
    \subfloat[Energy, \( N = 4 \)]{
        \includegraphics[width=0.3\textwidth]{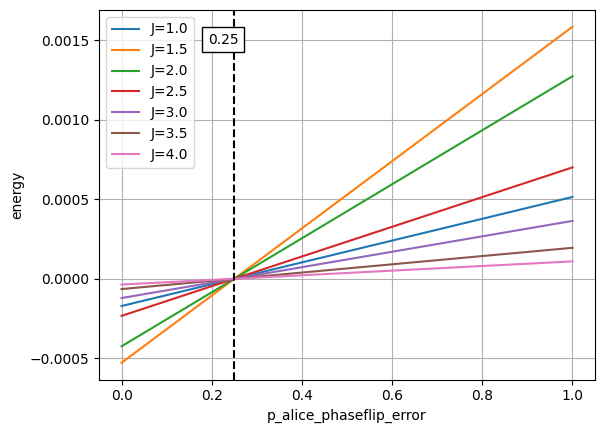}
    }

    \vspace{0.5em}

    \subfloat[Charge, \( N = 2 \)]{
        \includegraphics[width=0.3\textwidth]{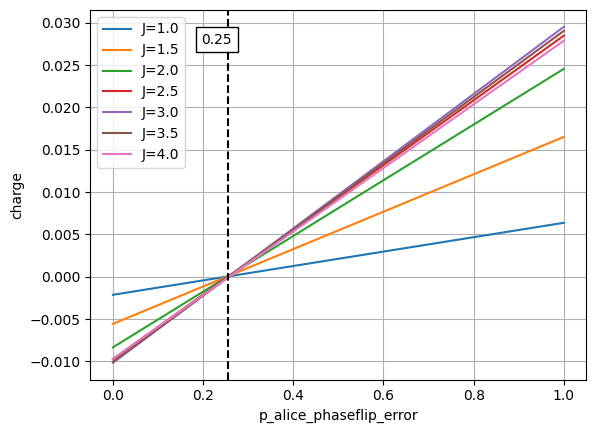}
    }
    \hfill
    \subfloat[Charge, \( N = 3 \)]{
        \includegraphics[width=0.3\textwidth]{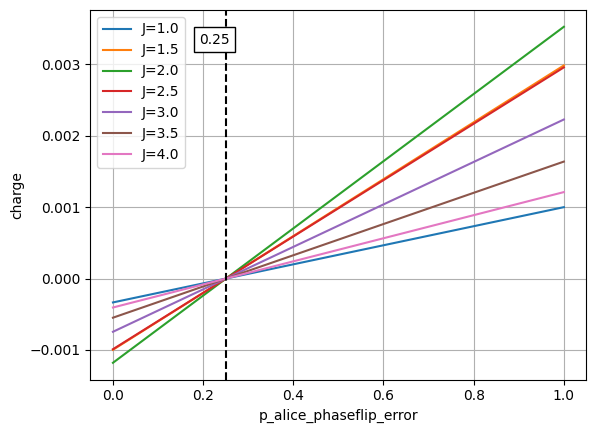}
    }
    \hfill
    \subfloat[Charge, \( N = 4 \)]{
        \includegraphics[width=0.3\textwidth]{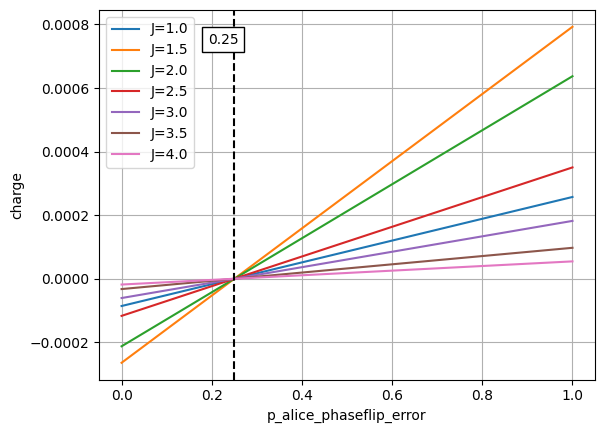}
    }
    
    \caption{Numerical simulation results for teleported expectation value vs. the probability $p$ of phaseflip at Alice's site for Nearest Neighbors Hamiltonian \(H^{(2)}\) with Alice base \(\sigma_A = Y_0\).}
    \label{fig:nn_Y_num_alice_phaseflip_error_appendix}
\end{figure*}

\begin{figure*}[tb!]
    \centering
    \subfloat[Energy, \( N = 2 \), \( \sigma_A = X_0 \)]{
        \includegraphics[width=0.225\textwidth]{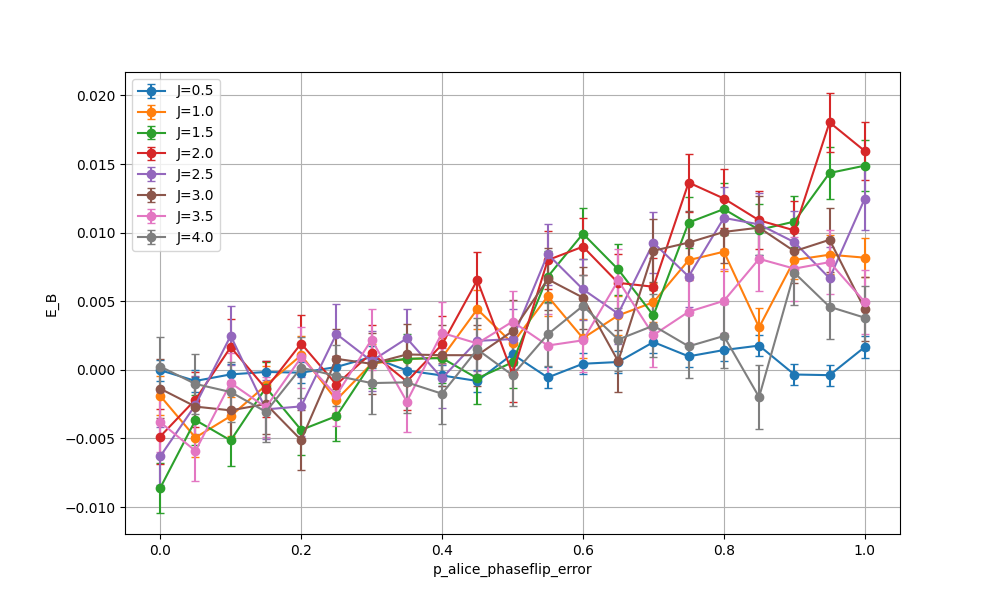}
    }
    \hfill
    \subfloat[Energy, \( N = 3 \), \( \sigma_A = X_0 \)]{
        \includegraphics[width=0.225\textwidth]{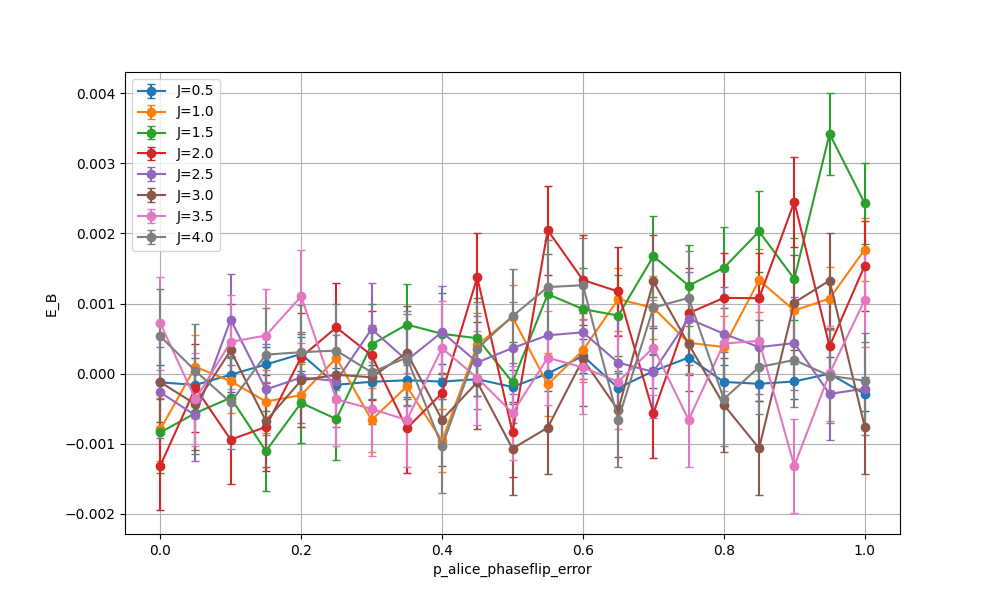}
    }
    \hfill
    \subfloat[Charge, \( N = 2 \), \( \sigma_A = X_0 \)]{
        \includegraphics[width=0.225\textwidth]{plots/errors/qiskit/N2_charge_vs_p_alice_phaseflip_error.png}
    }
    \hfill
    \subfloat[Charge, \( N = 3 \), \( \sigma_A = X_0 \)]{
        \includegraphics[width=0.225\textwidth]{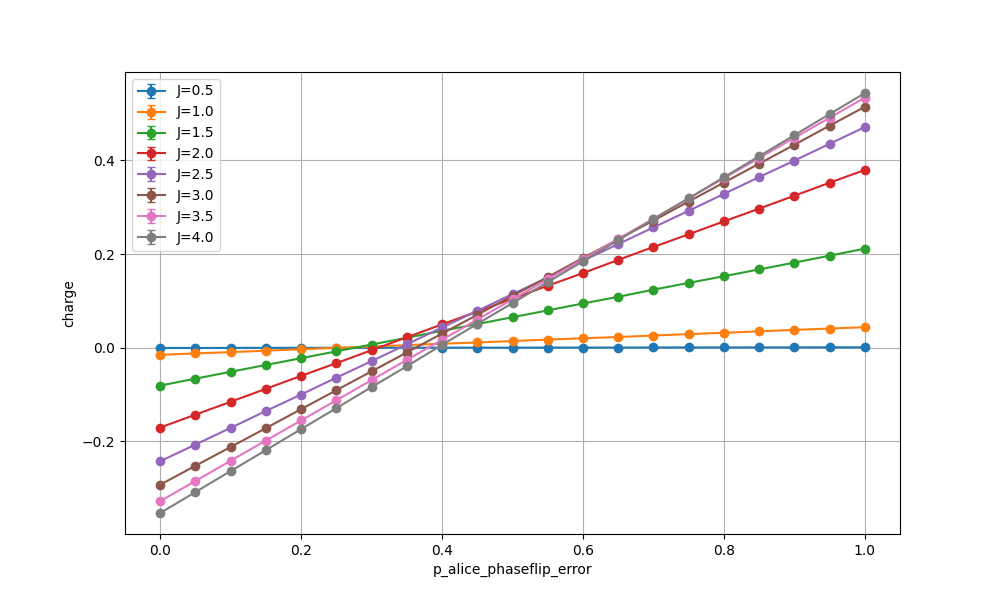}
    }

    \vspace{0.5em}

    \subfloat[Energy, \( N = 2 \), \( \sigma_A = Y_0 \)]{
        \includegraphics[width=0.225\textwidth]{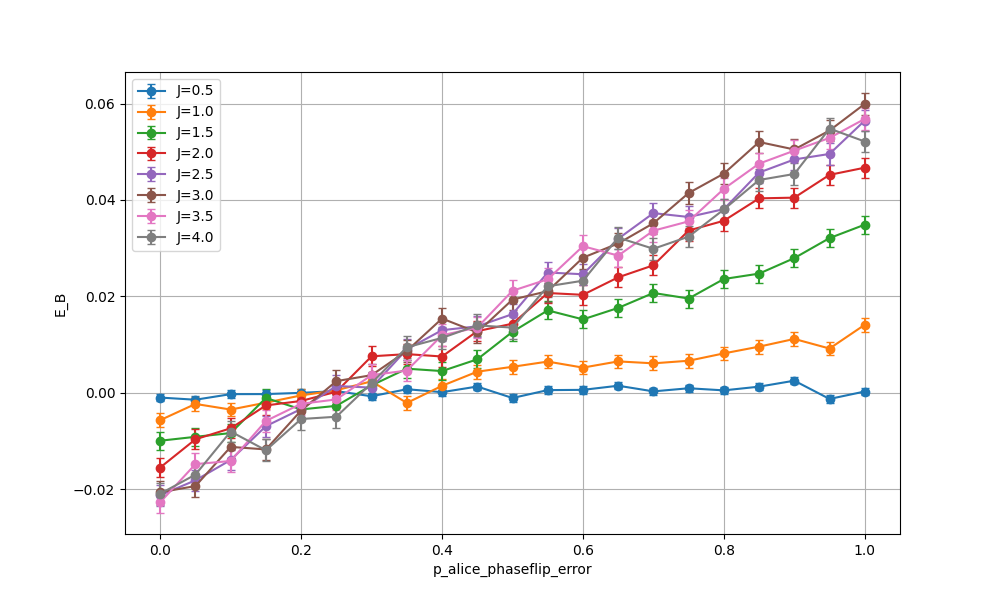}
    }
    \hfill
    \subfloat[Energy, \( N = 3 \), \( \sigma_A = Y_0 \)]{
        \includegraphics[width=0.225\textwidth]{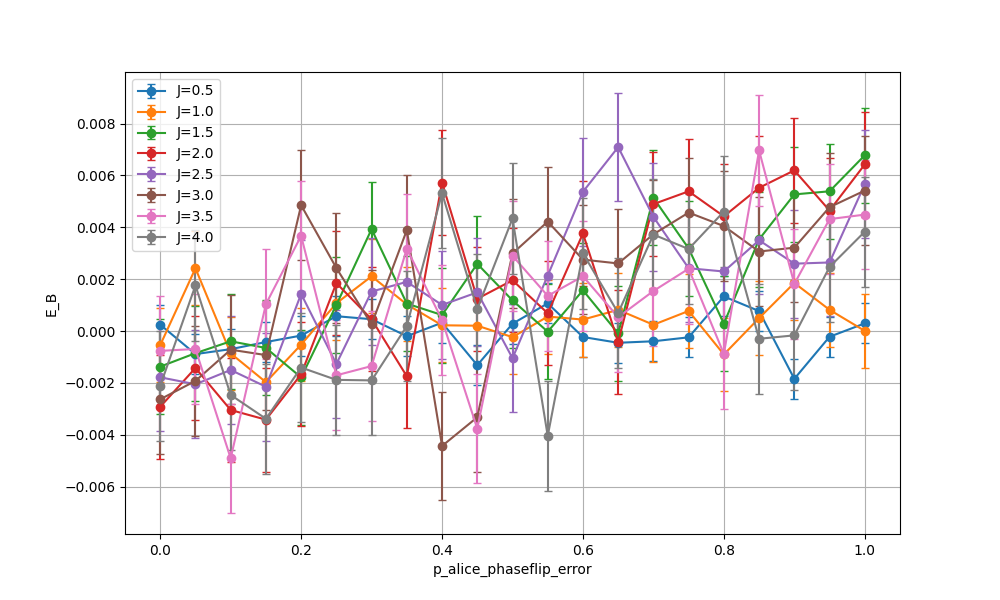}
    }
    \hfill
    \subfloat[Charge, \( N = 2 \), \( \sigma_A = Y_0 \)]{
        \includegraphics[width=0.225\textwidth]{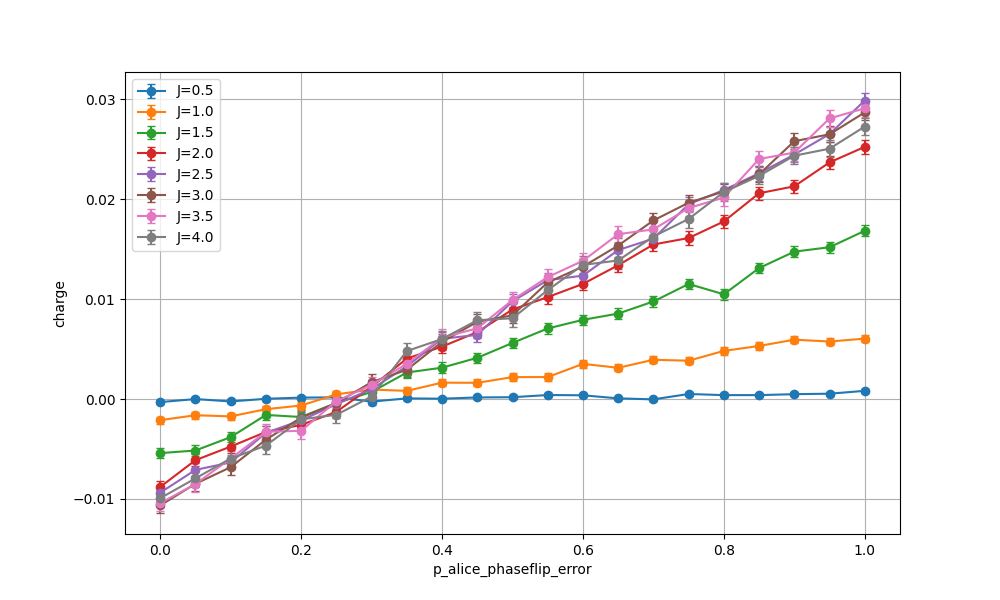}
    }
    \hfill
    \subfloat[Charge, \( N = 3 \), \( \sigma_A = Y_0 \)]{
        \includegraphics[width=0.225\textwidth]{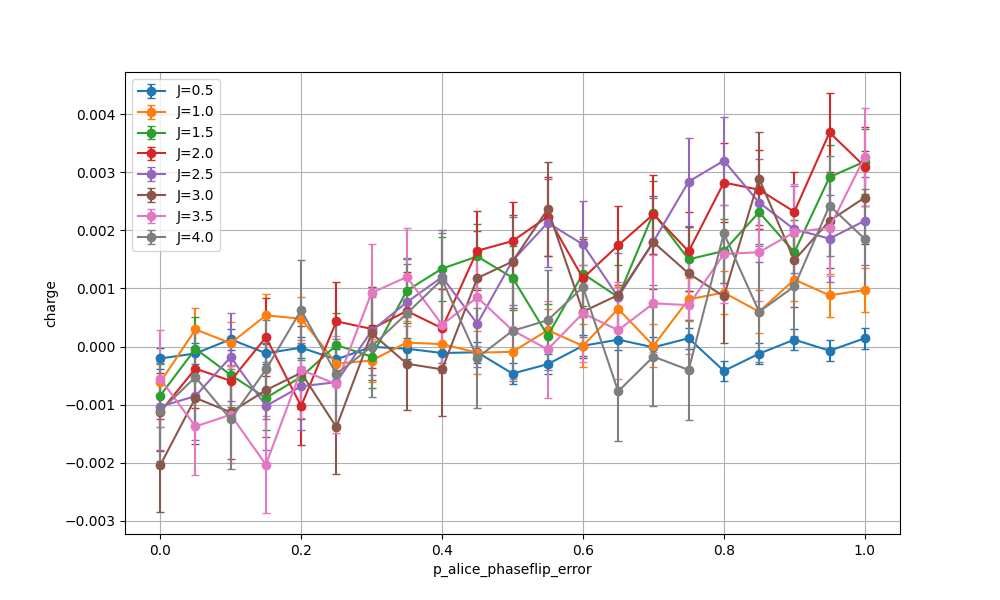}
    }
    
    \caption{Qiskit simulation results for teleported expectation value vs. the probability $p$ of phaseflip at Alice's site for Nearest Neighbors Hamiltonian \(H^{(2)}\) for both Alice bases.}
    \label{fig:nn_qiskit_alice_phaseflip_error_appendix}
\end{figure*}

A phase-flip error on Alice's qubit, modeled by the channel $\rho \rightarrow (1-p)\rho + pZ_{0}\rho Z_{0}$, represents a significant threat to the protocol's integrity. The impact of this error is fundamentally different from a bit-flip, as it directly interferes with the measurement process itself.

The physical mechanism behind this vulnerability is the decoherence of Alice's measurement basis \cite{QKDbyQET, Lydersen2010, Zhao2008}. Alice performs her measurement in a basis (e.g., $\sigma_A = X_0$ or $Y_0$) that anticommutes with the phase-flip operator $Z_0$. This error randomizes the phase relationship between the eigenstates of Alice's measurement, degrading her ability to accurately project the shared state. This directly suppresses the expectation value of the correlator that forms the "engine" of the teleportation, $\eta = i\langle[O_B, \sigma_B]\sigma_A\rangle$. As $\eta$ is responsible for generating the negative signal at Bob's side, its attenuation causes the teleported expectation value to degrade and shift towards positive values.

The simulation results presented in Figure \ref{fig:nn_qiskit_alice_phaseflip_error_appendix} confirm this behavior. 
\begin{itemize}
    \item Across all models and parameters, a phase-flip error on Alice's qubit leads to a roughly linear degradation of the teleported signal, for both energy and charge. Unlike the more benign bit-flip error, this degradation is severe enough to cause a sign-flip at a relatively low error probability, representing a failure of the QKD protocol.
    
    \item The critical threshold where the signal crosses zero is consistently found in the range of $p \approx 0.25-0.33$, as seen in the numerical plots. This indicates that both energy and charge teleportation protocols are similarly vulnerable to this type of noise.
    
    \item The Qiskit simulations in Figure \ref{fig:nn_qiskit_alice_phaseflip_error_appendix} validate the trends observed in the ideal numerical calculations. They reproduce the signal attenuation and eventual sign-flip, demonstrating that this vulnerability persists in circuit-level implementations, albeit with additional statistical noise.
\end{itemize}
In summary, a phase-flip error on Alice's qubit is a detrimental noise source that attacks the core mechanism of the teleportation protocol by decohering the measurement basis. Its impact is severe for both energy and charge observables, establishing a critical error threshold beyond which secure key distribution is not possible.

\subsection{Bob site Phase-Flip Error}

\begin{figure*}[tb!]
    \centering
    \subfloat[Energy, \( N = 2 \)]{
        \includegraphics[width=0.3\textwidth]{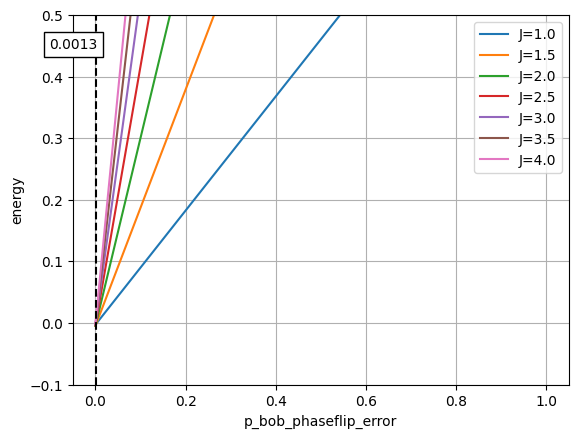}
    }
    \hfill
    \subfloat[Energy, \( N = 3 \)]{
        \includegraphics[width=0.3\textwidth]{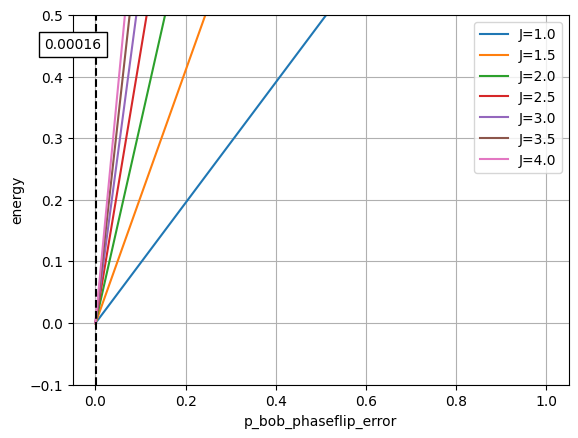}
    }
    \hfill
    \subfloat[Energy, \( N = 4 \)]{
        \includegraphics[width=0.3\textwidth]{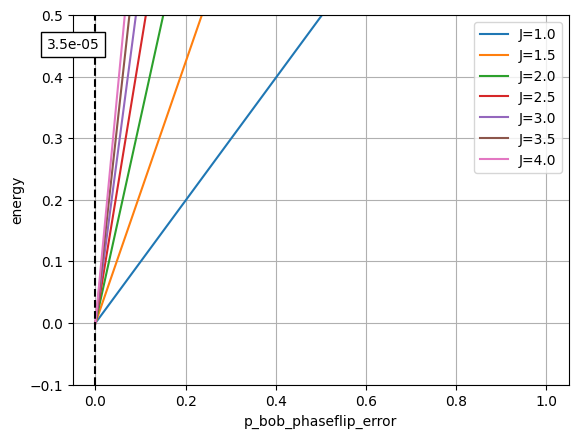}
    }

    \vspace{0.5em}

    \subfloat[Charge, \( N = 2 \)]{
        \includegraphics[width=0.3\textwidth]{plots/errors/numerical/N2_charge_vs_p_bob_phaseflip_error.png}
    }
    \hfill
    \subfloat[Charge, \( N = 3 \)]{
        \includegraphics[width=0.3\textwidth]{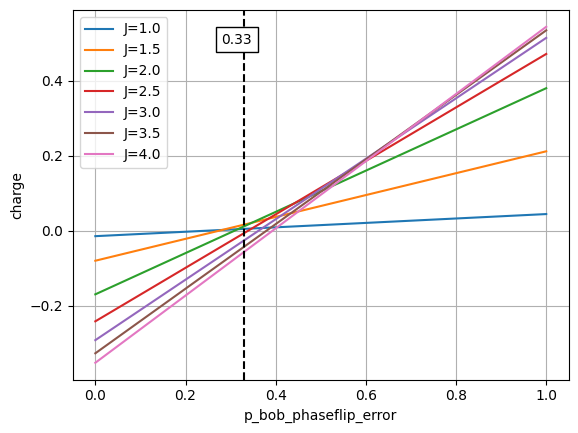}
    }
    \hfill
    \subfloat[Charge, \( N = 4 \)]{
        \includegraphics[width=0.3\textwidth]{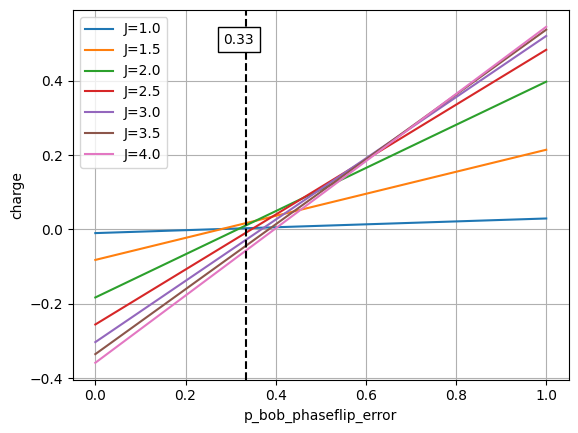}
    }
    
    \caption{Numerical simulation results for teleported expectation value vs. the probability $p$ of phaseflip at Bob's site for Nearest Neighbors Hamiltonian \(H^{(2)}\) with Alice base \(\sigma_A = X_0\).}
    \label{fig:nn_X_num_bob_phaseflip_error_appendix}
\end{figure*}

\begin{figure*}[tb!]
    \centering
    \subfloat[Energy, \( N = 2 \)]{
        \includegraphics[width=0.3\textwidth]{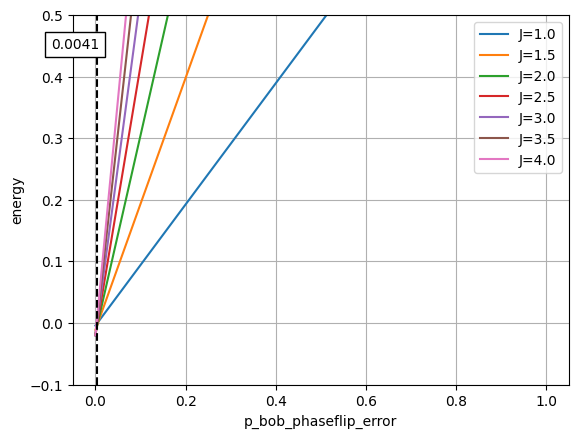}
    }
    \hfill
    \subfloat[Energy, \( N = 3 \)]{
        \includegraphics[width=0.3\textwidth]{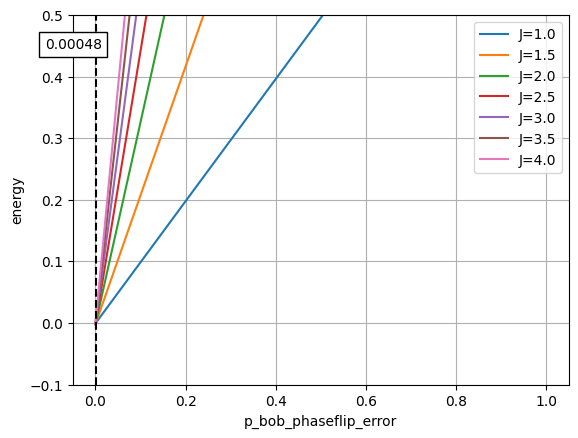}
    }
    \hfill
    \subfloat[Energy, \( N = 4 \)]{
        \includegraphics[width=0.3\textwidth]{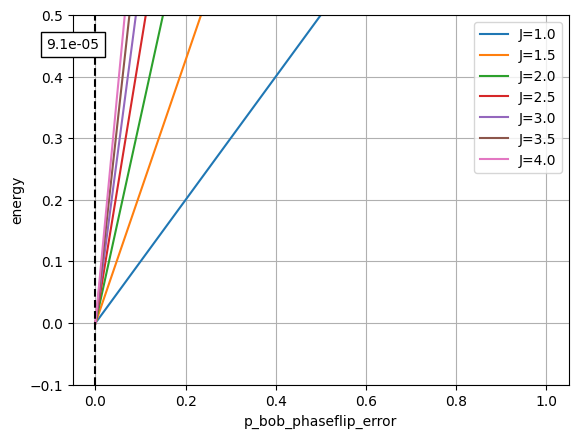}
    }

    \vspace{0.5em}

    \subfloat[Charge, \( N = 2 \)]{
        \includegraphics[width=0.3\textwidth]{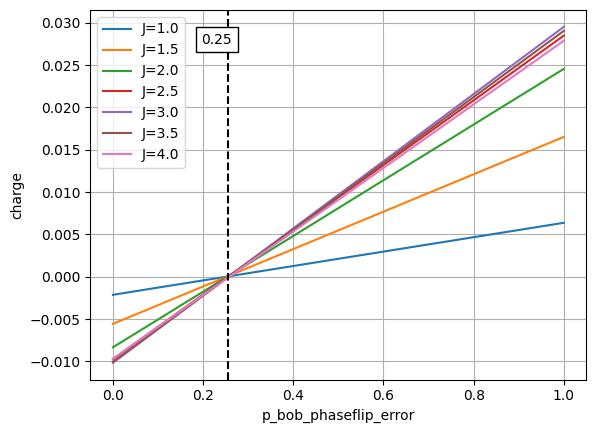}
    }
    \hfill
    \subfloat[Charge, \( N = 3 \)]{
        \includegraphics[width=0.3\textwidth]{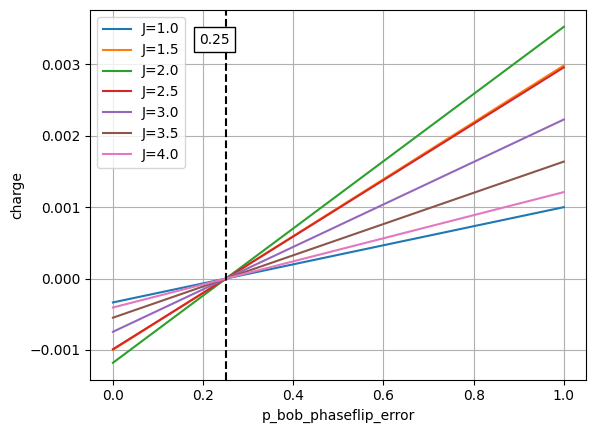}
    }
    \hfill
    \subfloat[Charge, \( N = 4 \)]{
        \includegraphics[width=0.3\textwidth]{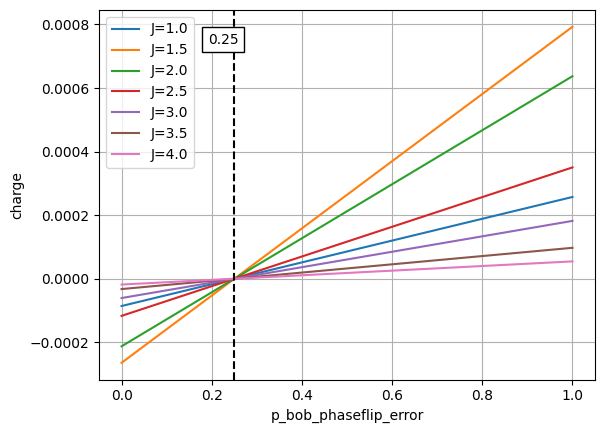}
    }
    
    \caption{Numerical simulation results for teleported expectation value vs. the probability $p$ of phaseflip at Bob's site for Nearest Neighbors Hamiltonian \(H^{(2)}\) with Alice base \(\sigma_A = Y_0\).}
    \label{fig:nn_Y_num_bob_phaseflip_error_appendix}
\end{figure*}

\begin{figure*}[tb!]
    \centering
    \subfloat[Energy, \( N = 2 \), \( \sigma_A = X_0 \)]{
        \includegraphics[width=0.225\textwidth]{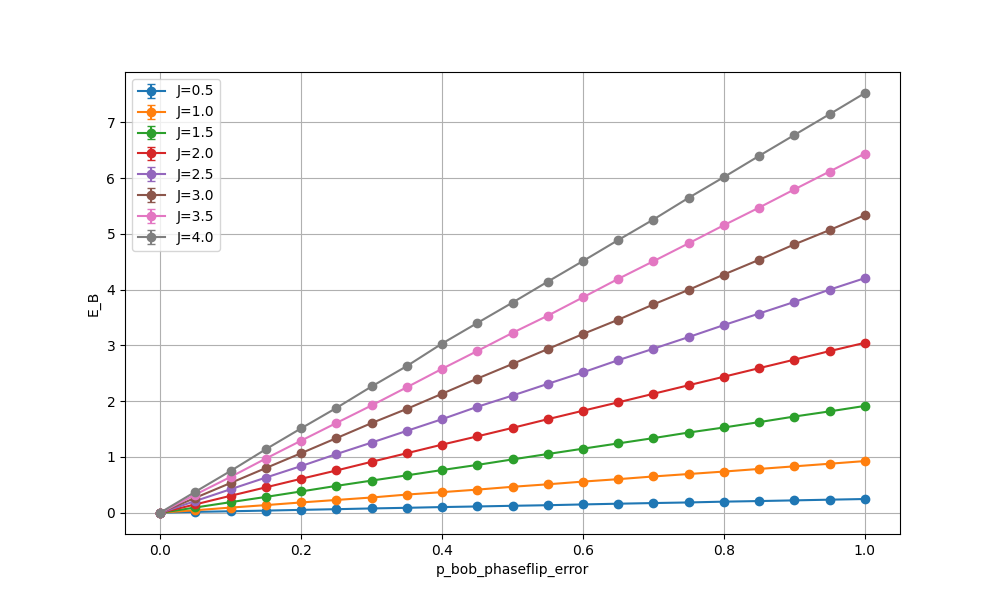}
    }
    \hfill
    \subfloat[Energy, \( N = 3 \), \( \sigma_A = X_0 \)]{
        \includegraphics[width=0.225\textwidth]{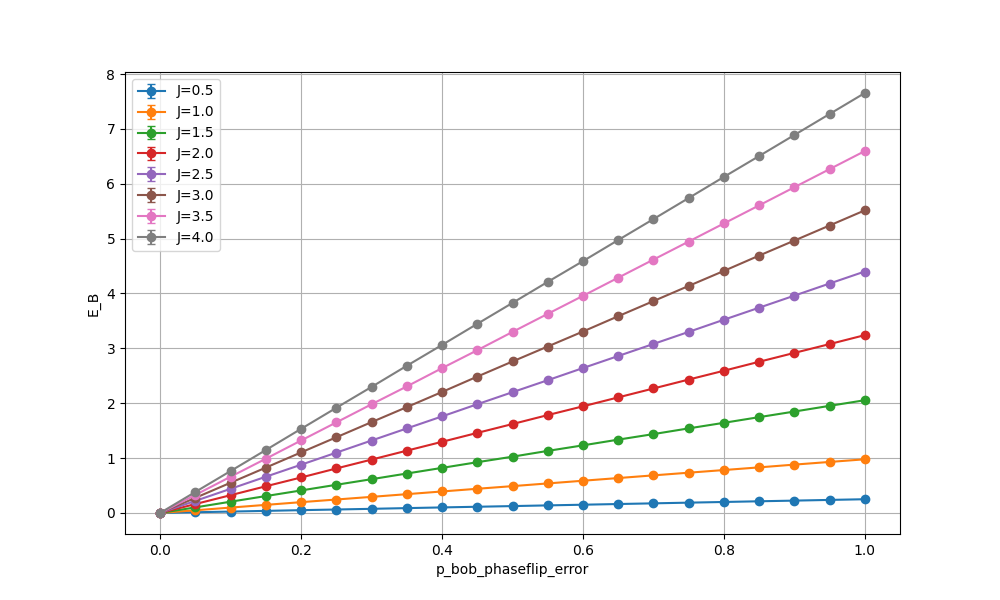}
    }
    \hfill
    \subfloat[Charge, \( N = 2 \), \( \sigma_A = X_0 \)]{
        \includegraphics[width=0.225\textwidth]{plots/errors/qiskit/N2_charge_vs_p_bob_phaseflip_error.png}
    }
    \hfill
    \subfloat[Charge, \( N = 3 \), \( \sigma_A = X_0 \)]{
        \includegraphics[width=0.225\textwidth]{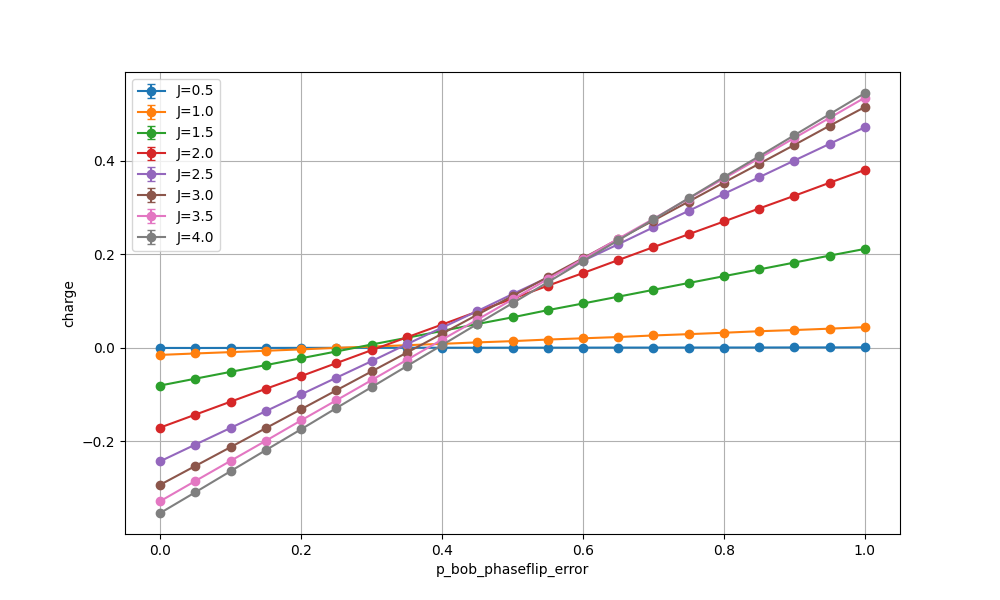}
    }

    \vspace{0.5em}

    \subfloat[Energy, \( N = 2 \), \( \sigma_A = Y_0 \)]{
        \includegraphics[width=0.225\textwidth]{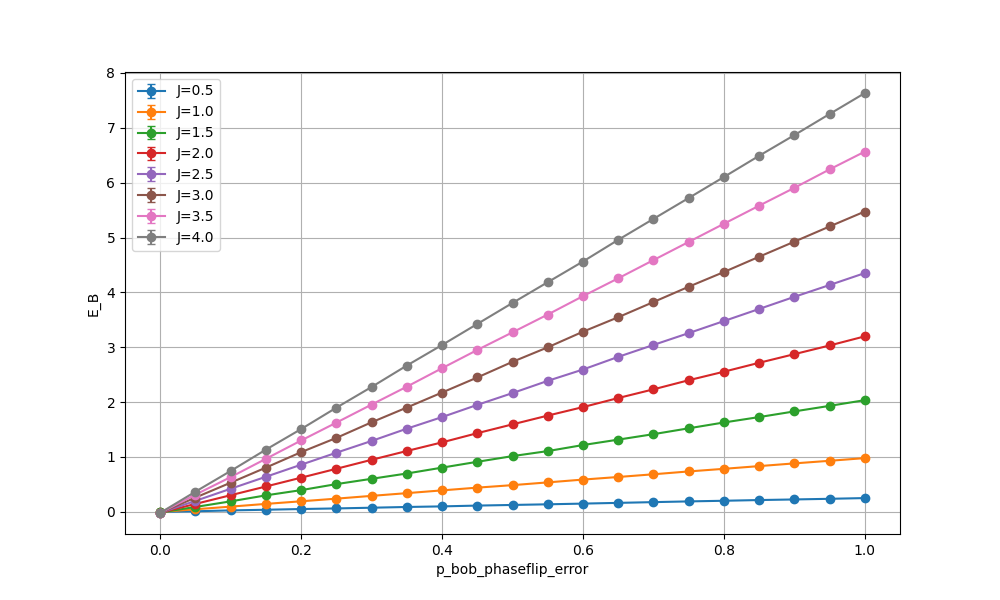}
    }
    \hfill
    \subfloat[Energy, \( N = 3 \), \( \sigma_A = Y_0 \)]{
        \includegraphics[width=0.225\textwidth]{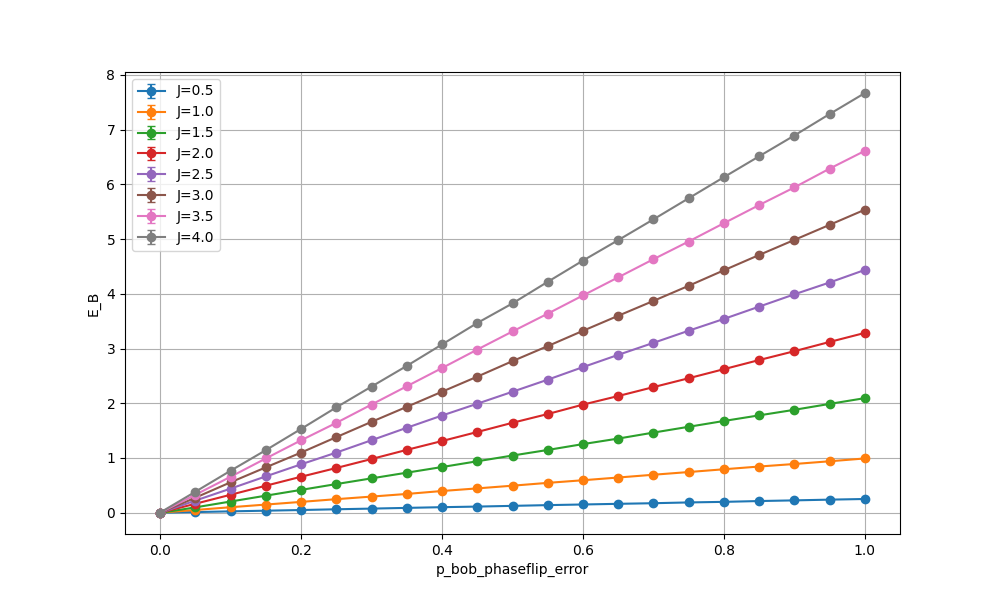}
    }
    \hfill
    \subfloat[Charge, \( N = 2 \), \( \sigma_A = Y_0 \)]{
        \includegraphics[width=0.225\textwidth]{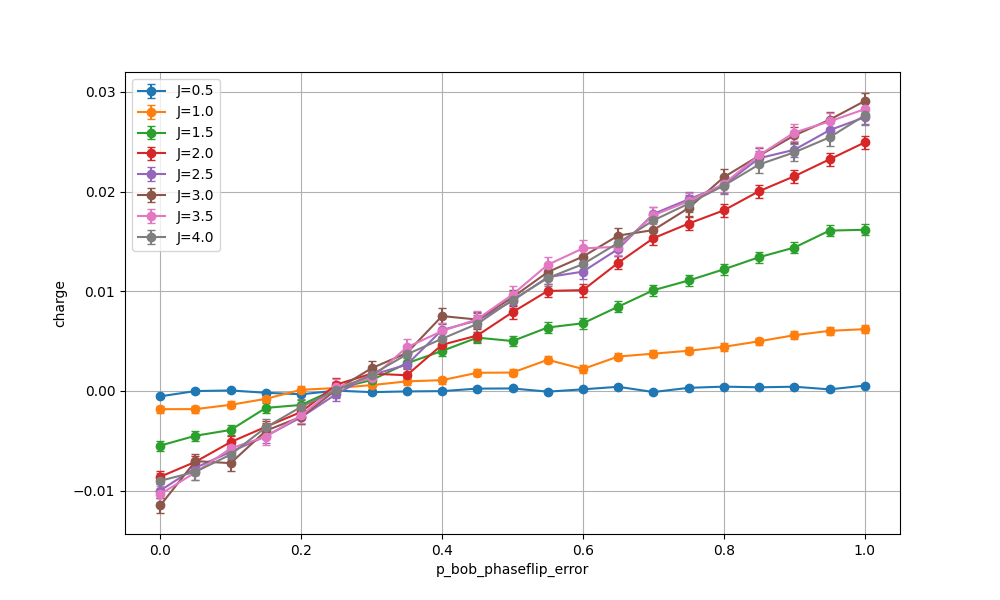}
    }
    \hfill
    \subfloat[Charge, \( N = 3 \), \( \sigma_A = Y_0 \)]{
        \includegraphics[width=0.225\textwidth]{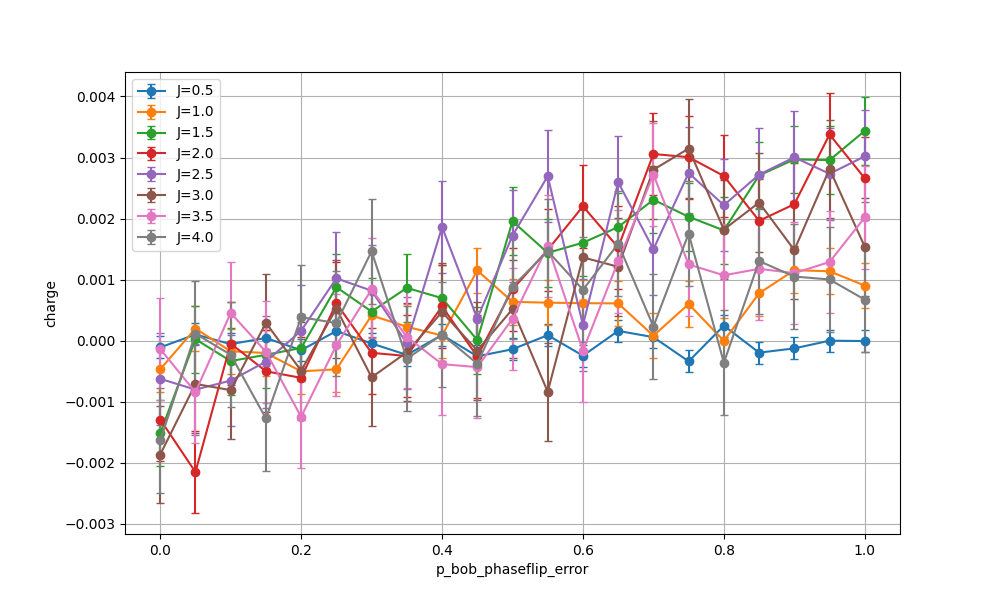}
    }
    
    \caption{Qiskit simulation results for teleported expectation value vs. the probability $p$ of phaseflip at Bob's site for Nearest Neighbors Hamiltonian \(H^{(2)}\) for both Alice bases.}
    \label{fig:nn_qiskit_bob_phaseflip_error_appendix}
\end{figure*}

A phase-flip error occurring at Bob's site ($n=N$), modeled by the channel $\rho \rightarrow (1-p)\rho + pZ_{N}\rho Z_{N}$, is particularly detrimental to the protocol. Unlike an error on Alice's qubit which degrades the measurement, an error on Bob's qubit directly sabotages his corrective operation.

The physical mechanism for this vulnerability lies in how the phase-flip operator interacts with Bob's conditional rotation. Bob's unitary operation is typically a rotation around the Y or X axis (e.g., $U_B(\theta) = e^{-i\theta\sigma_B}$ where $\sigma_B \in \{X_N, Y_N\}$). A phase-flip error transforms this rotation via conjugation:
\begin{equation*}
Z_N R_Y(\theta) Z_N^\dagger = R_Y(-\theta)
\end{equation*}
This transformation inverts the sign of Bob's rotation angle. The effect is physically equivalent to Bob receiving the wrong classical bit from Alice and applying the incorrect unitary based on that misinformation \cite{QKDbyQET}. This causes a direct and rapid mixing of the intended negative signal with the unwanted positive signal, leading to an accelerated decay towards failure.

This behavior is clearly demonstrated in the simulation results presented in Figure \ref{fig:nn_qiskit_bob_phaseflip_error_appendix}.
\begin{itemize}
    \item All plots show a strong, nearly linear degradation of the teleported signal as a function of the error probability $p$. This rapid decay is characteristic of an error that directly inverts the protocol's intended outcome.
    
    \item This error leads to a sharp sign-crossing, with a critical failure threshold consistently observed around $p \approx 0.32-0.33$. This low tolerance highlights the protocol's sensitivity to phase noise on the receiving qubit.
    
    \item Both energy and charge protocols are similarly vulnerable to this error, as it attacks the final step of the protocol (Bob's conditional rotation) which is fundamental to both. The Qiskit simulations (Figure \ref{fig:nn_qiskit_bob_phaseflip_error_appendix}) validate this severe impact, confirming that this vulnerability is a primary concern for practical hardware implementations.
\end{itemize}
In conclusion, a phase-flip error on Bob's qubit is one of the most damaging forms of local noise for this QKD protocol, as it effectively mimics a classical communication error and causes a rapid failure of the key generation process.



\clearpage

\bibliography{references}

\end{document}